\documentclass[twocolumn]{aa}
\usepackage[varg]{txfonts}
\usepackage{graphics,graphicx}
\usepackage{multirow}
\usepackage{eurosym}
\usepackage{enumitem}
\usepackage{natbib}      
\usepackage{dashrule}    
\usepackage{aalongtable}
\def\msun{M$_{\odot}$}

\def\asec{\ifmmode ^{\prime\prime}\else$^{\prime\prime}$\fi}
\def\amin{\ifmmode ^{\prime}\else$^{\prime}$\fi}

\def\farcs{\hbox{$.\!\!^{\prime\prime}$}}  
\def\lsim{\mathrel{\rlap{\lower4pt\hbox{\hskip1pt$\sim$}}
    \raise1pt\hbox{$<$}}}                
\def\gsim{\mathrel{\rlap{\lower4pt\hbox{\hskip1pt$\sim$}}
    \raise1pt\hbox{$>$}}}                

\newcommand{\ie}[0]{$\textnormal{i.e. }$}
\newcommand{\sub}[1]{_{\textnormal{#1}}}
\newcommand{\tn}[1]{\textnormal{#1}}

\def\degs{\ifmmode ^{\circ}\else$^{\circ}$\fi}

\newcommand{\gp}{\mbox{$g^\prime$}}
\newcommand{\rp}{\mbox{$r^\prime$}}
\newcommand{\ip}{\mbox{$i^\prime$}}
\newcommand{\zp}{\mbox{$z^\prime$}}
\newcommand{\yp}{\mbox{$y^\prime$}}

\begin{document}

\title{Quasar clustering at redshift 6}

\titlerunning{Quasar clustering at redshift 6}

\author{J. Greiner\inst{1}, J. Bolmer\inst{1}, R.M. Yates\inst{2, 3},
  M. Habouzit\inst{4, 5},
  E. Ba\~nados\inst{4}, P.M.J. Afonso\inst{6}, P. Schady\inst{7}
}

\authorrunning{Greiner et al.}

\offprints{J. Greiner, jcg@mpe.mpg.de}

\institute{
  Max-Planck Institut f\"ur extraterrestrische Physik, 
    Giessenbachstr. 1,  85748 Garching, Germany
  \and
  Max-Planck Institut f\"ur Astrophysik, Karl-Schwarzschildstr. 1, 85748 Garching, Germany
  \and
  University of Surrey, Guildford, Surrey GU2 7XH, United Kingdom
  \and
  Max-Planck Institut f\"ur Astronomie, K\"onigstuhl 17, 69117 Heidelberg, Germany
  \and
  Zentrum f\"ur Astronomie der Universit\"at Heidelberg, ITA,
   Albert-Ueberle-Str. 2, 69120 Heidelberg, Germany
  \and
  American River College, Physics \& Astronomy Dpt., 4700 College Oak Drive, Sacramento, CA 95841, USA
  \and
  Department of Physics, University of Bath, Bath BA2 7AY, United Kingdom
}

\date{Received 12 March 2021  / Accepted 20 July 2021 }

\abstract{Large-scale surveys over the last years have revealed 
  about 300 QSOs at redshift above 6. Follow-up observations
  identified surprising properties, such as the very high
  black hole (BH) masses, spatial correlations with surrounding cold gas
  of the host galaxy, or high C\ion{IV}-Mg\ion{II} velocity shifts.
  In particular, the discovery of luminous high-redshift quasars
  suggests that at least some black holes likely have large masses at
  birth and grow efficiently.
}
{
  We aim at quantifying quasar pairs at high redshift for a large sample of
  objects. This provides a new key constraint on a combination of parameters
  related to the origin and assembly for the most massive black holes:
  BH formation efficiency and clustering, growth efficiency and relative
  contribution of BH mergers.}
{We observed 116 spectroscopically confirmed QSOs around redshift 6
  with the simultaneous 7-channel imager GROND in order to search for
  companions. Applying identical colour-colour cuts as for those which led
  to the spectroscopically confirmed QSO, we perform Le\,PHARE fits to the
  26 best QSO pair candidates, and obtained spectroscopic observations
  for 11 of those.
}
{We do not find any QSO pair with a companion brighter than
  M$_{1450}$ (AB)$< -26$ mag within our 0.1--3.3 $h^{-1}$ cMpc search radius,
  in contrast to
  the serendipitous findings in the redshift range 4--5.
  However, a low fraction of such pairs at this luminosity and redshift is
  consistent with indications from present-day cosmological-scale galaxy
  evolution models.
  In turn, the incidence of L- and T-type brown dwarfs which occupy a
  similar colour space as $z \sim 6$ QSOs, is higher than
  expected, by a factor of 5 and 20, respectively.
}
{}


\maketitle

\section{Introduction}

Clustering of quasars provides one of the most important observational 
constraints to probe their physical properties, formation
and evolution \citep{Haiman+2001}
as well as the mass of their dark matter (DM) haloes \citep{Sheth+2001}.
As easily detectable objects throughout the Universe, quasars trace the
underlying dark matter distribution \citep[e.g.][]{HaehneltNusser1999},
and thus provide a powerful test of hierarchical structure formation
theory \citep[e.g.][]{Fang1989}.
How this structure formation in the early Universe proceeds in detail
is far from understood. How many massive galaxies form in a single massive
dark matter halo?
Do all of these massive galaxies have a massive black hole (BH)? Do all these
black holes accrete from the rich gas reservoir, or what are the duty cycles
of QSO activity? Do massive galaxies form sub-structures and/or multiple
massive black holes?

Particularly pressing questions over the last years are that of the formation channels and efficiencies of massive BHs, and of their growth rates with cosmic time \citep[e.g.][]{Johnson+2013, Valiante+2017, Valiante+2018, Inayoshi+2020}.
Quantifying the number of QSO pairs
at high-z is one observational avenue to constrain a combination of parameters of BH formation and BH growth rate.
High-z QSOs can be as massive as the most massive BHs found in the local Universe \citep[e.g.][]{Banados+2018,Yang+2020,Wang+2021},
and so must have grown several orders of
magnitude in less than 800 Myrs. Indeed, BH formation mechanisms predict the formation of BH seeds with at most masses of a few ${\sim}10^{5-6}\rm \, M_{\odot}$ \citep{Volonteri2010, Inayoshi+2020}.
If the growth of quasars was mostly through gas accretion, then they must have grown at the Eddington rate a large fraction of their lives. Finding QSO pairs, or the absence of, can help us understand if BH mergers play a role in their growth, and if dense environments can sustain efficient growth for several QSOs in the same region of the sky, or if baryonic processes (e.g., AGN feedback) prevent them from doing so \citep[][for QSO feedback preventing galaxy formation/growth in their surroundings]{Wyithe+2005, Kashikawa+2007, Utsumi+2010, Costa+2014, Habouzit+2019}. 
Radiation from a QSO can impact a region from sub Mpc to tens of Mpc, depending on the QSO mass, accretion rate, and properties of its surrounding environment, and could therefore also inhibit the accretion activity of nearby BHs.
In addition, quantifying the number of QSO pairs also provides us with constraints on BH formation efficiency, i.e. whether BH formation takes place in a large number of halos, but also on whether the formation of massive seeds could be clustered. The existence of many pairs of QSOs could indeed imply that massive seeds can form close to each other.
Constraints from QSO pairs on QSO origins and growth with time are likely degenerated, but are key to improve our understanding of these extreme objects.

Quasar clustering has been investigated at different scales and depths,
among others using the Sloan Digital Sky Survey
\citep{Shen+2007, Hennawi+2010}, the 2dF QSO Redshift 
and Dark Energy Survey \citep{Porciani+2004, Chehade+2016}, or the
Canada-France-Hawaii Telescope Legacy Survey (CFHTLS) \citep{McGreer+2016}.
Studies at different depth for a given redshift range are important
since a large luminosity dependence is theoretically predicted at
higher redshifts \citep{Hopkins+2007}.
Clustering at large scales, sampling Mpc or larger distances,
is stronger at high redshifts ($z>3.5$) than at low redshifts
($2.9 < z < 3.5$) \citep{Shen+2007}.
Clustering at small scales, sampling the 10 kpc - 1 Mpc scale
environment, also increases from smaller ($z\sim3$) to larger ($z\sim4$)
redshift, but shows a much stronger amplitude on kpc scales than on
Mpc scales, e.g. a factor 4 stronger at $z \sim 2$ \citep{Eftekharzadeh+2017}.

Clustering measurements at high redshift ($z > 3$) are confronted with
a problem: close quasar pairs are extremely rare.
For comparison, the mean quasar separation at $z>3$ is $\sim$ 150h$^{-1}$ Mpc
(given the standard luminosity function of $\Phi \sim 10^{-6}$ Mpc$^{-1}$ mag$^{-1}$ for $M{\rm (AB)}< -24$; e.g. \citealt{Willot2010, Ross+2013, Matsuoka+2018, Shen+2020})
and at small scales ($\lsim$ 1 Mpc), the correlation function does not
increase as fast as the volume decreases \citep{Hennawi+2010}.
However, direct observations of pairs might suggest differently - the
three hitherto
highest-redshift QSO pairs were all found serendipitously:
(i) \cite{Schneider+2000} reported a pair at z=4.26 with a separation of
  33\asec\ (corresponding to 160$h^{-1}$ kpc proper on the sky) which
  was found in the slit while spectroscopically confirming another QSO
  candidate;
  (ii) \cite{Djorgovski+2003} discovered a z=4.96 quasar at a separation of
  196\asec\ (or proper transverse separation of 0.9 h$^{-1}$ Mpc) from a
  z=5.02 quasar discovered by \cite{Fan+1999}, and
(iii) \cite{McGreer+2016}  reported a pair at z=5.02 with a separation of
  21\asec\ (90$h^{-1}$ kpc), discovered in a faint $z \sim 5$ quasar
  (but not pair)
  search down to $i = 23$ mag, and selected simultaneously with identical
  colour criteria.
Given the relatively small sample sizes which these
discoveries are based upon, they suggest that quasars are even stronger
clustered at redshifts $>4$ on small scales.
This in turn has been used to argue that feedback is very inefficient
at high redshift \citep{Willot2010, McGreer+2016}, since
inefficient feedback would correspond to the correlation length
to flatten out at high redshift, while concordant QSO and host halo mass
growth would imply a sharply rising correlation length.
In addition, merger models of galaxy formation predict
a large fraction of binary QSOs, at least over some time span
\citep[e.g.][]{Volonteri+2003}, with the fraction of binary (off-centre
and dual) QSOs increasing with redshift \citep{Volonteri+2016}.
This state of knowledge, and the importance of the conclusions which can
be drawn, certainly
warrants further searches for clustering properties at high redshift.

Here, we search a sample of 116 spectroscopically confirmed QSOs
at redshift 6 \citep[collected from the compilation of][]{Banados+2016}
for a companion.
Throughout this paper we use the best-fit $Planck$ cosmological model with 
$\Omega_m = 0.286$, $\Omega_\Lambda = 0.714$, and $h = 0.69$ \citep{Planck2015}.
At a mean sample redshift of 6, an angular separation of 1\asec\
corresponds to a proper (comoving) transverse separation of 4 $h^{-1}$ kpc
(28 $h^{-1}$ kpc).
All magnitudes are in the AB system.

\begin{figure}[th]
 \centering
 \includegraphics[width=0.90\columnwidth,viewport=1 1 525 525, clip]{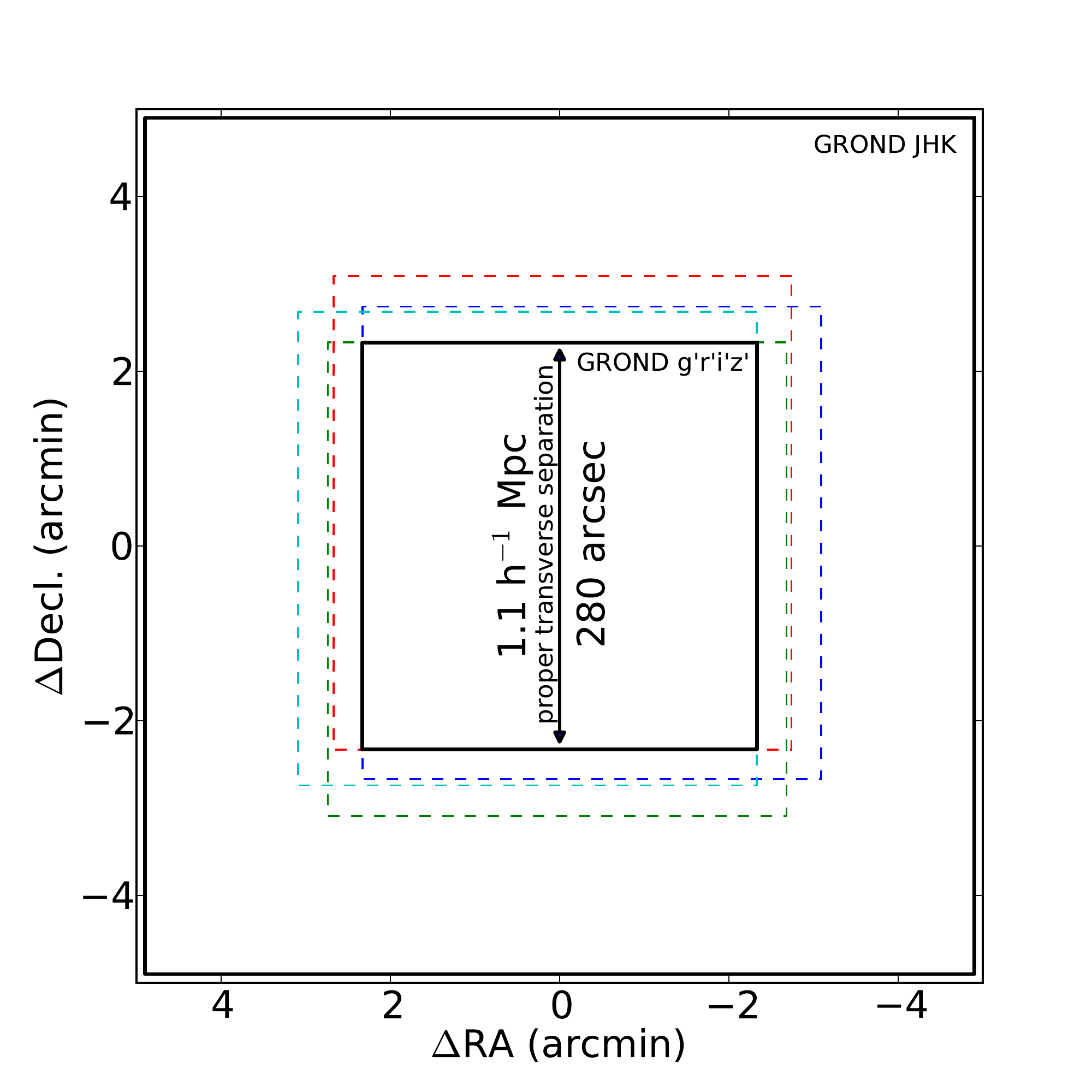}
 \vspace{-0.2cm}
 \caption{A schematic of the GROND optical ($g'r'i'z'$) and near-infrared
    ($JHK$) channel spatial coverage at a redshift of 6. The dashed boxes
    are the respective $g'r'i'z'$ FOV for each individual telescope 
    dither position; a similar pattern applies to the $JHK$ channels 
    (not shown). With most of our targets being within 1 arcmin from
    the centre of the FOV, 
    and since we require the \rp\ip\zp\ bands,
    our conservative search radius is 80\asec, corresponding to
    about 471 $h^{-1}$ kpc proper, or 3.3 $h^{-1}$ Mpc comoving.}
 \label{fov}
\end{figure}

\section{Selection criteria, observations and data analysis}

\subsection{Selection}

From the sample of 173 spectroscopically confirmed QSOs at $z>5.6$ compiled
in \cite{Banados+2016} we have selected all 144 sources south of
Decl. $< +30$\degs, thus being visible from La Silla (Chile). 
We have added 6 objects from \cite{Mazzucchelli+2017},
and 2 object from \cite{Jiang+2016}.
For the observing campaigns, we have prioritised
brighter sources, so that it is easier (less time consuming) to reach
limiting magnitudes fainter than the spectroscopically confirmed QSOs.
After weather and technical losses, we obtained useful new GROND observations
for 77 sources, in addition to those 42 objects from the \cite{Banados+2016}
sample, which had already been observed with GROND in 2013--2015 during
the selection process towards their spectroscopic sample.
Excluding in the following two QSOs at $z>6.6$,
our sample of 116 QSOs spans a redshift range from 5.6--6.6 and a \zp-band
brightness of 18.6--23.1 mag.

\subsection{Imaging Observations and Data Analysis}

\subsubsection{Strategy, Data Analysis and Source Detection}

Simultaneous imaging in $g'r'i'z'JHK_{\rm s}$ with GROND \citep{gbc08} 
of the 77 new fields was performed at six different time slots: 
2016 August 14 -- September 3, 
2016 October 11 -- 25, 
2017 January 24 -- February 7,
2017 April 22 -- May 2,
2017 September 18 -- 21, and
2018 March 23.
Since most fields have SDSS or Pan-STARRS coverage, no further observations for
photometric calibration were obtained. The remaining three fields were
calibrated against the SkyMapper Survey (DR1.1).

\begin{figure*}[ht]
\includegraphics[width=1.0\textwidth,viewport=25 1 1205 350 clip]{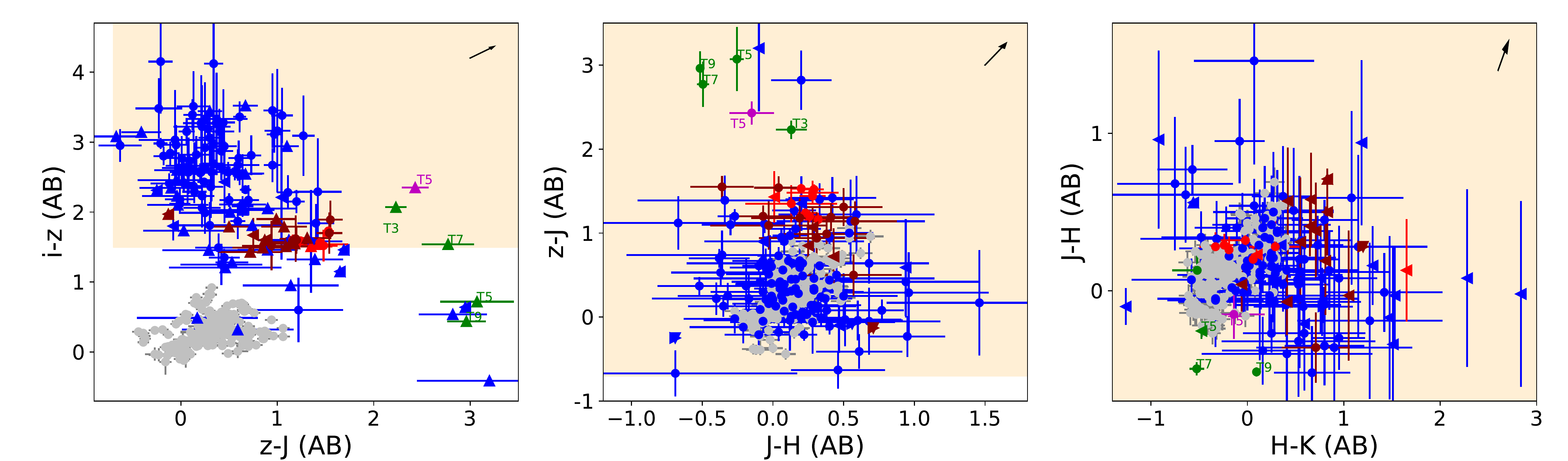}
\caption{Colour-colour diagram in the GROND-intrinsic $i'z'JHK_s$ bands
  of the prime quasars (blue; triangles denote limits) compared to
  normal foreground stars and galaxies (grey) of the same fields,
  with selected brown dwarfs from earlier GROND observations indicated in
  green and purple, and labeled with their spectral type.  Our 26 candidates
  are shown in red, with light red those 9 objects which have ALLWISE
  detections.
  The yellow-shaded region indicates the search region for the QSO pair.
  The arrow at the top right of each panel shows the amount by which
  a correction for $A_{\rm V} = 1$ would shift an object. Except for the grey
  objects, all are foreground-$A_{\rm V}$ corrected. 
  }
 \label{cc}
\end{figure*}

Since the field of view of the 4 visual channels $g'r'i'z'$ is smaller
than that of the NIR channels (Fig. \ref{fov}), the former
define the search distance from the spectroscopically identified QSO to
80\asec\ distance. Thus, our maximum search radius for a QSO companion
is 0.471 $h^{-1}$ Mpc proper or 3.3 $h^{-1}$ Mpc comoving (see Fig. \ref{fov}).
This accounts for the fact, that most of the primary QSOs are not
exactly centred in the GROND field of view.

The question to answer before the start of our observing campaign was:
how much deeper to look for a companion, relative to the prime QSO?
We have picked a 2 mag difference as threshold, based on three reasons:
(i) given the QSO luminosity function \citep{Willot2010, Ross+2013, Shen+2020},
this corresponds to a factor more than 10 in the number of objects, and
should imply a large enough sample depth,
(ii) optical variability is rare above $\sim$0.5 mag amplitude, ensuing that
the effective difference is not diminished substantially,
and (iii) the pair at $z=5$, found by \cite{McGreer+2016} also has such
an  \ip-band magnitude difference (19.4 vs. 21.4 mag).
Thus, exposure lengths of typically 20 minutes was decided upon, ensuring
that the 3$\sigma$ limiting magnitude in the $z'$-band was at least 2 mag
deeper than
the brightness of the spectroscopically identified QSOs (see below).
While this 2 mag difference might suggest that the search
could have been done just on Pan-STARRS data with its typical 5$\sigma$
depth of \zp $<$22.3 mag, only 11 PS1 QSOs are bright enough to also
survive the colour cut of \ip -- \zp\ $>$ 1.5 mag at that faint \zp\ brightness
(see below).

GROND data have been reduced in the standard manner \citep{kkg08}
using pyraf/IRAF \citep{Tody1993, kkg08b}.
The optical/NIR imaging was calibrated against the Sloan Digital Sky Survey
(SDSS)\footnote{http://www.sdss.org} \citep{Eisenstein+2011},
Pan-STARRS1\footnote{http://pan-starrs.ifa.hawaii.edu} \citep{Chambers+2016}
or the SkyMapper Survey\footnote{http://skymapper.anu.edu.au} \citep{Wolf+2018}
catalogs for $g^\prime r^\prime i^\prime z^\prime$, 
and the 2MASS catalog \citep{Skrutskie+2006} for the $JHK_{\rm s}$ bands.
This results in typical 
absolute accuracies  of $\pm$0.03~mag in $g^\prime r^\prime i^\prime 
z^\prime$ and $\pm$0.05~mag in $JHK_{\rm s}$. 
Since the GROND dichroics were built after the Sloan filter system
  \citep{gbc08}, the colour
  terms are very small, below 0.01 mag, except for the \ip-band
  which is substantially narrower than the SDSS i'-band. All our analysis
  is done in the SDSS system, with the following transformations used:
  $\gp_{SDSS} - \gp_{GROND} = (-0.006\pm0.014) + (0.015\pm0.025) * (\gp_{SDSS} - \rp_{SDSS})$,
  $\rp_{SDSS} - \rp_{GROND} = (-0.004\pm0.004) + (0.012\pm0.015) * (\rp_{SDSS} - \ip_{SDSS})$,
  $\ip_{SDSS} - \ip_{GROND} = (-0.023\pm0.010) + (0.216\pm0.054) * (\ip_{SDSS} - \zp_{SDSS})$, 
  $\zp_{SDSS} - \zp_{GROND} = (-0.003\pm0.005) - (0.009\pm0.027) * (\ip_{SDSS} - \zp_{SDSS})$.
For the fields calibrated against Skymapper and PS1, we use their conversion to SDSS \citep{Scolnic+2015}. 
The PS1 \zp-band has a substantial colour term
(\zp$_{GROND}$ = \zp$_{PS1}$ - 0.214$\times$(\zp$_{PS1}$ - \yp$_{PS1}$) due to
the missing tail beyond 920 nm, but this becomes important only for
redshift above 6.6 when Ly-$\alpha$ moves beyond the \zp-band limit.)

The calibrated \zp- and $J$-band images for each field were stacked after
re-sampling due to their different pixel sizes (0\farcs15 for the visual,
and 0\farcs59 for the near-infrared);  for the four objects at $z>6.5$ we
just used the $J$-band images. A source detection procedure on
each stacked image then provides a list of all objects per field which
then is used as input for 'forced' photometry on the individual 7-band
images. This provides upper limits for the bands where the sources are
not detected; otherwise it performs PSF photometry for \gp\rp\ip\zp,
  and aperture photometry in $JHK$ (with a radius of 1$\times$FWHM)
  at the position of the source in the
master catalog which is allowed to re-centre by $\pm$0.3 of the PSF width.
Table \ref{log} contains the details of the observation for each QSO,
including the 3$\sigma$ limiting magnitude in the $z^\prime$-band.
Table \ref{phot} contains the foreground-A$_{\rm V}$ corrected AB magnitudes
of all 116 QSOs in the individual GROND filter bands.

\subsubsection{Colour-colour selection of Candidates}

With these GROND data in hand, the goal was to search the field around
each spectroscopically confirmed QSO for a nearby source with a similar colour,
taking advantage of the simultaneous multi-filter
optical/NIR spectral energy distributions (SEDs) of all objects inside the
field of view (FOV).

We implicitly assume that the searches leading to the discovery of
the QSOs have picked the brighter of the pairs; this is corroborated
by modelling which suggests that current flux-limited surveys of QSOs
at high redshift preferentially detect objects at their peak luminosity,
and thus miss a substantial population of similarly massive black holes
accreting at lower accretion rates \citep{Costa+2014}.
Thus, here we are looking for objects fainter than the original QSO.

In all 116 fields we apply the following two colour selections (all in AB): 
\ip -- \zp\ $>$ 1.5 mag, and \zp -- $J$ $>$ --0.7 mag.
We checked that there are no detections in the \gp\ and \rp\ bands.
For the $z>6.5$ objects, we adjust the \zp -- $J$ colour to --0.5,
add $J-H$ $<$ --0.5, and require a \ip\ non-detection.
This returns 74 new high-z quasar candidates.
We excluded 48 candidates through visual
inspection of the 7-channel image cut-outs of these candidates, based on
either source confusion, i.e. either a nearby source with overlapping PSF,
or an extended source which could either be an interloper galaxy, or a
non-resolved equally bright pair of objects. The remaining 26 candidate
sources are shown in Fig. \ref{cc} as red (light and dark) dots,
overplotted over the distribution of the spectroscopically confirmed
quasars (shown in blue), which by itself show that our 
  colour cut criteria recover all but 13 objects (with most of these
  failing due to non-detections, i.e. upper limits, in the \ip-band,
  i.e. deeper \ip-band exposure would likely recover those as well).

For 9 of the remaining 26 candidate sources we found catalog entries
in the ALLWISE all-sky catalog \citep{Wright+2010, Mainzer+2011}
within 1 arcsec distance, and
after visual checking of the images accepted these as matching objects.
Tables \ref{photWise} and \ref{photnoWise} contain the photometry of these
26 candidates, separately for those with and without ALLWISE matches.

We then performed Le\,PHARE \citep{Arnouts+1999, Ilbert+2006}
fits to the photometry of the 26 candidates (including the ALLWISE magnitudes
when available).
We included an alternative IGM opacity model \citep{Songaila+2004}
in addition to the default one \citep{Madau+1999}, but otherwise used
the default spectral templates except for adding about 200 cloned
high-z QSO templates based on stacked observed SDSS spectra of low-z
QSOs, similar to \cite{Willott+07}.
The goal of these Le\,PHARE fits was to
distinguish late-type dwarfs from QSO candidates, where the distinction
is based on two observational features: the steepness of the SED in the
\ip-\zp-$J$ range, and the SED shape in the $HK$ range which is much
redder for dwarfs. Since dwarfs in our sensitivity range would be
at distances of order 100 pc, their photometry would not be affected
by interstellar dust. We thus created two SEDs for each candidate, one
without A$_{\rm V}$-correction (for Le\,PHARE fits with dwarf templates)
and one with full galactic A$_{\rm V}$-correction (for Le\,PHARE fits with
QSO templates). For 14 (out of the 26) candidates the best-fit template is
that of a dwarf (see last column of Tabs. \ref{photWise}, \ref{photnoWise}),
for 10 candidates that of a QSO at high redshift (labeled in boldface
in Tabs. \ref{photnoWise}),
for one a galaxy at intermediate redshift, and for 1 candidate
(SDSSJ2054-0005\_4) the reduced $\chi^2$ does not provide a clear
favorite. The Le\,PHARE fits of the 10 high-z quasar candidates
(i.e. with A$_{\rm V}$-correction applied) are shown in Fig. \ref{sed},
which also contains the best-fit photo-$z$ (6th field in the red line
of labels), all within 13\% of the spectroscopic redshift except
for PSO J055.4244-00.8035 (the source we did not get a spectrum of).

\subsection{Spectroscopic Observations}

We obtained spectroscopy of 11 objects (9 of the 10 QSO candidates,
the galaxy, and the unidentified object), mostly with the GMOS instruments
of the Gemini Observatory (under proposal IDs GN-2019B-FT-201 and
GS-2019B-FT-202). One object was observed with the FIRE (Folded-port
InfraRed Echellette) spectrograph in Prism mode at the Magellan Baade
telescope.
A log of the observations is given in Tab. \ref{speclog}.
At Gemini-North (South), a Hamamatsu detector was used with the
R150 grating and filter G5308 (G5326).
The spectra cover the range from 615 (750) to 1080 nm.
Wavelength calibration was provided by CuAr comparison lamps.
Flux calibration was done against the standard stars G191B2B (LTT7987).
For the FIRE spectrum, we used archival A0V stars for telluric and
flux calibration.
The data were reduced in the standard manner, using GMOS specific
routines provided within the IRAF \citep{Tody1993} package,
and a custom-made python code for the FIRE observation.

For object PSOJ071.0322-04.5591  no trace is visible in the 2D spectrum;
for the other
sources the optimally-extracted 1D spectra are shown in Fig. \ref{spectra}.
None shows a sign of the Ly-$\alpha$ 'step' at the wavelengths
suggested by the photometric redshifts (vertical lines in Fig. \ref{spectra}).

\section{Results and Discussion}

\subsection{Pair candidates}

We do not find a single QSO pair candidate brighter than
M$_{1450}$ (AB)$< -26$ mag
in our GROND data of 116 spectroscopically confirmed redshift 6 quasars.
Except for two candidate object (near PSOJ055.4244-00.8035 and PSOJ071.0322-04.5591) we have obtained
optical spectroscopy for all other Le\,PHARE-derived QSO pair candidates,
but could not confirm the quasar nature as suggested by the colour
selection and SED fitting. This is somewhat surprising,
since the (post-facto) photometric redshifts we derived with Le\,PHARE 
for the prime QSOs where in good agreement with the spectroscopic
redshifts reported by \cite{Banados+2016}, with only a 5\% fraction
mismatch. We consider it very unlikely
that there is an unrecognised quasar among our identified
candidates in Tabs. \ref{photWise} and \ref{photnoWise}.
This is in contrast to our empirical expectation
(at the start of the project) of 3 pairs
in our sample, based on the serendipitously discovered pair at
z=4.26 \citep{Schneider+2000} and the two pairs at
z=5.02 \citep{Djorgovski+2003, McGreer+2016} (see details in the
introduction), which when averaged together
(1/100, 1/14, 1/47) suggest one pair every 40 QSOs, or 2.5\%.
This line of reasoning includes a few simplifications.
  First, it uses post-facto arguments and guesses on the search
  volume of the above serendipitous discoveries.
  Next, it assumed that the quasar clustering fraction does not
  evolve with redshift --  the question to be answered by the project.
  Also, it ignored the different luminosity ranges sampled at different
  redshifts - the integral over the luminosity function is certainly
  the most sensitive component contributing to the $w_{\rm p}$ statistic
  (see below).

\begin{figure}[ht]
\includegraphics[width=0.95\columnwidth,viewport=1 1 420 310, clip]{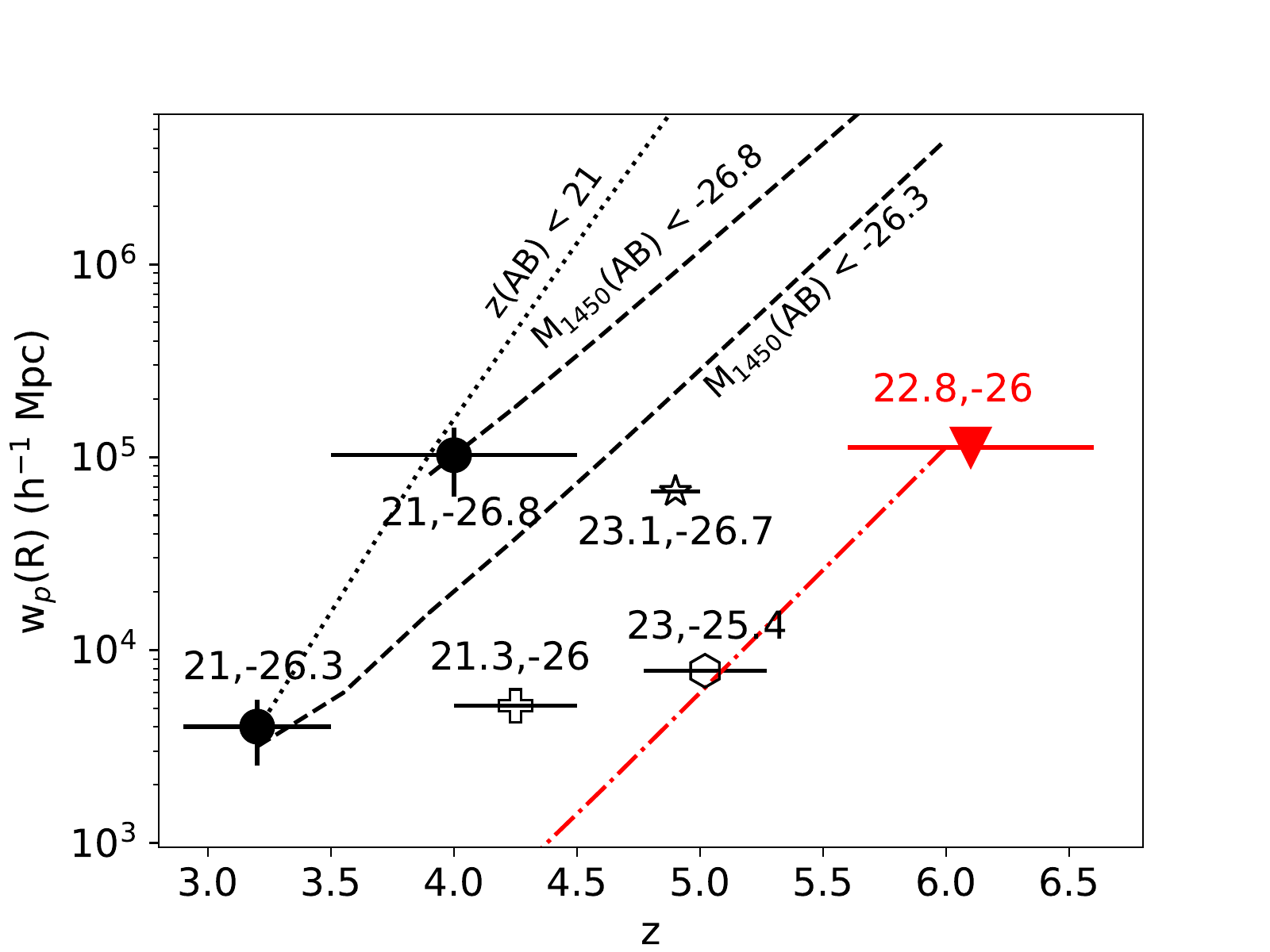}
\caption{Our upper limit (red triangle) on the projected correlation
  function of $z \sim 6$ QSOs, as derived from the GROND observations.
  The two filled circles are the corresponding findings by
  \cite{Shen+2010} of 7 and 8 pairs, respectively,
  modified for the different
  luminosity function (and cosmology) used over time, as well as adapted to our
  search scale of 0.1--3.3 h$^{-1}$ Mpc according to the $-2$ powerlaw
  scale dependence.
     Open symbols indicate the serendipituous discoveries from
  \cite{Schneider+2000}, \cite{Djorgovski+2003}, and
  \cite{McGreer+2016},
  also extrapolated to our search scale.
  Horizontal 'error' bars visualise the corresponding redshift
    range of the studies.  The labels at the data
  points report the limiting i/z-band AB magnitude of the search
  and the corresponding absolute M$_{1450}$(AB) limiting magnitude. 
  The two black dashed lines depict w$_p$(R) for the
  same rate of (7 and 8, respectively)
  pairs as in \cite{Shen+2010} extrapolated to higher redshift
  with their same M$_{1450}$(AB), and unchanged luminosity function,
  suggesting higher w$_p$(R) at larger redshift.
  Similarly, the dotted line shows the
  extrapolation for the case of constant limiting $z$-band magnitude.
  The red line depicts the backwards extrapolation of our upper limit 
  to lower redshifts, at our M$_{1450}$(AB)$< -26$ mag sensitivity, 
  demonstrating the mismatch with respect to the findings at lower redshift.
    All the lines are computed via integration over the luminosity function
    of \cite{Shen+2020} in redshift-steps of 0.25, and proper redshift-dependent
    transformation between observed \zp\ and absolute magnitudes.
 \label{ProjCorr}}
\end{figure}

In order to put this in context, we follow \cite{Hennawi+2006}
and \citet[][in particular their eq. 4]{Shen+2010} in estimating the
projected correlation function, or $w_{\rm p}$ statistics, as defined by e.g.,
\cite{DavisPeebles1983}. Given the mean seeing in our images and
correspondingly the ability to separate two sources, we take an inner
radius of 100 $h^{-1}$ kpc (corresponding to 3\asec) for the
projected comoving area of the cylindrical annulus (the outer one
being our search radius of 3.3 $h^{-1}$ Mpc).
We use the latest luminosity function description of \cite{Shen+2020},
as well as their publicly provided tools (https://github.com/gkulkarni/QLF)
  to convert our observed \zp(AB)
  magnitude limit and mean redshift to  M$_{1450}$, and applying
  the corresponding bolometric correction (using their 'global fit A';
  we verified our procedure against their Fig. 5).
This results in a number density of 9$\times$10$^{-10}$ Mpc$^{-3}$ for
quasars with M$_{1450} < -26$ mag, for our redshift range $5.6<z<6.6$.
We obtain an upper limit of $w_{\rm p} < 110\,000$ $h^{-1}$ Mpc as shown in
Fig. \ref{ProjCorr}, where we use a formal upper limit of our search of 1 pair
out of the 116 objects searched for, given the missing
spectroscopy for PSOJ055.4244-00.8035.
The figure also includes measurements at lower redshifts from
\cite{Schneider+2000}, \cite{Shen+2010} and \cite{McGreer+2016}
(modified by the updated luminosity function and corrected for
our search scale according to the $\gamma = 2$ powerlaw dependence of
$w_{\rm p}(R) \propto R^{-\gamma}$ from \cite{Shen+2010}).
Overall, our upper limit corresponds to a similar $w_{\rm p}$ as
previous findings  at lower redshift,
suggesting that there is no strong rise of clustering towards high redshift.

Using eq. 5 in \cite{Shen+2007} and $\gamma = 2$, our upper limit
  on $w_{\rm p}$ corresponds to
  a limit of the auto-correlation length $r_o < 1000 h^{-1}$ Mpc.

\begin{table*}[ht]
  \caption{Comparison of search volumes and depth for previous,
    serendipituously found high-z QSO pairs.}
   \vspace{-0.2cm}
     \begin{tabular}{lccccc}
     \hline
     \noalign{\smallskip}
     Publication & No. of pairs & z-range & separation      & depth & comments \\
                 &             &          & (arcsec)/(cMpc) & (AB mag) & \\
     \noalign{\smallskip}
     \hline
     \noalign{\smallskip}
     \cite{Shen+2010}      & 7 & 2.9--3.5  & ~~~-- / 1.0 & $i<21.0$ & 1 \\
                           & 8 & 3.5--4.5  & ~~~-- / 1.2 & $i<21.0$ & 1 \\
     \cite{Schneider+2000} & 1 & 4.0--4.5  & ~~33 / 1.2 & $i<21.0$ & 2 \\
     \cite{McGreer+2016}   & 1 & 4.77--5.27& ~~21 / 0.8 & $z<21.4$ & 3 \\
     \cite{Djorgovski+2003}& 1 & 4.8--5.0  & 160 / 6.2  & $z<23.1$ & 4 \\
     this paper            & 0 & 5.6--6.6  & 3--80 / 3.3  & $z<22.8$ & \\
     \noalign{\smallskip}
     \hline
     \noalign{\smallskip}
     \end{tabular}
     
    \small{ \noindent Notes on comments column: \\
     $^1$
     Out of 319 binary candidates, follow-up spectroscopy 
     ``is about half finished'' \citep{Hennawi+2010}.
     Based on the spectroscopic follow-up, \cite{Shen+2010} derive
     a search completeness of 0.38 and 0.52 for the low-z and high-z
     sub-sample, respectively. These completeness fractions are used
     in computing $w_p$ for Fig. \ref{ProjCorr}. \\
     $^2$ The second QSO of the pair was serendipituously in the slit,
     so we use the pair separation of 33\asec\ depite that this
     does not correspond to an azimuthally complete search cone.\\
     $^3$ The second QSO of the pair was serendipituously in the slit,
     so we use the pair separation of 21\asec. \\
     $^4$ The redshift depth is taken as the highest redshift QSO found
     in the parent sample. The search radius is estimated from the
     imaging field of view
     and guessing a 1 arcmin loss in the side-length due to dithering.
    }
   \label{searchvol}
\end{table*}

\subsection{Completeness and biases}

In the past, clustering studies have typically relied on selecting
pairs from large catalogs (e.g. SDSS), and the most important question
is about the completeness of the underlying catalog
\citep[e.g.][]{Hennawi+2006, Eftekharzadeh+2017}. Here, we have
obtained follow-up observations of known quasars at known redshift.
Thus, in our case the question of completeness reduces to the questions
of {\bf (i)} whether or not the selection criteria which led to the sample
of these known redshift quasars had a bias against selecting
binary quasars?
and {\bf (ii)} whether our search strategy misses QSOs? 
As to item (i), one potential bias is that the seeing limit in the
  original survey data
  affects the discovery of QSO on 1-2\asec\ scales due to nearby
  contaminating sources. As this is a statistical problem, this is not
  expected to affect the fraction of binary QSO in the spectroscopically
  confirmed QSO sample.

 As to item (ii), this splits into two sub-questions: First, what
  is the completeness of our colour-colour cuts? Since we ran our colour-colour
  selection blindly on all sources in each observation, we can answer
  this with the recovery
  rate of the spectroscopically confirmed QSO: out of the 116 QSO,
  we miss 9, with 3 of those not detected in \zp, and the other 6
  due to insufficiently deep \ip-band limits. Since the 3 non-detections
  are due to variability, they are not counted here (see below on the
  variability bias), so our incompleteness of the colour-colour selection
  is about 5\%.
  Second, what is the completeness of the Le\,PHARE fits in recognizing
  the QSO nature? Again, this can be answered with the sample of the
  spectroscopically confirmed QSOs: ignoring sources with detections
  in less than 3 photometric bands, we ran Le\,PHARE fits on the remaining
  97 spectroscopically confirmed QSOs in the same manner as with the
  candidates (i.e. two runs, one with and one without A$_V$ correction).
  Only 5 sources are classified as dwarfs, all other 92 as QSOs. We
  double-checked that the colours and photometric errors of those 5
  are not outliers with respect to the sample distribution of the 92
  correctly classified QSOs. So this corresponds to only $<$5\%
  incompleteness.

Further, we note that it is unlikely that even substantial optical
variability of the QSOs   
would affect the fraction of pairs, since statistically
one expects the same number of ``risers'' vs. ``faders''.
The only exception would be effects due to Malmquist bias,
but since only a handful of objects is detected with a \zp-band
uncertainty larger than 0.2 mag, this imbalance is unimportant.

Lastly, a potential bias could be due to dust obscuration of AGN,
in our case obscuring the companion of the spectroscopically confirmed QSO.
Again, this is not expected to be a major issue. First, previous suggestions
were that obscuration could primarily happen in the highest mass accretiion
phases, one would first expect the brighter object of the pair to be obscured,
rather than the lower-luminosity companions of the unobscured spectroscopicall
confirmed QSOs. Second, while ALMA observations revealed substantial dust
at high redshift, there is no evidence that  these dusty objects are
obscured AGN (see sect. 3.4 below). Finally, 
theory of dust production fails to explain dust in large amounts at $z>5$ 
\citep{LesnMichal2019}, preventing quantitative estimates or simulations.

Another potential bias could come in if black holes of similar mass
have different Eddington ratios at different redshift, such that
only the optically brightest QSOs are found at high redshift.
However, \cite{Mazzucchelli+2017} have shown  that
(at least for the biased high-z sample they considered)
BH masses and Eddington ratios of the $z>6.5$ and
SDSS $0.35 < z < 2.25$ samples are consistent in a luminosity-matched
sample.
Thus, our assumption was that 
the clustering fraction at redshift 4-5 \citep{Schneider+2000, McGreer+2016}
is not much different than at redshift 6, leading to our expectation
of more pairs as compared to our findings.

Further, our search comes with the biases, that
the luminosity difference between the two members of the pair
can of course be larger than our selected 2 mag, due to at least two
different reasons: firstly, the BH masses and/or accretion rates
can differ substantially, and secondly the Ly$\alpha$ emission which 
contributes to the $z$-band magnitude, can
be drastically different between the two members of the pair,
replicating the differences in the spectroscopically confirmed
QSO sample \citep[see e.g.][]{Banados+2016}.
As to the first reason, simulations suggest differently.
Based on the large-scale cosmological hydrodynamical Horizon-AGN
simulations, \cite{Volonteri+2016} find
that the luminosity ratio between the central BH (the more massive one)
and an off-centre BH is within a factor of ten (rms = 5x)
for the cases of $L_{bol} > 10^{43}$ erg/s and BH masses up to 10$^8$ \msun.
While this corresponds to the ratio we employed, it should be noted
that this ratio depends on selected input parameters for the simulation,
among others a mass ratio $<$1:6 for the merging galaxies
at which AGN activity is triggered in both black holes.
Similarly, \cite{Bhowmick+2019} find that in the MassiveBlack II simulations
the satellite quasar luminosity is similar to that of central quasars.
Finally, we note that for 41 of our targets the 3$\sigma$ limiting
\zp-magnitude reached in
our observations is actually 3 mag fainter than that of the
spectroscopically confirmed QSO, which still would lead to
an expectation of one pair in our sample.

\subsection{Comparison to simulations}

The upper limit to the projected correlation function for bright QSO pairs at $z\sim{}6$ presented in Section 3.1 serves as an important constraint for galaxy evolution models. Currently, cosmological-scale simulations typically either focus on quasar properties below $z\sim{}5$ due to improved statistics (e.g. \citealt{DeG17,Bhowmick+2019,H21}) or above $z\sim{}7$ due to reduced computational limitations (e.g. \citealt{DiMatteo+2017,T19,M20}). Zoom-in simulations of haloes containing QSOs at $z\sim{}6-7$ have been performed (e.g. \citealt{Costa+2014,Lupi+19}), however these do not readily provide statistics on large scales. Consequently, model predictions for the total number of bright QSO pairs at the redshift of our sample are still lacking in the literature.

At lower redshift, models suggest that bright QSO pairs are rare. For example, in the \textsc{MassiveBlack-II} simulation box of volume $(100\,h^{-1}\,\tn{Mpc})^{3}$, \citet{Bhowmick+2019} find no pairs of QSOs with $g < 20.85$ within a search radius of 4 Mpc at $z=2$, and only three such systems by $z=0.6$.

At higher redshift, work on the \textsc{BlueTides} simulation has shown that QSOs bright enough to match observations are also very rare, even when accounting for large black hole seeds \citep{DiMatteo+2017}. There is only one black hole of mass above $10^{8}$ \msun\ present in the \textsc{BlueTides} $(400\,h^{-1}\,\tn{Mpc})^{3}$ box at $z=8$, which has a luminosity of $L\sub{bol}\sim{}2.2\times{}10^{46}\tn{erg/s}$ \citep{DiMatteo+2017,T19}.

However, despite the lack of pairs, galaxy evolution models still find that brighter QSOs cluster more strongly than their fainter counterparts, particularly on smaller scales. For example, on scales of $\sim{}10$ kpc at $z=4$, \citet{DeG17} find the two-point auto-correlation function, $\xi(r)$, for AGN with $L\sub{bol} \geq 10^{44}$ erg/s in the \textsc{Illustris} simulation is higher by a factor of three compared to that for all AGN. On the larger scales probed by our observational measurement ($\sim{}0.1-3.3\,h^{-1}\,\tn{Mpc}$), this enhancement is reduced but still remains present for the brightest objects, as also seen in the \textsc{L-Galaxies} semi-analytic model \citep{B09} and the \textsc{Horizon-AGN} simulation \citep{Volonteri+2016}. All three of these models report a value of $\xi(r=1.2\,h^{-1}\,\tn{Mpc}) \sim{} 10$ at $z=4$ for AGN with $L\sub{bol} \geq 10^{44}$ erg/s, with the expectation that this value will be higher at $z=6$, as also seen in the observations discussed in Section 1.

In order to better probe the exact epoch and search volume of our observational QSO sample, we have carried-out a census of bright AGN pairs at $z=6$ in the \textsc{Illustris-TNG300} (hereafter, TNG300) and \textsc{L-Galaxies 2020} galaxy evolution models. TNG300 \citep{P18,Springel+18,Naiman+18,Marinacci+18,Nelson+18} is a magneto-hydrodynamical simulation run in a $(205\,h^{-1}\,\tn{Mpc})^{3}$ box, using a Bondi-Hoyle-Lyttleton BH accretion model with no boost factor but an assumed BH seed mass of $8\times{}10^{5}$ \msun\ (see \citealt{H21}). \textsc{L-Galaxies 2020} \citep{He20} is a semi-analytic model run on the larger $(480\,h^{-1}\,\tn{Mpc})^{3}$ \textsc{Millennium-I} box \citep{S05}, but with a more simplified phenomenological model of quasar- and radio-mode BH accretion and no BH seed mass (see \citealt{C06}). In both cases, we assume that all AGN are radiatively efficient when calculating their luminosities, which provides an upper limit on the expected number of bright pairs. Therefore, the AGN bolometric luminosity for both models is calculated here as
\begin{equation}
	L\sub{bol} = \frac{\epsilon\sub{r}}{1-\epsilon\sub{r}}\,\dot{\tn{M}}\sub{BH}\,c^{2}\ \ ,
\end{equation}
where $\epsilon\sub{r}$ is the assumed radiative efficiency of the accretion disc, $\dot{\tn{M}}\sub{BH}$ is the BH accretion rate, and $c$ is the speed of light. Given that the detection limit for the fainter of the two QSOs in our observational search is $M\sub{B} \sim{} -26$ mag (\ie{}$L\sub{B} \sim{} 8.25\times{}10^{45}$ erg/s), we assume a minimum bolometric luminosity of $L\sub{bol,min} = 4\times{}10^{46}$ erg/s when searching for BH pairs in the models, given the bolometric conversion factor of 5.13 recommended by \citet{D20}.

\begin{figure}[ht]
\includegraphics[width=0.48\textwidth, viewport=40 150 560 610, clip]{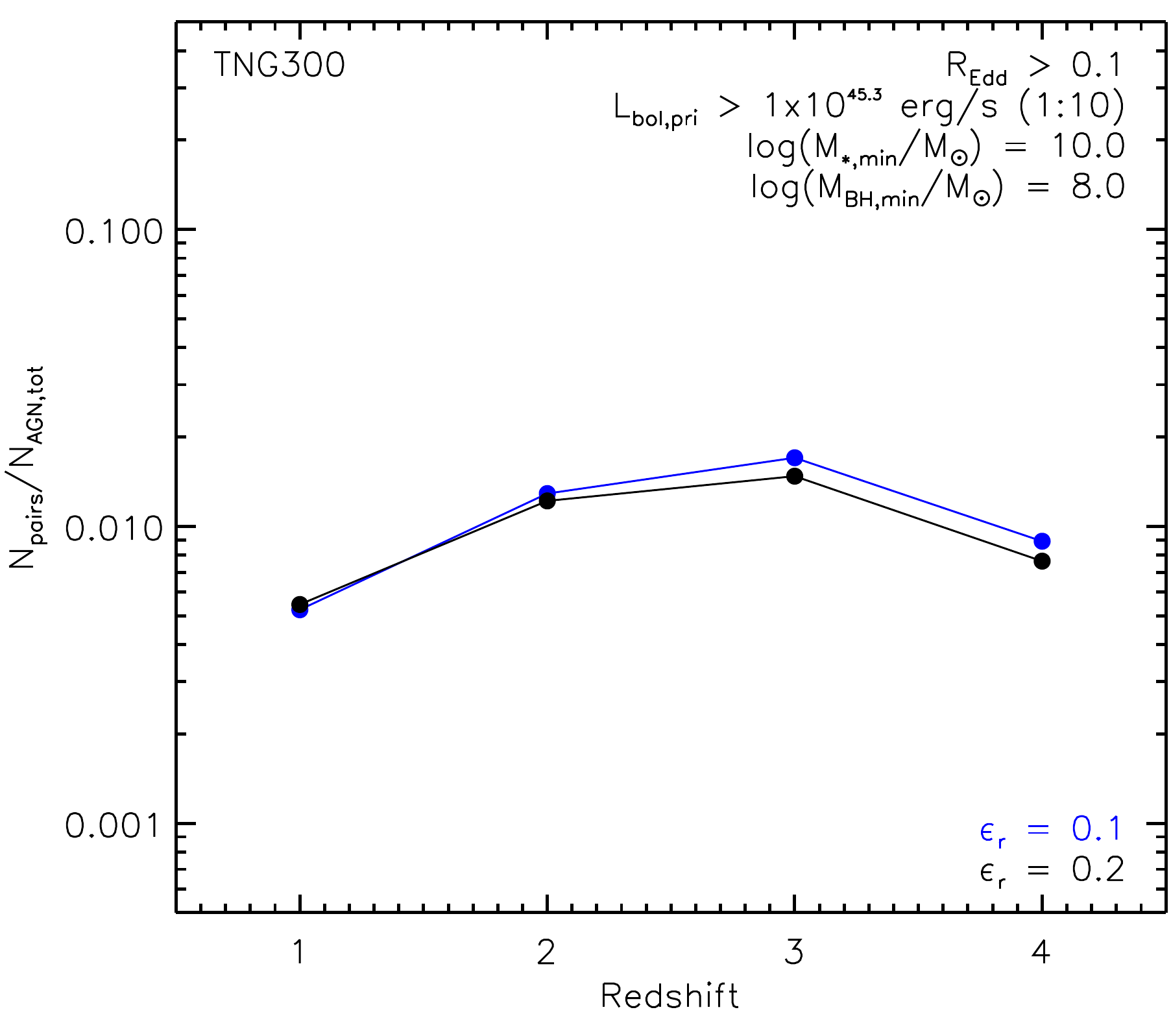}
\vspace{-0.3cm}
\caption{Fraction of BH pairs over redshift in TNG300, for two values of
  $\epsilon\sub{r}$. Following \citet{S20}, pairs are selected to have
  primary BH luminosities of $L\sub{bol,pri} \geq{} 10^{45.3}$ erg/s,
  secondary BH luminosities of $L\sub{bol,sec} \geq{} L\sub{bol,pri}/10$,
  a minimum BH mass of $10^{8}$ \msun, and a minimum host-galaxy stellar
  mass of $10^{10}$ \msun. At redshifts $>$5 there are AGN, but no pairs.
  }
 \label{pairfract}
\end{figure}

For the same search area as our observational sample (\ie{}an inner radius
of $125\,h^{-1}\,\tn{kpc}$ and outer radius of  $3.3\,h^{-1}\,\tn{Mpc}$ comoving, see Section 3.1), we find no BH pairs with $L\sub{bol} \geq 4\times{}10^{46}$ erg/s in either TNG300 or \textsc{L-Galaxies 2020} at $z=6$. This holds true even when doubling the assumed radiative efficiency from $\epsilon\sub{r}=0.1$ to $0.2$. Indeed, there are no BHs at that luminosity at all within the TNG300 box above $z=5$ when assuming $\epsilon\sub{r}=0.2$. This theoretical result is nicely consistent with the null result returned by our observational search at $z=6$, especially when considering the relatively large box sizes of \textsc{L-Galaxies 2020} and TNG300.

We do not find any BH pairs at $z=5$ in TNG300 or \textsc{L-Galaxies 2020} that match the serendipitous discovery by \citet{McGreer+2016} of a bright pair with $M_{1450} \leq{} -25.4$ (\ie{}$L\sub{bol} \geq 2.4\times{}10^{46}$ erg/s) separated by 21\asec{} (\ie{}816 ckpc) at $z=5.02$. These simulations are therefore somewhat in tension with this discovery, although the exact search area to use is not well constrained. However, we do find pairs in TNG300 that match the serendipitous 
discovery on a BH pair with $M\sub{B} \leq -23.7$ mag (\ie{}$L\sub{bol} \geq 5.1\times{}10^{45}$ erg/s) by \citet{Schneider+2000} at $z=4.26$. When assuming $\epsilon\sub{r}=0.1$, and a standard search volume around the mean deprojected separation of $r = 1.92$ cMpc with inner radius $r-(r/4) = 1.44$ cMpc and outer radius $r+(r/4) = 2.40$ cMpc, we find 14 BH pairs with $L\sub{bol} \geq{} 5.1\times
{}10^{45}$ erg/s at $z=4$ in TNG300. This equates to an upper limit on the pair fraction (\ie{}the number of BH pairs divided by the total number of BHs at that luminosity) of $\sim{}3.1$\%. We are also able to reproduce a fairly redshift invariant pair fraction of $\sim{}1$\% between $z\sim{}4$ and 1 (see Fig. \ref{pairfract}) for BH pairs with a primary BH luminosity of $L\sub{bol,pri} \geq 1\times{}10^{45.3}$ erg/s, when selecting systems as \citet{S20} do for the \textsc{Horizon-AGN} simulation (see their section 5.1). No such pairs were found in TNG300 above $z=4$.

In conclusion, the observed number of bright QSO pairs seen at high redshift appears largely consistent with that predicted by current cosmological-sized galaxy evolution models.
Although, further modelling efforts at the redshift of our observational sample are still required to confirm this.

\subsection{The number of QSO pairs as a constraint on massive BH origins and assembly}

The origin of $M_{\rm BH} \sim 10^{8-9}$ \msun\ black holes in massive galaxies at $z \sim 6-7$ is largely unconstrained
\citep[e.g.][]{Johnson+2013, Valiante+2017, Valiante+2018, Inayoshi+2020}.
The cosmic time since the Big Bang is likely too short for light seeds with masses of ${\sim}100$ \msun\ \citep[e.g., PopIII remnant model,][]{MadauRees2001, BrommLoeb2003, Volonteri+2003b} to grow to such masses even if the seeds accrete at the Eddington rate most of their lifetime.
Instead, more massive seeds with masses of order $\sim 10^4$ \msun\ would have a somewhat less challenging growth history to reach the masses of the high-z QSOs. 

Several theoretical models have been put forward to explain the formation of such massive BH seeds. By analogy with the nuclear clusters that we commonly observe in low-redshift galaxies, compact nuclear clusters in metal-poor environments are predicted to form very massive stars by runaway stellar collisions \citep{2008ApJ...686..801O,2009ApJ...694..302D,2009MNRAS.393..858R}. These very massive stars would collapse onto BH seeds of $M_{\rm BH}\sim 10^{2}-10^{3}\, \rm M_{\odot}$. Seeds of about the same masses can be formed by runaway black hole mergers in metal-poor star clusters in the centre of high-redshift galaxies. The direct collapse model predicts the most massive seeds, called direct-collapse BHs (DCBHs): atomic cooling halos form single supermassive stars, which can collapse onto massive seeds of $M_{\rm BH}\sim 10^{4}-10^{6}\, \rm M_{\odot}$ \citep{1994ApJ...432...52L, BrommLoeb2003, 2006ApJ...652..902,2006MNRAS.370..289B,2006MNRAS.371.1813L,2008MNRAS.391.1961D,Visbal+2014,2015MNRAS.452.1026L,Habouzit+2016}. In order to avoid the fragmentation of the gas into multiple and less massive stars, the presence or formation of efficient gas coolants (i.e., molecular hydrogen and metals) must be prevented. Therefore, the formation of the first DCBHs depends on the number of surrounding haloes
as the Lyman-Werner radiation (photons with energy in the 11.2--13.6 eV range)
needed to prevent H2 formation comes from surrounding star-forming galaxies \citep{2016MNRAS.456.1901}.

These mechanisms yield different galaxy occupation fractions, i.e. the probability for a galaxy to host a BH, and therefore could {\it a priori} lead to different predictions on the number of QSO pairs at high redshift. While the compact nuclear stellar cluster model and the model of runaway black hole mergers in clusters 
predict potentially large occupation fractions, i.e. the presence of BHs in many galaxies, the number density of direct collapse BHs is predicted to be small \citep[e.g.,][]{2008MNRAS.391.1961D,Habouzit+2016}. 
However, DCBHs could appear clustered.
Indeed, whether the large Lyman-Werner radiation is produced by several or a single star-forming halo, the high intensity required could irradiate several nearby halos, potentially forming DCBHs in some of these halos if metal-poor conditions are met. 
The need for clustered halo environments is not required for all the variations of the direct collapse model \citep[e.g., the synchronised pair model,][]{Visbal+2014}.
It has also been proposed that DCBHs themselves could ignite a runaway process of further DCBH formation \citep{Yue2014}. 
Indeed, 
DCBHs are expected to be Compton thick, and their reprocessed radiation outshines small high-redshift galaxies by more than a factor 10. Thus, once the first DCBHs form, their Lyman-Werner radiation could help providing surrounding halos with the needed conditions to form new DCBHs \citep{Yue2014}.
This process comes to an end when the host atomic-cooling halos are rapidly photo-evaporated by ionizing photons. 
An analytical model of these processes \citep{Yue2014} shows that a highly clustered spatial distribution of haloes is the pre-requisite to triggering the formation of several DCBHs in the same region.

The different mechanisms of BH formation are not {\it a priori} mutually
exclusive in the Universe, which therefore makes the interpretation of the
QSO clustering difficult. Moreover, successive mergers of halos with time
alter the initial halo occupation fraction and clustering of BHs at birth. 

Observing several QSO pairs in our sample would likely suggest that massive BH seeds can form in the same regions (on scales of up to a Mpc), and that they can grow efficiently in the dense environments in which we often find high-redshift QSOs.
The fact that we find no QSO pairs, i.e. no bright companion to any of our 116 spectroscopically confirmed QSOs could mean that (i) most QSOs are isolated objects with no massive galaxy in their surroundings, (ii) QSOs are often located in dense environments but nearby galaxies are devoid of BHs or devoid of massive BHs, or (iii) QSOs are often located in dense environments and nearby galaxies also host massive BHs, however these BHs have high variability/low duty cycle and no companion BHs were active when observed.

While the question of whether QSOs are embedded in the densest environments
is still subject to debate in observations
\citep[][and references therein]{Kim+2009, Husband+2013, Mazzucchelli+2016, McGreer+2016, Habouzit+2019}, the  hypothesis (i) of a large fraction of QSOs being isolated objects is unlikely. Theoretical models suggest that the first massive galaxies formed through mergers of gas-rich galaxies at very high redshift.
Recently, \cite{Decarli+2017} and \cite{Neeleman+2019} 
identified six $z>6$ QSOs with close,
gas-rich companions through ALMA observations of [CII] and dust emission;
5 of these are in our sample: SDSSJ0842+1218, PSOJ167.6415-13.4960,
SDSSJ1306+0356, PSOJ308.0416-21.2339 and CFHQSJ2100-1715.
These QSO-galaxy pairs prove that $z>6$ QSOs are not living isolated from the surrounding matter in their halos. This QSO-galaxy clustering is consistent with the quasar/LBG clustering
at $z\sim4$, see Fig. 3 in \cite{Decarli+2017}.
Recent deep X-ray observations of some of the above sources did
  not reveal potentially obscured quasars. So far, there is no strong
  evidence of these obscured companions of $z\sim6$ quasars to be
  obscured AGN \citep{Connor+2020, Vito+2021}.

Therefore, our non-detection of bright QSO-QSO pairs could hint towards an evolutionary scenario which prevents the formation of a massive accreting BH in the potentially gas-rich QSO companion. This could be due to tidal interaction of the primary QSO being more rapid than the accretion on the central BH of the companion. Feedback from the QSO could also reduce the gas reservoir of the companion galaxy through its lifetime, therefore diminishing the ability of a BH to efficiently accrete and become as massive as the BH powering the QSO.
Additionally, efficient stripping of gas from companion galaxies of massive systems is seen out to several virial radii at low redshift in cosmological simulations \citep{Ayromlou+2021a, Ayromlou+2021b}. If this effect is present in the neighbourhood of the brightest QSOs at high redshift, it could also contribute to a reduction in the gas reservoir of companion BHs.
The absence of QSO pairs could also be explained in case of low efficiency of BH formation, so that early massive BH formation happens only in a small fraction of galaxies, or 
alternatively if the formation of light seeds is dominant in the Universe. 
Due to their low masses, light seeds would have a hard time growing and even sinking efficiently to the potential well of their host galaxies. 
Indeed, searching for companions of the ten most massive BHs in TNG300 reveals
  none with masses $M_{\rm BH}$  $\geq 10^{7.7}$ \msun\ at z=5 and above. There is
  just one companion BH with a mass of $M_{\tn BH} \sim{} 10^{7.3}$ \msun\ at
  z=6, but with
  a luminosity of only $L\sub{bol} = 10^{43.3}$ erg/s, which is not even the
  brightest luminosity in the companion sample (which is $10^{45.1}$ erg/s),
  and below our search sensitivity. With its limited volume of
  (205 $h^{-1}$ cMpc)$^{3}$ 
  there is no environment in TNG300 that was able to build up two
  extremely massive and luminous BHs in the same region by $z \sim{} 6$.

Finally, this leads us to our hypothesis (iii) of very short coincident activity periods of both QSOs (see, e.g.,
\cite{Haiman+2001} for an early account of the lifetime-dependence of QSO clustering). In that case, QSO companion galaxies would also host massive BHs, but these BHs would not be efficiently accreting at the time of observation. In that case, the companion massive BHs would have grown by several orders of magnitude, potentially releasing a significant amount of energy in their host galaxies through cosmic time, which could explain an accretion history with a succession of ``on and off'' phases. In the observations presented here, we can only detect very bright objects. Further investigations at larger telescopes (e.g. with the GROND-descendant SCORPIO instrument at Gemini Observatory) will shed new light onto the degenerate scenarios to explain the absence of QSO pairs with $M_{1450} (AB)< -26$ mag.

\subsection{Incidence of Brown dwarfs}

The exposure for the spectroscopic observations was determined to
provide $>$5$\sigma$ detections of flux redwards of Ly-$\alpha$
(if it existed) binned at 200 \AA.
Thus, the signal-to-noise ratio in these spectra
is good enough to distinguish brown dwarfs and low-redshift galaxies
from QSOs, but way too small
to perform a spectral classification of the dwarfs or galaxies.
We therefore revert to the results of the Le\,PHARE fits, and provide
the spectral types of the best-fitting template in the last
columns of Tabs. \ref{photWise}, \ref{photnoWise}. For the conversion
from the Le\,PHARE best-fit temperature to spectral type we used
the ``mean'' value from the compilation of \cite{Kellogg2017}.
The associated error is $\pm$2 sub-types over most of the range.

Since after the selection of our targets for spectroscopic follow-up
  the unWISE \citep{Schlafly+2019} and CatWISE \citep{Marocco+2021}
  catalogs became available, we run a new
  cross-correlation of our targets against these catalogs.
  From our list of 17 candidates without ALLWISE counterparts
  (Tab. \ref{photnoWise}) we find 6 objects with W1 and W2 entries.
  Re-running Le\,PHARE fits with this extended photometry yields
  consistent results for three dwarf identifications (within the
  above mentioned  $\pm$2 sub-type error), but deviates for the
  other three objects:
  (i) two SEDs, previously best fit with a QSO or M8 dwarf template
  are now best fit with a low-redshift ($z \sim 1.2$ in both cases)
  galaxy, similar to ULASJ0148+0600 (Tab. \ref{photWise}),
  (ii) for another source, SDSSJ2054-0005\_3, the best-fit template
  is that of a galaxy at redshift 6, which is inconsistent with
  our absolute photometry. The second-best template (nearly identical
  $\chi^2$) is that of a QSO at redshift 6, which is unlikely
  given our GMOS spectrum, and the third-best template is that of
  a L2 dwarf, but with substantially worse $\chi^2$ - we therefore
  do not assign an identification in Tab. \ref{photunWise}.

For the candidate of PSOJ055.4244-00.8035, which also had a QSO
  template as best fit in Le\,PHARE but for which we did not
obtain a spectrum,
we refrain from changing the ID, though Occam's razor would argue
against a QSO.

\begin{figure}[ht]
 \vspace{-0.4cm}
 \centering
 \includegraphics[width=0.95\columnwidth,clip]{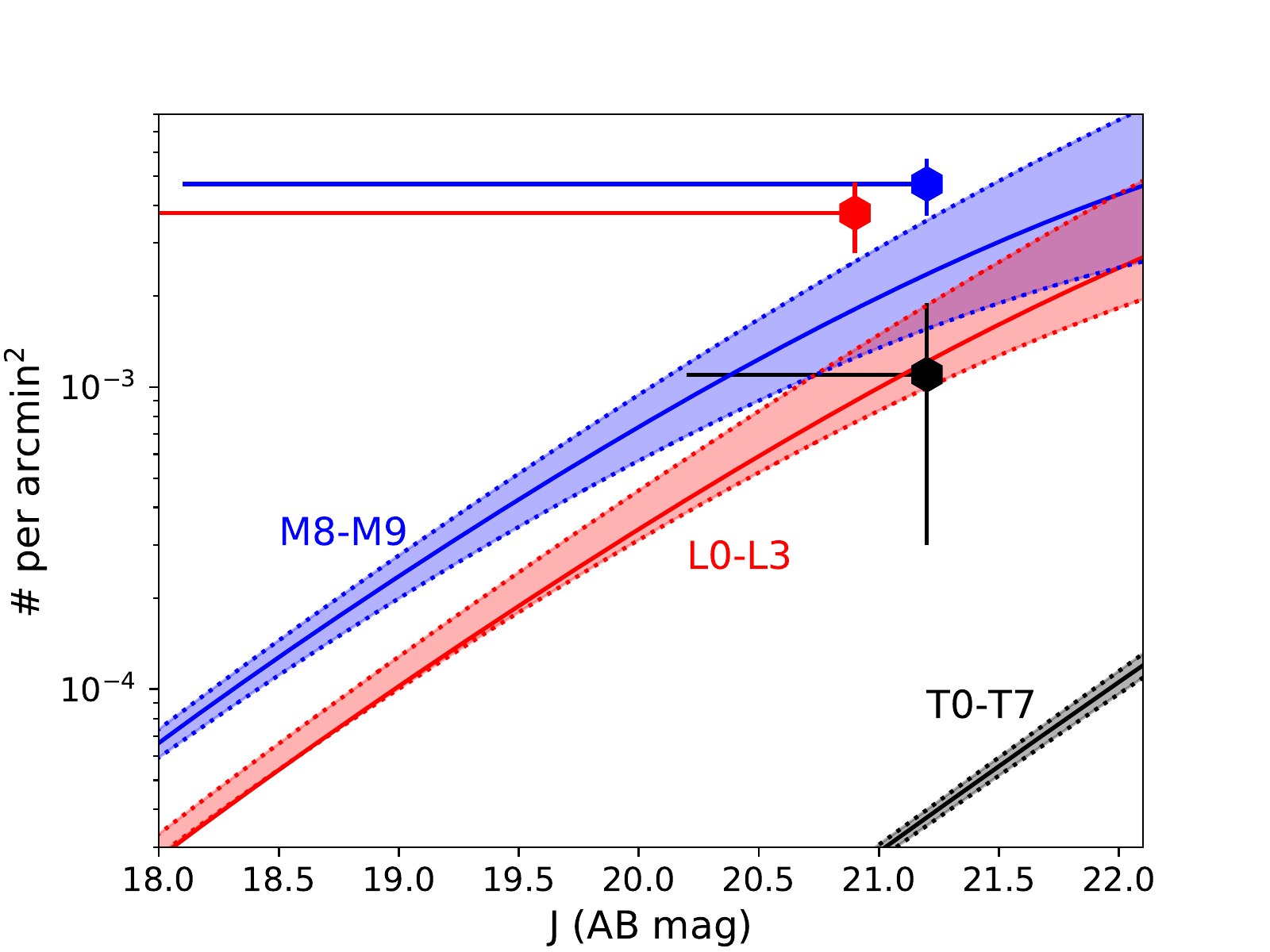}
 \caption{The number of brown dwarfs among our sample of 26 candidates
  per total area of our search region over 116
  QSOs, compared to the cumulative surface density as modeled by
  \cite{RyanReid2016},
  plotted here for their COSMOS sample field (central line) as well as
  the fields with the smallest and largest numbers
  (corresponding to different galactic latitudes)
  providing the covered range (shaded area). 
  Our data points are plotted at the faintest magnitude per spectral type range,
  and the horizontal bar covers the magnitude range up to the brightest.
  While the observed number of M dwarfs is compatible with the known
  surface density, that of T dwarfs is substantially larger.
 \label{BD_Density}}
\end{figure}

\begin{figure*}[ht]
  \vspace{-1.5cm}
  \includegraphics[width=0.32\textwidth,clip]{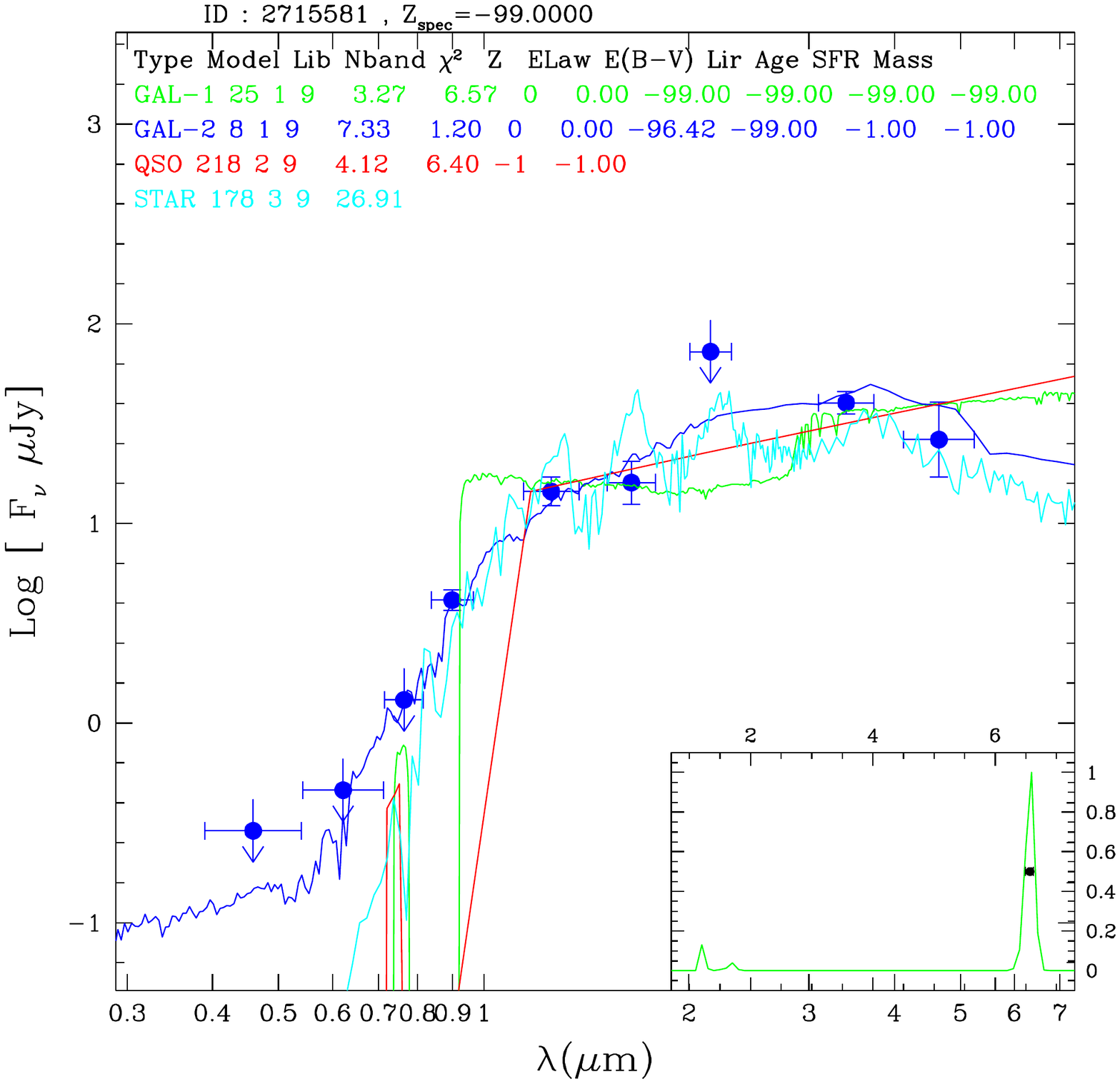}
  \includegraphics[width=0.32\textwidth,clip]{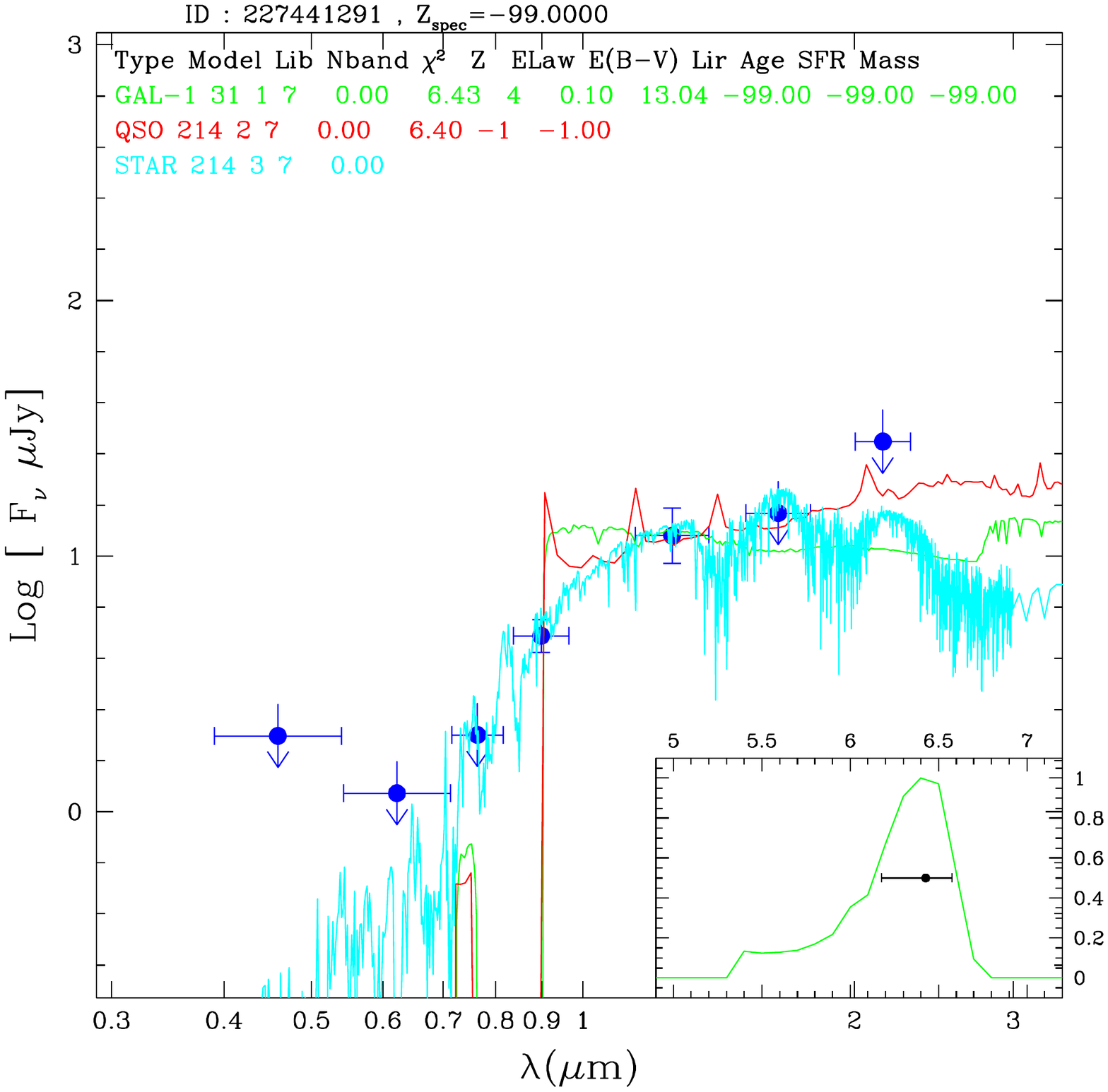}
  \includegraphics[width=0.32\textwidth,clip]{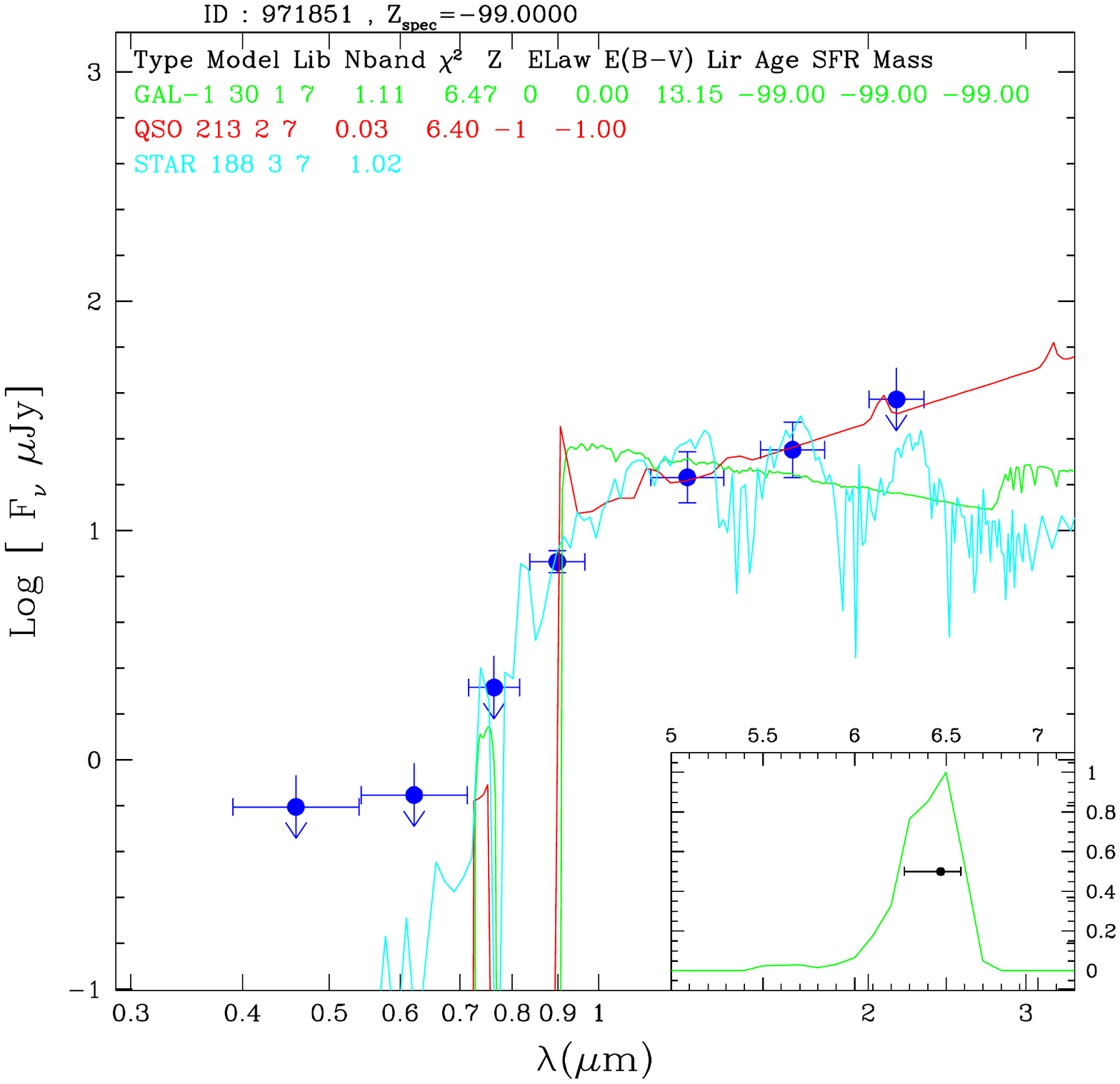}\vspace*{-2.5cm}
  
  \includegraphics[width=0.32\textwidth,clip]{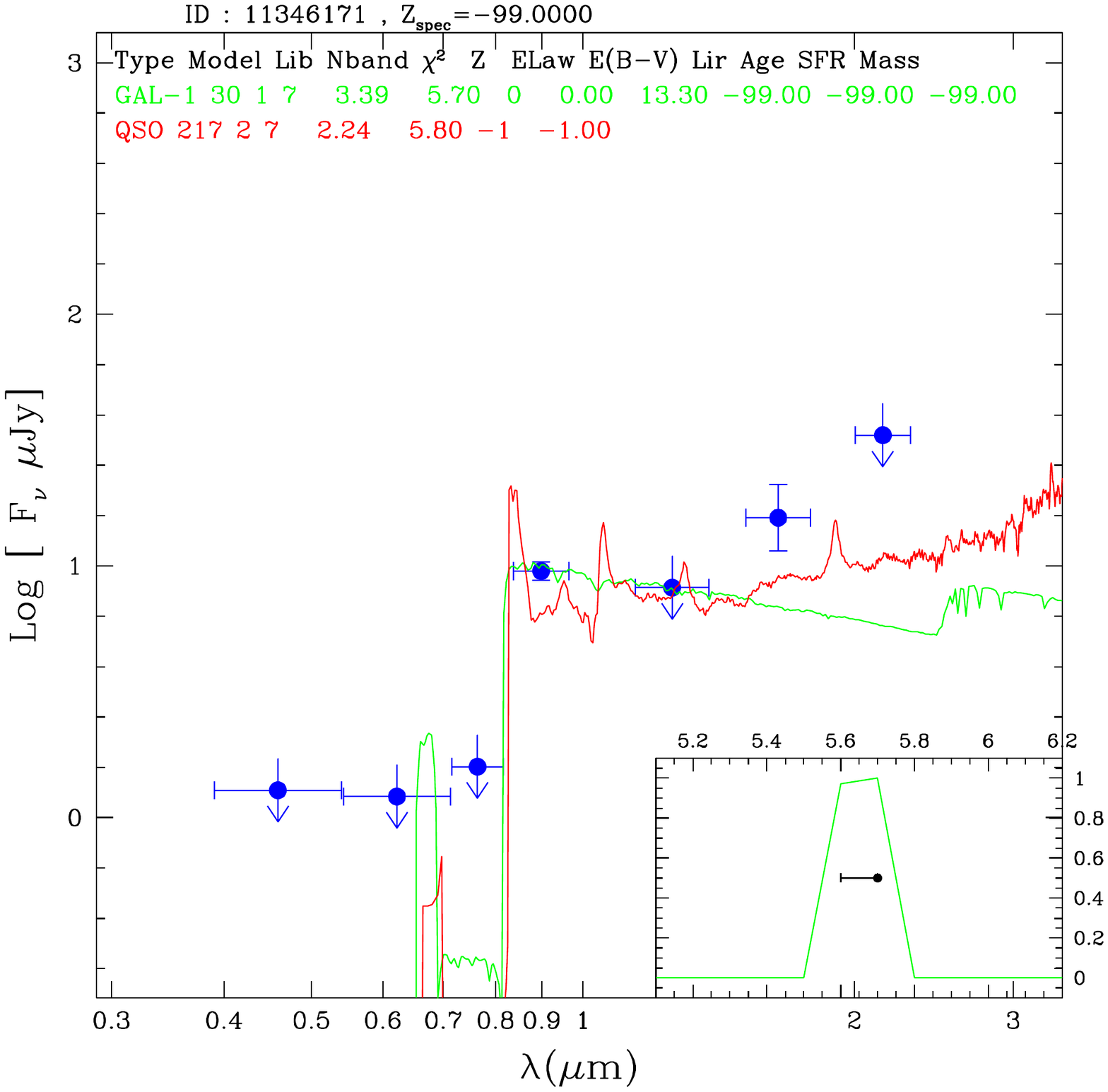}
  \includegraphics[width=0.32\textwidth,clip]{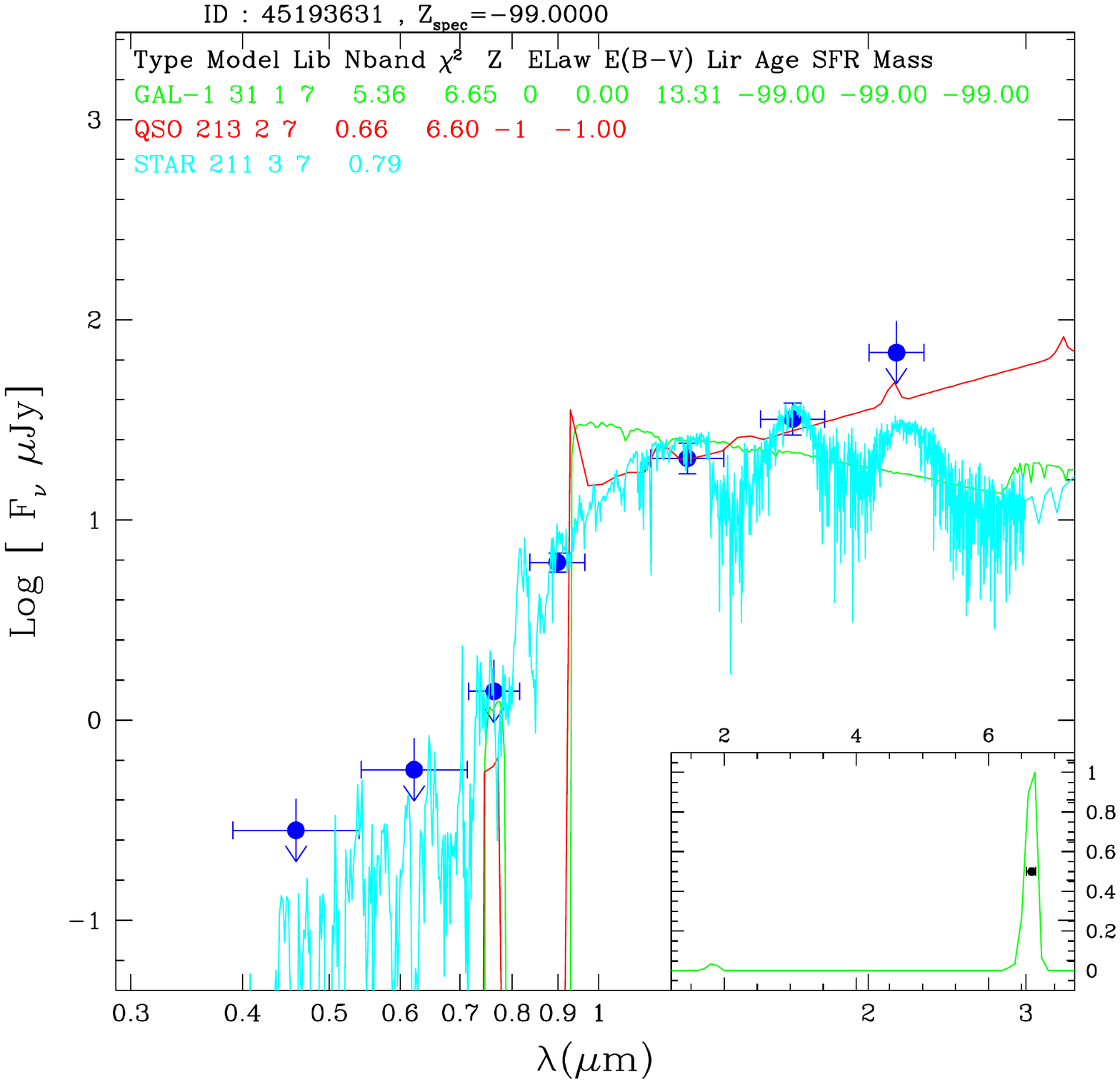}
  \includegraphics[width=0.32\textwidth,clip]{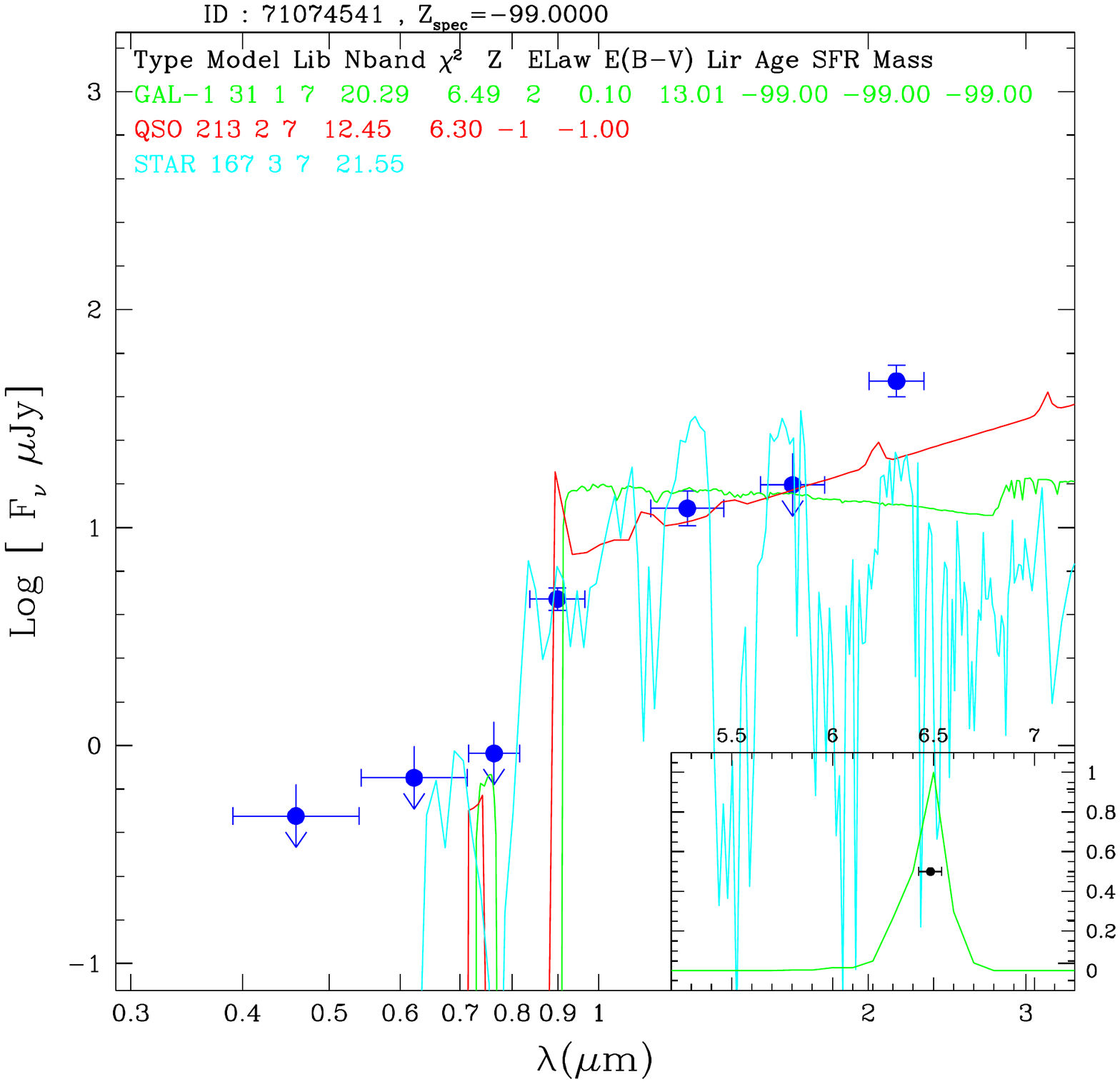}\vspace*{-2.5cm}
  
  \includegraphics[width=0.32\textwidth,clip]{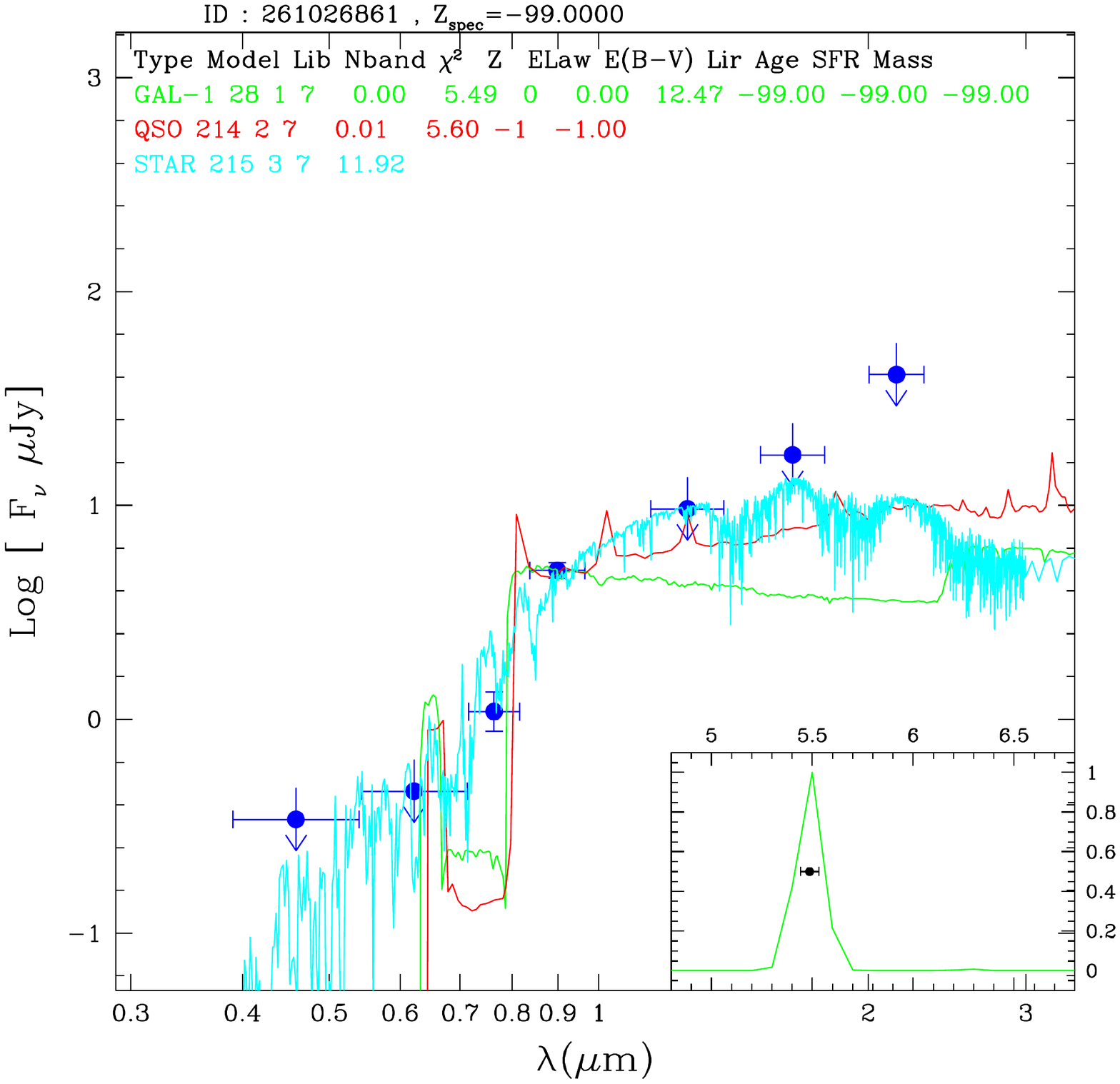}
  \includegraphics[width=0.32\textwidth,clip]{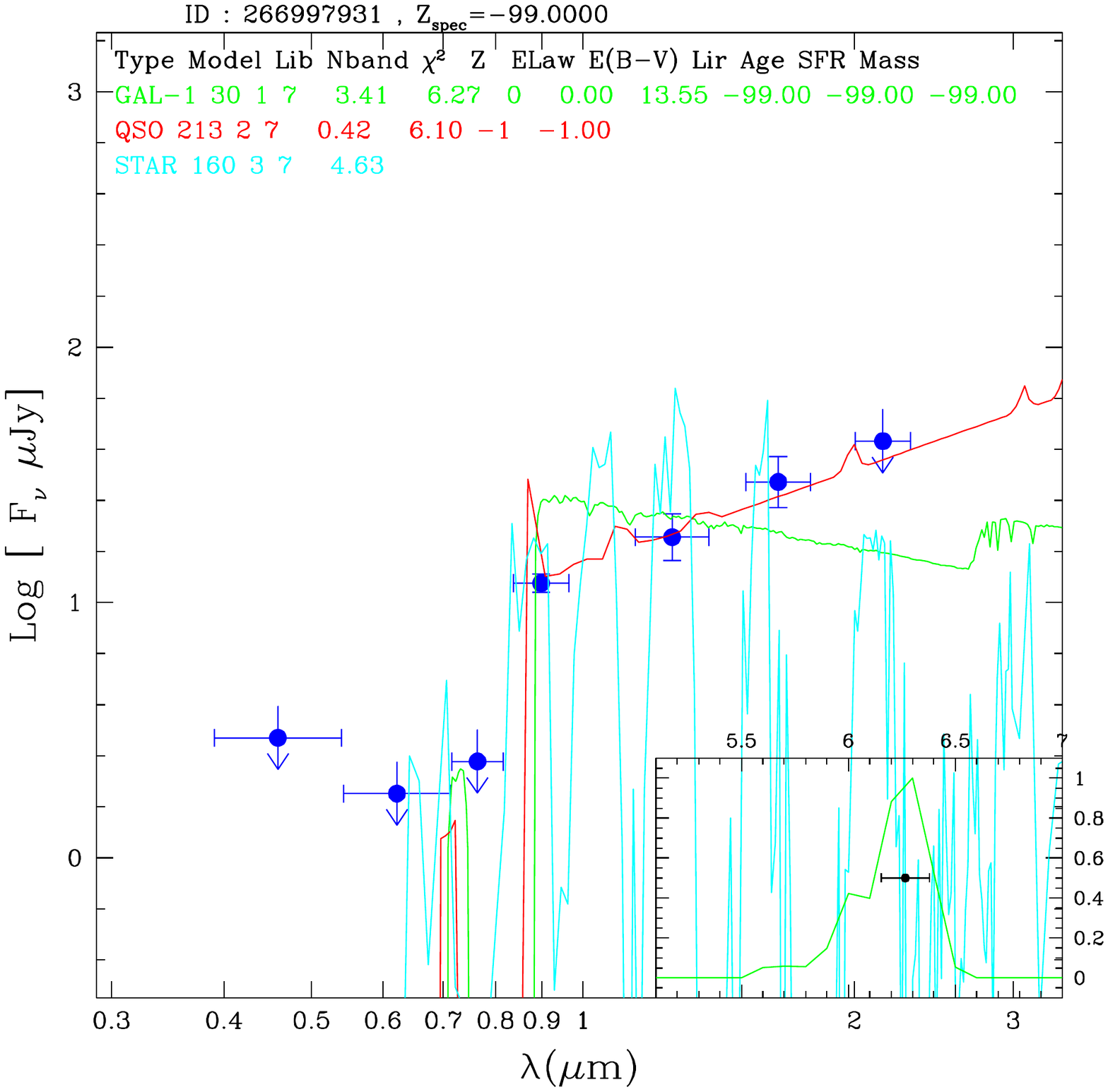}
  \includegraphics[width=0.32\textwidth,clip]{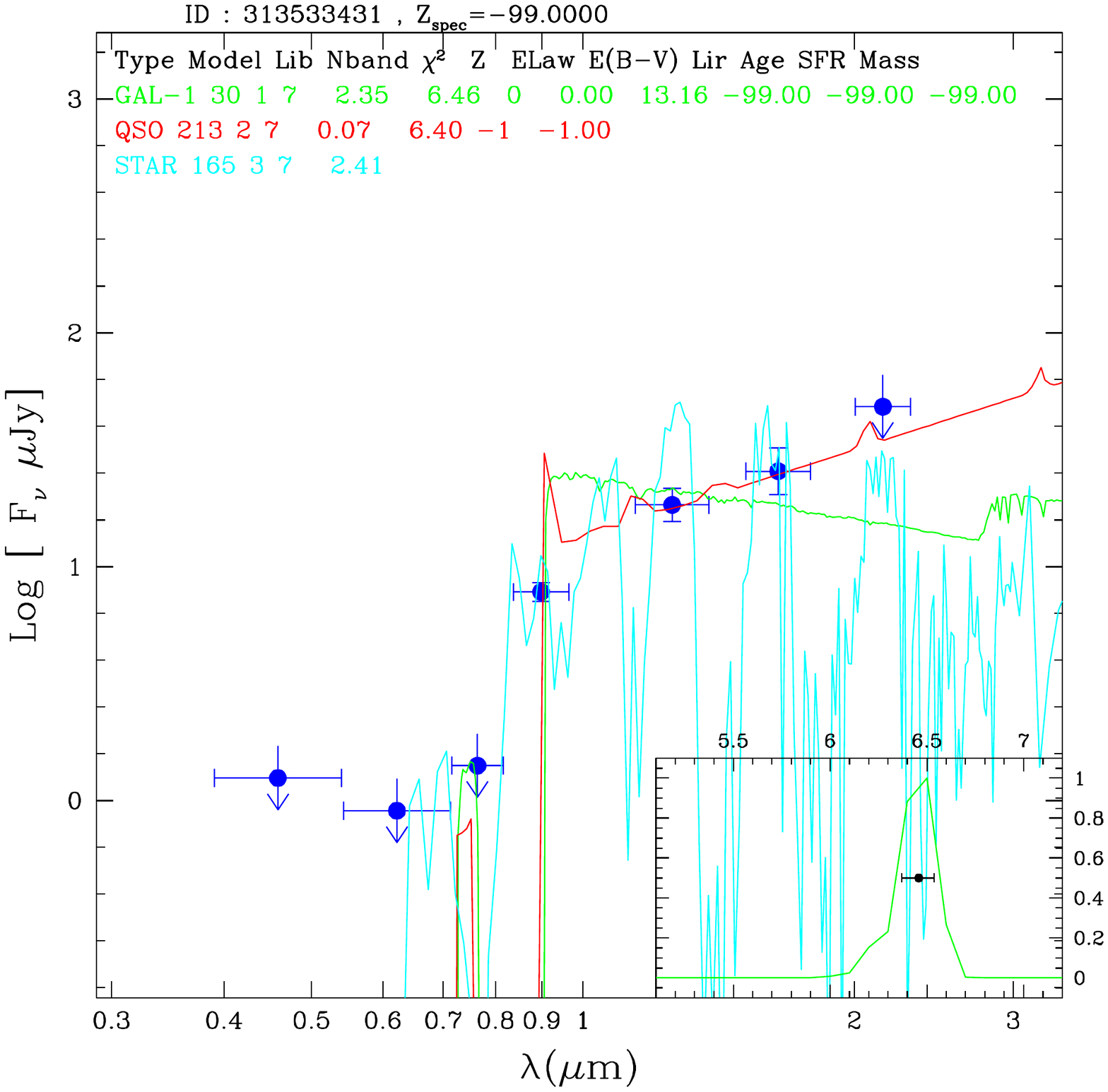}\vspace*{-2.5cm}
  
  \includegraphics[width=0.32\textwidth,clip]{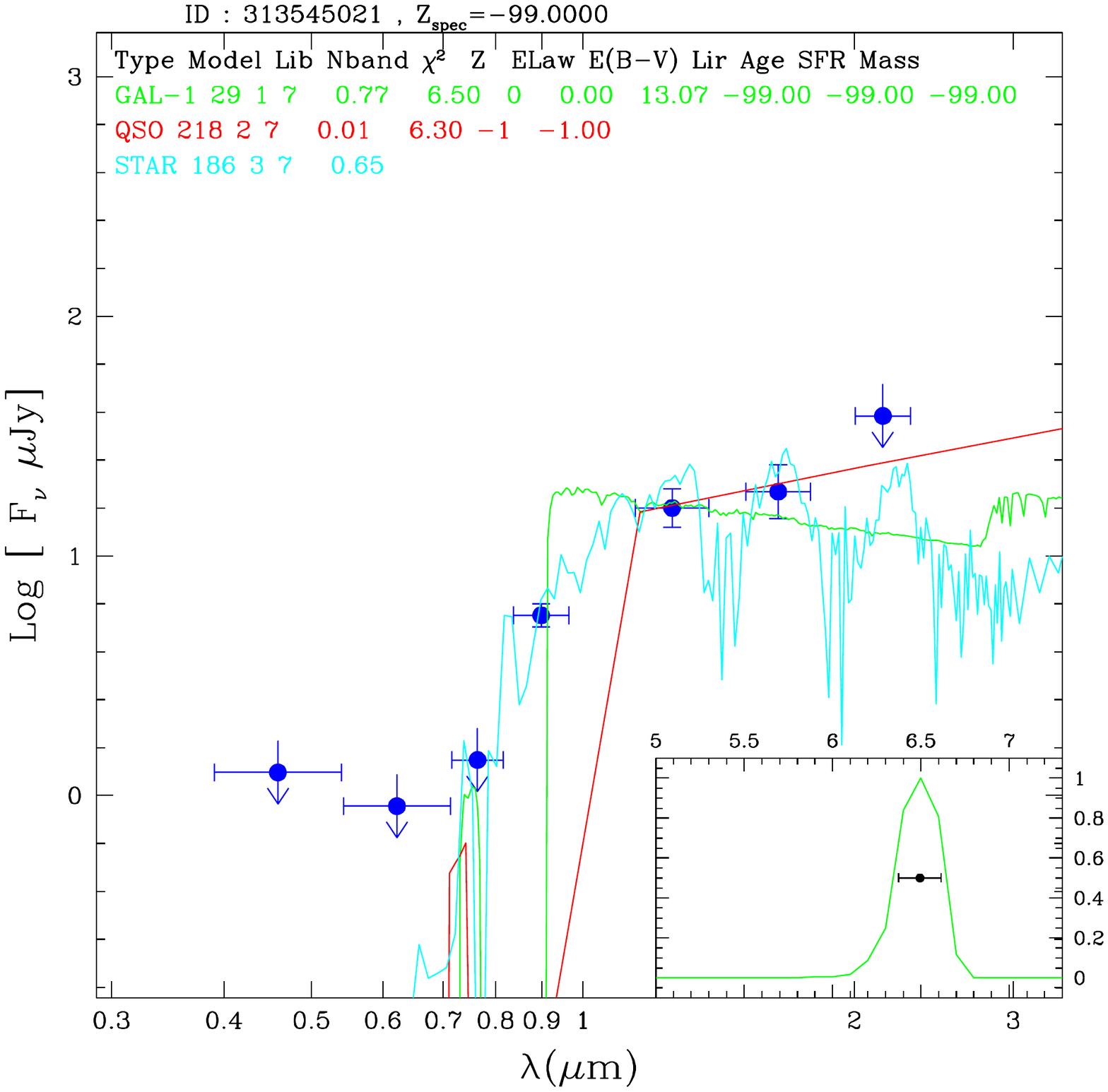}
  \includegraphics[width=0.32\textwidth,clip]{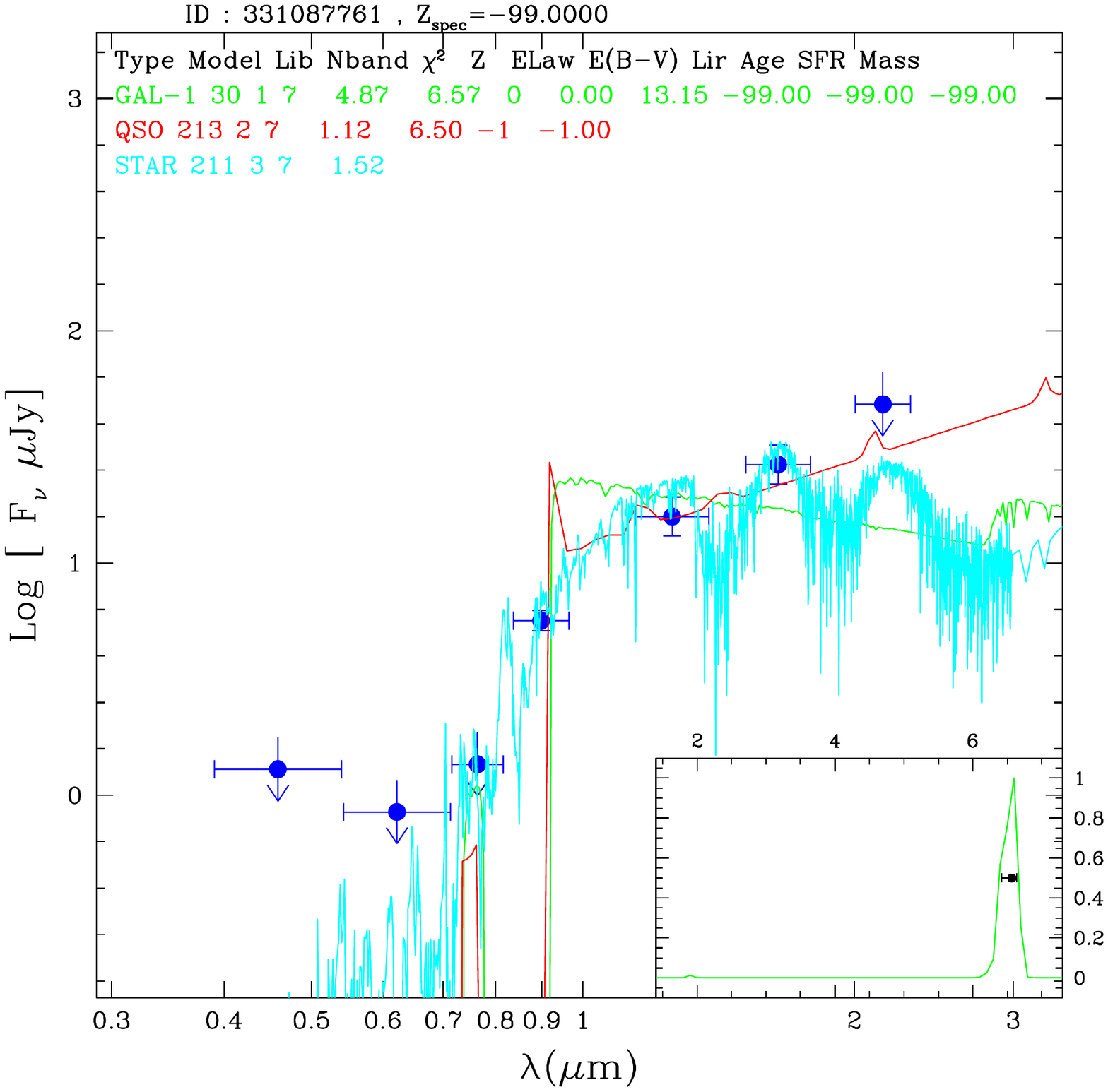}
  \vspace*{-1.5cm}
  \caption{Le\,PHARE fits for the 10 objects from Tabs. \ref{photWise} \& \ref{photnoWise}, for which a QSO template provides the better fit, plus the one
    with a galaxy template fitting best;
    from top left to right bottom:
     ULASJ0148+0600,
     CFHQSJ1509-1749, 
     PSOJ009.7355-10.4316, 
     PSOJ011.3899+09.0325\_C2, 
     PSOJ045.1840-22.5408, 
     PSOJ071.0322-04.5591, 
     PSOJ261.0364+19.0286\_C1, 
     PSOJ267.0021+22.7812\_C2,
     SDSSJ2054-0005\_C3,
     SDSSJ2054-0005\_C4,
     IMSJ2204+0012. 
     These are the SEDs after correction of the galactic foreground-A$_{\rm V}$,
     and most of the dwarf fits return an even worse reduced $\chi^2$
     for the no-A$_{\rm V}$ SEDs than shown here.
  }
 \label{sed}
\end{figure*}

\begin{figure*}[ht]
  \vspace*{-1.cm}\includegraphics[angle=270, width=0.9\textwidth,clip]{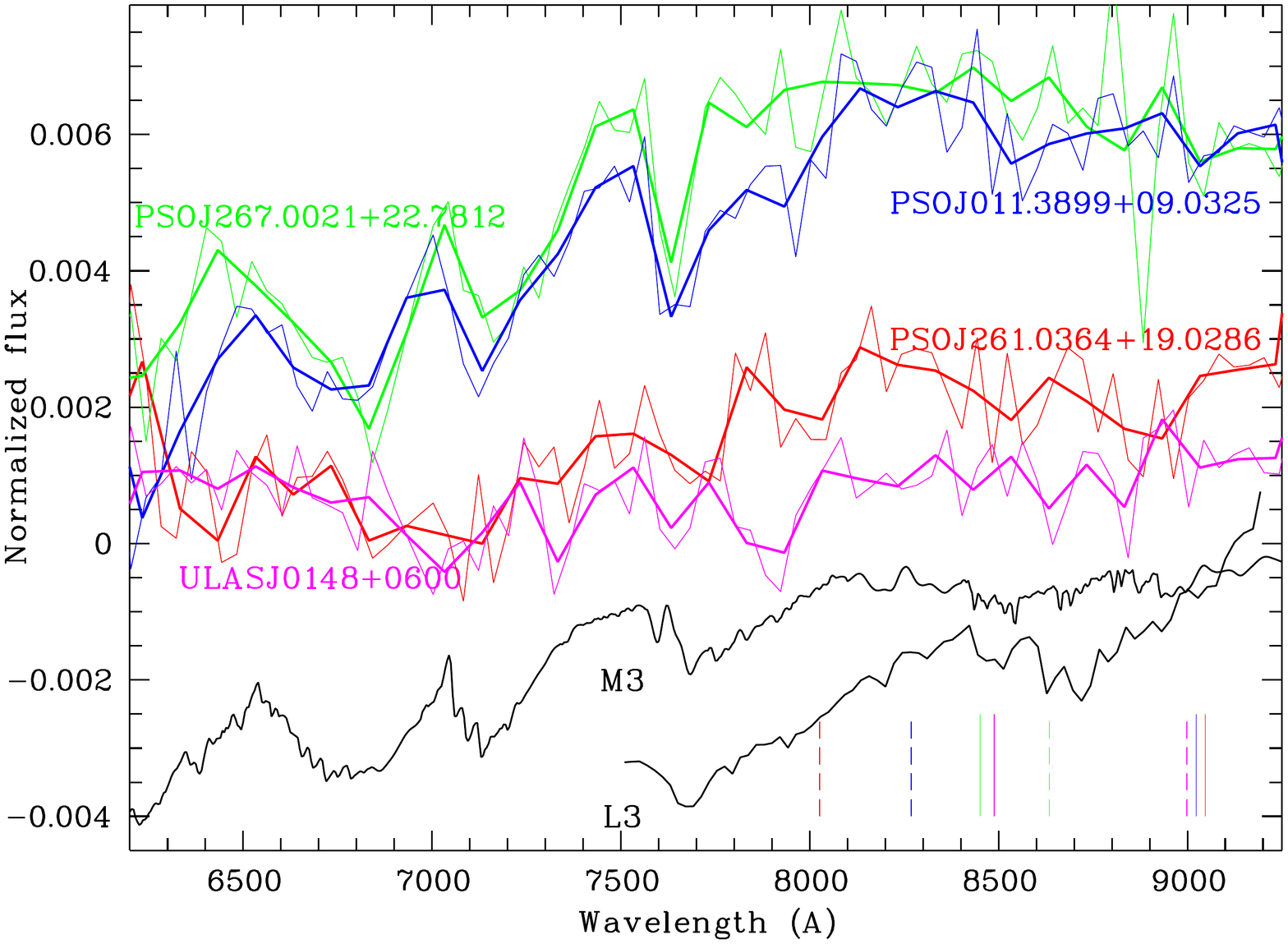}
  \vspace*{-1.cm}\includegraphics[angle=270, width=0.9\textwidth,clip]{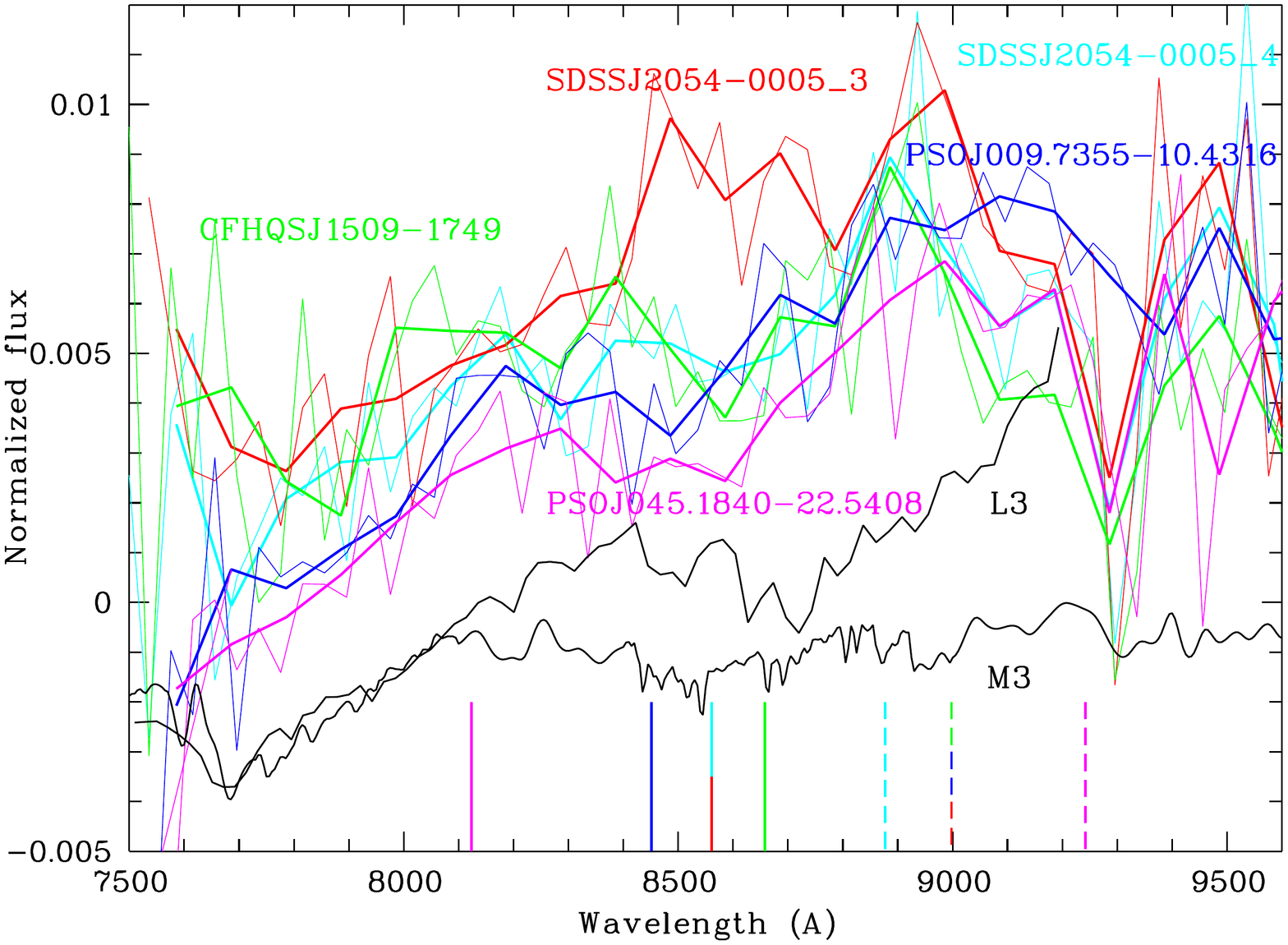}
  \caption{Long-slit spectra binned at 160/400 \AA\ (thin/thick lines)
    obtained with the GMOS spectrographs
    at the Gemini North (top panel) and South (lower panel) Observatories.
    The vertical solid lines denote the position of Ly$\alpha$ 
    of its ``mother'' QSO, the vertical dashed line the position
    of Ly$\alpha$ based on the Le\,PHARE fits. The dwarf templates
    are at arbitrary normalizations.
  }
 \label{spectra}
\end{figure*}

The combination of  our spectroscopic results and the Le\,PHARE
  fits results in identifications as follows: 20 dwarfs, 3 low-redshift
  galaxies, 1 QSO, and 2 unsettled cases (Tabs. \ref{photWise},
  \ref{photnoWise}, \ref{photunWise}).
Their spectral range is rather wide, from M5--T8,
though accurate spectral typing remains to be done.
The large incidence rate of such dwarfs is a surprise,
both, because our search was not tuned towards
  late dwarfs, and also given our small search size of just 0.52 square degrees
(summing over all 116 candidates). For instance,
\cite{Kakazu+2010} found 6 faint
($19 \lsim J \lsim 20$) ultracool T dwarfs over a 9.3 deg$^2$ area
with a limiting magnitude of \zp(AB) $\lsim$ 23.3 mag. This is
even slightly deeper than our mean depth of \zp(AB) $\lsim$ 23.1 mag,
yet we find 2 T-dwarfs per 0.52 deg$^2$, as compared to their 0.6 deg$^{-2}$,
a factor 6 higher. Given that we pointed at known high-z QSOs, our
search area similarly avoids the galactic plane. 
In a more recent compilation, \cite{RyanReid2016} used thick/thin
exponential disk models and the luminosity function of ultracool
dwarfs to predict their surface density for per spectral type.
While this depends strongly on the galactic coordinates, the prediction 
for T-dwarfs has only a small dependence on galactic latitude,
suggesting an even higher excess in our data,
about a factor of 20, with even the \cite{Kakazu+2010} surface density
being higher by a factor of $\sim$3.
For early L-type dwarfs, the excess is smaller,
and for the M8-9 dwarfs we find consistency. 
Fig. \ref{BD_Density} summarises the findings.

Since we observe a substantial excess of T dwarfs, search completeness 
  cannot be an issue. Also foreground extinction correction should not
  be an issue, since most of these dwarfs are within of order 100 pc distance.
  Even if we adopt a twice as large (systematic?) error in the temperature
  estimates from the Le\,PHARE fits, this would not change our numbers.
  One potential reason could be the appropriateness of
  the templates in Le\,PHARE, i.e. that dwarfs are more numerous
  than e.g. low-redshift red galaxies.  This is
 contrary to the case of high-z QSOs, for which late dwarfs form a numerous
 class of false positives. This finding of a larger than expected T dwarf
 number density is a surprising puzzle, which warrants further study.
 Existing GROND data could provide a first step in this direction,
 though spectroscopic typing of the candidate dwarfs is the
 ultimate step.

\begin{longtable}{lcccccc}
  \caption{\label{log} Details of the GROND observations of selected known QSOs, sorted by Right Ascension.
    The redshift and the observed \zp-magnitude are taken from the original
    discovery papers, i.e. predominantly \cite{Banados+2016} unless marked
    as follows:
    $^a$ = \cite{Willott+07}, $^b$ = \cite{Venemans+2013},
    $^c$ = \cite{Jiang+2016}, $^d$ = \cite{Mazzucchelli+2017},
    $^e$ = \cite{Zeimann+2011}, $^f$ = \cite{Matsuoka+2016},
    $^g$ = \cite{Kim+2015}, $^h$ = \cite{Reed+2015},
    $^i$ = \cite{Venemans+2015}, $^j$ = \cite{Jiang+2009},
    $^k$ = \cite{Willott+09}, $^l$ = \cite{Carnall+2015},
    $^m$ = \cite{Willot2010}.
    The 4th column is the date at the start of the observing night, and the
    5th column indicates the nth observing night of that target (first
    number) and which observation blocks (OB)s were used in the analysis
    (numbers after the underscore): more than one number indicates stacking
    of multiple OBs. Column six provides the 3$\sigma$ limiting \zp\ magnitude
    of the (stacked) observation, not corrected for galactic foreground
    extinction (which is given in column 7, based on \citep{Schlafly+11}).
    The letter attached to the \zp\ limiting magnitude denotes the catalog
    used for photometric calibration of the optical channels \gp\rp\ip\zp:
    S=SDSS (DR12), P=PanSTARRS1, SM=SkyMapper (DR1.1).
  }\\
      \hline\hline
      \noalign{\smallskip}
      Name &  z & z$^\prime_{cat}$ (AB) & Obs.-date & OB & 
        limiting z$^\prime$ calib & A$_{\rm V}$\\
           &      &  (mag)    &   &   &  &  (mag)  \\
      \noalign{\smallskip}
      \hline
      \noalign{\smallskip}
      \endfirsthead
      \caption{continued.}\\
      \hline\hline
      \noalign{\smallskip}
      Name &  z & z$^\prime_{cat}$ (AB) & Obs.-date & OB & 
        limiting z$^\prime$ calib & A$_{\rm V}$\\
           &      &  (mag)    &   &   &  &  (mag)  \\
      \noalign{\smallskip}
      \hline
      \noalign{\smallskip}
      \endhead
      \hline
      \endfoot
  PSOJ000.3401+26.8358  &  5.75 & 19.28$\pm$0.02 &  2016-10-25     &    2\_1  &   22.6S & 0.11 \\ 
        SDSSJ0005-0006  &  5.85 & 20.44$\pm$0.04 &  2016-08-29     &    1\_1  &   23.3S & 0.09 \\ 
  PSOJ002.1073-06.4345  &  5.93 & 20.25$\pm$0.03 &  2013-09-09     &  1\_123  &   22.8S & 0.09 \\ 
  PSOJ004.3936+17.0862  &  5.80 & 20.82$\pm$0.07 &  2016-10-25     &    1\_1  &   22.6S & 0.17 \\ 
  PSOJ004.8140-24.2991  &  5.68 & 19.62$\pm$0.02 &  2013-09-09     &   1\_12  &   22.4P & 0.04 \\ 
  PSOJ007.0273+04.9571  &  6.00 & 20.56$\pm$0.05 &  2013-01-14     &   2\_12  &   22.9S & 0.04 \\ 
    CFHQSJ0033-0125$^a$ &  6.13 & 22.44$\pm$0.08 &  2016-08-26     &    1\_1  &   23.0S & 0.06 \\ 
  PSOJ009.3573-08.1190  &  5.72 & 19.91$\pm$0.03 &  2015-11-05     &    1\_1  &   23.2S & 0.11 \\ 
  PSOJ009.7355-10.4316  &  5.95 & 20.82$\pm$0.04 &  2014-12-14     &   2\_12  &   23.1S & 0.08 \\
  PSOJ011.3899+09.0325$^d$ & 6.42 & $>$22.33 &  2016-09-20     &   1\_12  &   23.1S & 0.16 \\
   CFHQSJ0055+0146$^k$  &  6.01 & 22.19$\pm$0.06 &  2016-09-01     &    1\_1  &   24.1S & 0.06 \\ 
        SDSSJ0100+2802  &  6.30 & 18.61$\pm$0.01 &  2016-08-31     &    1\_1  &   23.3S & 0.15 \\ 
   CFHQSJ0102-0218$^k$  &  5.95 & 22.30$\pm$0.08 &  2016-08-18     &    1\_1  &   23.0S & 0.11 \\ 
  PSOJ021.4213-25.8822  &  5.79 & 19.66$\pm$0.03 &  2014-09-21     &    1\_1  &   22.7P & 0.04 \\ 
    SDSSJ0129-0035$^j$  &  5.78 & 22.16$\pm$0.11 &  2016-09-02     &    1\_3  &   23.4S & 0.09 \\ 
  PSOJ023.0071-02.2675  &  5.90 & 20.19$\pm$0.03 &  2013-09-10     &   2\_12  &   23.4S & 0.09 \\ 
       CFHQSJ0136+0226  &  6.21 & 22.06$\pm$0.16 &  2016-09-02     &    1\_1  &   23.4S & 0.11 \\ 
  PSOJ025.2376-11.6831  &  5.85 & 20.18$\pm$0.02 &  2013-09-08     &  1\_123  &   22.9P & 0.06 \\ 
ATLASJ025.6821-33.4627$^l$ & 6.31 & 19.63$\pm$0.08 &  2016-08-18     &    1\_1  &  22.3SM & 0.07 \\ 
        ULASJ0148+0600  &  5.98 & 19.45$\pm$0.01 &  2016-08-31     &    1\_1  &   23.7S & 0.16 \\ 
  PSOJ029.5172-29.0886  &  5.99 & 19.48$\pm$0.02 &  2013-09-10     &   1\_12  &   22.8P & 0.04 \\ 
ATLASJ029.9915-36.5658$^l$ & 6.02 & 19.54$\pm$0.08 &  2016-08-18     &    1\_1  &  22.4SM & 0.04 \\ 
        ULASJ0203+0012  &  5.72 & 20.74$\pm$0.06 &  2016-08-29     &    1\_1  &   23.1S & 0.07 \\ 
    CFHQSJ0210-0456$^m$ &  6.43 & 22.67$\pm$0.05 &  2016-09-02     &    1\_1  &   23.4S & 0.05 \\ 
    CFHQSJ0221-0802$^m$ &  6.16 & 22.63$\pm$0.05 &  2016-08-18     &    1\_1  &   23.0S & 0.08 \\ 
  PSOJ036.5078+03.0498  &  6.54 & 21.44$\pm$0.12 &  2016-08-29     &    1\_1  &   23.1S & 0.09 \\
   CFHQSJ0227-0605$^k$  &  6.20 & 21.71$\pm$0.06 &  2016-09-01     &    1\_1  &   23.6S & 0.09 \\ 
  PSOJ037.9706-28.8389  &  6.00 & 20.73$\pm$0.06 &  2013-01-17     &  1\_123  &   23.4P & 0.04 \\ 
    SDSSJ0239-0045$^j$  &  5.82 & 22.08$\pm$0.11 &  2016-08-18     &    1\_1  &   22.8S & 0.08 \\ 
  PSOJ040.0159+17.5458  &  5.68 & 20.60$\pm$0.05 &  2016-08-30     &    2\_1  &   23.8P & 0.23 \\ 
  PSOJ042.6690-02.9174  &  5.89 & 20.49$\pm$0.05 &  2013-09-10     &  1\_123  &   23.4S & 0.12 \\ 
  PSOJ045.1840-22.5408  &  5.68 & 20.34$\pm$0.05 &  2013-01-14     &  1\_123  &   23.2P & 0.07 \\ 
        SDSSJ0303-0019  &  6.08 & 20.99$\pm$0.06 &  2016-08-30     &    1\_1  &   23.9S & 0.29 \\ 
      VIKJ0305-3150$^b$ &  6.60 & 22.12$\pm$0.07 &  2016-09-02     &   12\_1  &   23.7SM & 0.03 \\ 
       CFHQSJ0316-1340  &  5.99 & 21.57$\pm$0.12 &  2016-08-30     &    1\_1  &   24.1S & 0.14 \\ 
  PSOJ049.2934-26.5543  &  5.94 & 20.77$\pm$0.06 &  2016-08-31     &    2\_1  &   23.4P & 0.05 \\ 
      VIKJ0328-3253$^i$ &  5.86 & 19.83$\pm$0.02 &  2016-08-21     &    1\_1  &   23.1P & 0.03 \\ 
  PSOJ053.9605-15.7956  &  5.87 & 20.34$\pm$0.04 &  2016-08-21     &    1\_1  &   23.3P & 0.24 \\ 
  PSOJ055.4244-00.8035  &  5.68 & 20.19$\pm$0.04 &  2014-02-02     &    1\_1  &   22.6S & 0.25 \\ 
  PSOJ056.7168-16.4769  &  5.99 & 20.00$\pm$0.04 &  2016-08-19     &    1\_1  &   22.8P & 0.14 \\ 
        SDSSJ0353+0104  &  6.07 & 20.81$\pm$0.07 &  2016-08-15     &    1\_1  &   23.4S & 0.78 \\ 
  PSOJ060.5529+24.8567  &  6.18 & 20.18$\pm$0.03 &  2014-12-16     &   1\_12  &   23.0P & 0.64 \\ 
  PSOJ065.4085-26.9543  &  6.14 & 20.48$\pm$0.05 &  2014-02-05     &    2\_1  &   22.8P & 0.11 \\ 
  PSOJ065.5041-19.4579  &  6.12 & 19.79$\pm$0.03 &  2014-12-17     &    1\_1  &   22.8P & 0.10 \\ 
  PSOJ071.0322-04.5591  &  5.89 & 20.30$\pm$0.04 &  2015-11-06     &   1\_12  &   23.6S & 0.12 \\ 
  PSOJ071.4507-02.3332  &  5.69 & 19.18$\pm$0.01 &  2016-08-20 + 30 &   12\_1  &   23.6P & 0.10 \\ 
     DESJ0454-4448$^h$  &  6.10 & 20.20$\pm$0.01 &  2016-08-21     &    1\_1  &  23.4SM & 0.03 \\ 
  PSOJ075.9356-07.5061  &  5.88 & 20.33$\pm$0.05 &  2017-09-18     &    2\_1  &   23.6P & 0.52 \\ 
  PSOJ089.9394-15.5833  &  6.05 & 19.66$\pm$0.03 &  2015-11-08     &   1\_12  &   23.4P & 0.79 \\ 
  PSOJ108.4429+08.9257  &  5.92 & 19.45$\pm$0.02 &  2015-11-08     &    1\_1  &   22.8P & 0.23 \\ 
        SDSSJ0818+1722  &  6.02 & 19.55$\pm$0.02 &  2017-01-26     &    1\_1  &   22.9S & 0.09 \\ 
        ULASJ0828+2633  &  6.05 & 20.72$\pm$0.06 &  2017-02-07     &    1\_1  &   22.5S & 0.21 \\ 
  PSOJ127.2817+03.0657  &  5.85 & 20.69$\pm$0.05 &  2014-12-15     &   1\_12  &   22.8S & 0.04 \\ 
        SDSSJ0836+0054  &  5.81 & 18.70$\pm$0.01 &  2012-01-05     &   1\_12  &   23.3S & 0.13 \\ 
     VIKJ0839+0015$^i$  &  5.84 & 21.09$\pm$0.05 &  2017-01-26     &    1\_1  &   23.3S & 0.12 \\ 
        SDSSJ0842+1218  &  6.07 & 19.83$\pm$0.03 &  2017-05-02     &    1\_2  &   21.9S & 0.18 \\ 
     HSCJ0859+0022$^f$  &  6.39 & 22.77$\pm$0.01 &  2017-01-25     &    1\_1  &   23.4S & 0.08 \\ 
  PSOJ135.3860+16.2518  &  5.63 & 20.67$\pm$0.04 &  2014-02-04     &    1\_1  &   22.8S & 0.11 \\ 
  PSOJ135.8704-13.8336  &  5.91 & 20.31$\pm$0.04 &  2014-12-17     &  12\_12  &   23.2P & 0.18 \\ 
  PSOJ157.9070-02.6599  &  5.88 & 20.24$\pm$0.03 &  2017-01-25     &    2\_1  &   23.1S & 0.12 \\ 
  PSOJ159.2257-02.5438  &  6.38 & 20.46$\pm$0.04 &  2017-01-26     &    1\_1  &   23.5S & 0.13 \\ 
        SDSSJ1044-0125  &  5.78 & 19.31$\pm$0.01 &  2017-01-26     &    1\_1  &   24.1S & 0.14 \\ 
       CFHQSJ1059-0906  &  5.92 & 20.84$\pm$0.06 &  2017-01-24     &    1\_1  &   23.6P & 0.09 \\ 
  PSOJ167.6415-13.4960$^d$ & 6.51 & $>$22.94       &  2017-05-02     &    1\_1  &   23.4S & 0.15 \\ 
  PSOJ174.7920-12.2845  &  5.81 & 20.04$\pm$0.04 &  2017-01-28     &    1\_1  &   23.4P & 0.09 \\ 
  PSOJ175.4091-20.2654  &  5.69 & 20.17$\pm$0.04 &  2017-01-24     &    1\_1  &   23.6P & 0.14 \\ 
     VIKJ1152+0055$^f$  &  6.37 & 21.83$\pm$0.01 &  2014-02-05     &   1\_123 &   23.3S & 0.06 \\ 
  PSOJ183.2991-12.7676  &  5.86 & 19.47$\pm$0.02 &  2017-04-29     &   1\_3   &   23.3P & 0.13 \\ 
  PSOJ184.3389+01.5284  &  6.20 & 21.20$\pm$0.07 &  2017-05-01     &    1\_1  &   23.2P & 0.06 \\ 
  PSOJ187.1047-02.5609  &  5.77 & 20.92$\pm$0.05 &  2014-01-29     &  1\_123  &   23.1S & 0.09 \\ 
  PSOJ187.3050+04.3243  &  5.89 & 20.92$\pm$0.04 &  2017-04-30     &    1\_1  &   23.4S & 0.05 \\ 
        SDSSJ1306+0356  &  6.02 & 19.76$\pm$0.03 &  2017-04-30     &    1\_1  &   23.2S & 0.07 \\ 
        ULASJ1319+0950  &  6.13 & 20.13$\pm$0.02 &  2017-04-29     &    1\_1  &   23.6S & 0.05 \\ 
  PSOJ209.2058-26.7083  &  5.72 & 19.35$\pm$0.01 &  2014-02-02     &    1\_12 &   23.2P & 0.17 \\ 
  PSOJ210.8297+09.0474  &  5.88 & 20.31$\pm$0.03 &  2016-08-23     &    1\_1  &   22.4S & 0.07 \\ 
  PSOJ210.8722-12.0094  &  5.84 & 21.09$\pm$0.07 &  2018-03-23     &   23\_1  &   23.3S & 0.19 \\ 
  PSOJ212.2974-15.9865  &  5.83 & 20.98$\pm$0.06 &  2015-05-21     &  1\_123  &   23.2S & 0.26 \\ 
        SDSSJ1411+1217  &  5.90 & 20.58$\pm$0.02 &  2016-08-22     &    1\_1  &   22.2S & 0.06 \\ 
  PSOJ213.3629-22.5617  &  5.92 & 19.55$\pm$0.02 &  2016-08-22     &    2\_1  &   22.9P & 0.23 \\ 
  PSOJ213.7329-13.4803  &  5.78 & 20.86$\pm$0.05 &  2016-08-23     &    2\_1  &   23.0P & 0.21 \\ 
  PSOJ215.1514-16.0417  &  5.73 & 19.08$\pm$0.02 &  2011-03-05     &   9\_123 &   22.8P & 0.21 \\ 
  PSOJ217.0891-16.0453  &  6.11 & 20.46$\pm$0.04 &  2016-08-22     &    2\_1  &   22.8P & 0.25 \\ 
  PSOJ217.9185-07.4120  &  6.14 & 21.10$\pm$0.08 &  2015-05-19     &    1\_123  &   23.1P & 0.18 \\ 
  CFHQSJ1509-1749$^a$ &  6.12 & 20.26$\pm$0.02 &  2016-08-17     &    1\_1  &   23.2P & 0.24 \\ 
  PSOJ228.6871+21.2388  &  5.92 & 20.92$\pm$0.06 &  2017-04-29     &    2\_1  &   23.4S & 0.13 \\ 
  PSOJ235.9450+17.0079  &  5.82 & 20.23$\pm$0.03 &  2016-08-21     &    2\_1  &   23.1S & 0.09 \\ 
  PSOJ236.2912+16.6088  &  5.82 & 20.69$\pm$0.05 &  2016-08-20     &    1\_1  &   23.0S & 0.09 \\ 
  PSOJ238.8510-06.8976  &  5.81 & 20.43$\pm$0.05 &  2015-05-21     &   1\_12  &   22.7P & 0.45 \\ 
  PSOJ239.7124-07.4026  &  6.11 & 19.78$\pm$0.03 &  2016-08-21     &  12\_1   &   23.4P & 0.45 \\ 
  PSOJ242.4397-12.9816  &  5.83 & 19.76$\pm$0.04 &  2014-02-04     &    1\_1  &   23.0P & 0.80 \\ 
  PSOJ245.0636-00.1978  &  5.68 & 21.15$\pm$0.07 &  2015-05-19     &    1\_1  &   23.1S & 0.26 \\
  PSOJ247.2970+24.1277$^d$  &  6.48 & $>$22.77 &  2017-05-01     &   12\_1  &   23.7S & 0.14 \\
  PSOJ261.0364+19.0286$^d$  &  6.44 & $>$22.92 &  2017-04-22     &    2\_1  &   23.8P & 0.12 \\
  PSOJ267.0021+22.7812  &  5.95 & 20.89$\pm$0.06 &  2016-08-17     &    1\_1  &   23.0P & 0.27 \\ 
  PSOJ308.0416-21.2339  &  6.24 & 21.12$\pm$0.08 &  2014-09-24     &   1\_12  &   23.0S & 0.14 \\ 
  PSOJ308.4829-27.6485  &  5.80 & 19.71$\pm$0.02 &  2013-09-08     &  12\_12  &   23.0P & 0.13 \\ 
    SDSSJ2053+0047$^j$  &  5.92 & 21.41$\pm$0.06 &  2016-08-21     &    1\_1  &   23.3S & 0.22 \\ 
        SDSSJ2054-0005  &  6.04 & 21.03$\pm$0.09 &  2016-08-22     &    1\_1  &   23.5S & 0.26 \\ 
       CFHQSJ2100-1715  &  6.09 & 21.55$\pm$0.09 &  2016-08-21     &    1\_1  &   23.6S & 0.17 \\
  PSOJ319.6040-10.9326  &  5.90 & 19.94$\pm$0.02 &  2013-09-10     &    1\_123  &   23.4P & 0.13 \\
  SDSS211951.9-004020$^c$ & 5.87 & 21.68$\pm$0.10 &  2017-09-21     &    1\_1  &   23.2S & 0.12 \\
  PSOJ320.8703-24.3604  &  5.73 & 20.17$\pm$0.04 &  2013-09-05     &  1\_123  &   22.8P & 0.10 \\
   PSOJ323.1382+12.2986$^d$ & 6.59 & 21.56$\pm$0.10 &  2017-04-27     &    1\_1  &   23.3S & 0.28 \\
    SDSSJ2147+0107$^j$  &  5.81 & 21.61$\pm$0.08 &  2016-08-22     &    1\_1  &   23.3S & 0.30 \\ 
  PSOJ328.7339-09.5076  &  5.92 & 20.82$\pm$0.06 &  2016-08-22     &    1\_1  &   23.2P & 0.12 \\ 
     IMSJ2204+0012$^g$  &  5.94 & 22.95$\pm$0.07 &  2016-08-23     &    1\_1  &   23.4S & 0.13 \\ 
     HSCJ2216-0016$^f$  &  6.10 & 22.78$\pm$0.02 &  2016-08-31     &    1\_1  &   23.7S & 0.20 \\ 
        SDSSJ2220-0101  &  5.62 & 20.36$\pm$0.05 &  2016-10-20     &    2\_1  &   23.4S & 0.19 \\ 
     SDSSJ2228+0110$^e$ &  5.95 & 22.28$\pm$0.20 &  2016-10-20     &    2\_1  &   23.4S & 0.18 \\ 
       CFHQSJ2229+1457  &  6.15 & 21.84$\pm$0.12 &  2016-09-03     &    1\_1  &   23.8S & 0.15 \\ 
  PSOJ340.2041-18.6621  &  6.01 & 20.14$\pm$0.03 &  2012-05-22     &  1\_123  &   23.7P & 0.09 \\ 
  SDSSJ2310+1855$^c$  &  6.00 & 19.21$\pm$0.09 &  2016-10-25     &    1\_1  &   22.7S & 0.46 \\ 
  CFHQSJ2329-0403$^k$  &  5.90 & 21.87$\pm$0.08 &  2016-08-26     &    1\_1  &   23.0S & 0.12 \\ 
  PSOJ357.8289+06.4019  &  5.81 & 21.31$\pm$0.10 &  2014-09-25     & 1\_1234  &   23.2S & 0.21 \\ 
  PSOJ359.1352-06.3831  &  6.15 & 19.97$\pm$0.03 &  2014-09-25     &    1\_1  &   22.6S & 0.09 \\ 
  SDSSJ2356+0023$^j$  &  6.00 & 21.66$\pm$0.08 &  2016-08-29     &    1\_1  &   23.2S & 0.10 \\
\end{longtable}

\begin{longtable}{rccccccc}
  \caption{\label{phot} Full GROND photometry of all 116 QSOs observed.
    All magnitudes are in the AB system, and corrected for the Galactic
    extinction (see Tab. \ref{log}). Upper limits are at the 3$\sigma$
    confidence level.}\\
      \noalign{\vspace{-0.2cm}}
      \hline\hline
      \noalign{\smallskip}
      Name & \gp & \rp & \ip & \zp & $J$ & $H$ & $K_s$ \\
           & (mag) & (mag) & (mag) & (mag) & (mag) & (mag) & (mag)  \\
      \noalign{\smallskip}
      \hline
      \noalign{\smallskip}
      \endfirsthead
      \caption{continued.}\\
      \noalign{\vspace{-0.2cm}}
      \hline\hline
      \noalign{\smallskip}
      Name & \gp & \rp & \ip & \zp & $J$ & $H$ & $K_s$ \\
           & (mag) & (mag) & (mag) & (mag) & (mag) & (mag) & (mag)  \\
      \noalign{\smallskip}
      \hline
      \noalign{\smallskip}
      \endhead
      \hline
      \endfoot
PSOJ000.3401+26.8358 & $>$24.66 & 22.92$\pm$0.15 & 21.55$\pm$0.11 & 19.43$\pm$0.04 & 19.44$\pm$0.10 & 19.10$\pm$0.13 & $>$19.19 \\
      SDSSJ0005-0006 & $>$24.95 & $>$24.55 & 22.72$\pm$0.15 & 20.54$\pm$0.04 & 20.55$\pm$0.15 & 20.35$\pm$0.19 & 19.81$\pm$0.23 \\
PSOJ002.1073-06.4345 & $>$23.62 & $>$23.62 & $>$22.70 & 20.24$\pm$0.05 & 19.82$\pm$0.11 & 19.91$\pm$0.17 & $>$19.03 \\
PSOJ004.3936+17.0862 & $>$24.62 & $>$24.09 & $>$22.59 & 20.40$\pm$0.06 & 20.26$\pm$0.20 & 19.99$\pm$0.29 & $>$19.45 \\
PSOJ004.8140-24.2991 & $>$23.36 & $>$22.85 & 22.22$\pm$0.22 & 19.55$\pm$0.04 & 18.96$\pm$0.22 & 19.01$\pm$0.13 & $>$17.87 \\
PSOJ007.0273+04.9571 & $>$23.75 & $>$23.85 & 22.64$\pm$0.20 & 20.25$\pm$0.04 & 19.91$\pm$0.13 & 19.69$\pm$015 & $>$19.45 \\
     CFHQSJ0033-0125 & $>$24.45 & $>$23.95 & $>$23.33 & 22.45$\pm$0.22 & 21.10$\pm$0.24 & $>$20.90 & $>$19.05 \\
PSOJ009.3573-08.1190 & $>$24.56 & 22.94$\pm$0.26 & 22.20$\pm$0.11 & 20.09$\pm$0.04 & 19.45$\pm$0.24 & $>$18.42 & $>$19.03 \\
PSOJ009.7355-10.4316 & $>$24.42 & $>$24.08 & $>$23.35 & 20.83$\pm$0.06 & 19.56$\pm$0.10 & 19.41$\pm$0.12 & 19.43$\pm$0.20 \\
PSOJ011.3899+09.0325 & $>$23.44 & $>$23.51 & $>$23.20 & $>$22.94 & $>$20.63 & $>$20.08 & $>$19.13 \\
     CFHQSJ0055+0146 & $>$25.50 & $>$25.19 & $>$24.26 & 22.41$\pm$0.09 & $>$21.96 & $>$21.44 & $>$20.04 \\
      SDSSJ0100+2802 & $>$24.80 & $>$24.53 & 20.59$\pm$0.04 & 18.27$\pm$0.03 & 17.60$\pm$0.05 & 17.49$\pm$0.05 & 17.20$\pm$0.08 \\
     CFHQSJ0102-0218 & $>$22.83 & $>$23.34 & $>$22.98 & 22.66$\pm$0.29 & $>$21.57 & $>$21.13 & $>$20.30 \\
PSOJ021.4213-25.8822 & $>$23.91 & 22.89$\pm$0.18 & 22.36$\pm$0.19 & 19.44$\pm$0.10 & 19.19$\pm$0.07 & 19.20$\pm$0.09 & 19.20$\pm$0.17 \\
      SDSSJ0129-0035 & $>$25.13 & $>$24.80 & 23.49$\pm$0.26 & 22.24$\pm$0.14 & $>$21.42 & $>$20.54 & $>$19.70 \\
PSOJ023.0071-02.2675 & $>$24.99 & $>$24.37 & 23.39$\pm$0.28 & 20.58$\pm$0.05 & 19.85$\pm$0.09 & 19.81$\pm$0.12 & 19.28$\pm$0.30 \\
     CFHQSJ0136+0226 & $>$25.00 & $>$24.65 & $>$23.67 & 22.21$\pm$0.14 & $>$21.46 & $>$20.63 & $>$21.88 \\
PSOJ025.2376-11.6831 & $>$23.79 & $>$24.05 & 22.57$\pm$0.16 & 19.73$\pm$0.04 & 19.84$\pm$0.09 & 19.44$\pm$0.10 & $>$19.35 \\
ATLASJ025.6821-33.4627 & $>$22.38 & $>$22.84 & 21.32$\pm$0.25 & 19.45$\pm$0.05 & 18.86$\pm$0.06 & 18.67$\pm$0.08 & 18.57$\pm$0.21 \\
      ULASJ0148+0600 & $>$25.26 & $>$24.80 & 22.51$\pm$0.09 & 19.53$\pm$0.03 & 19.19$\pm$0.06 & 19.03$\pm$0.07 & 18.78$\pm$0.14 \\
PSOJ029.5172-29.0886 & $>$23.24 & $>$23.54 & 21.94$\pm$0.09 & 19.21$\pm$0.04 & 19.12$\pm$0.07 & 19.21$\pm$0.09 & 18.42$\pm$0.16 \\
ATLASJ029.9915-36.565 & $>$22.35 & $>$22.95 & 21.71$\pm$0.20 & 19.91$\pm$0.05 & $>$19.99 & 19.43$\pm$0.13 & $>$19.99 \\
      ULASJ0203+0012 & $>$24.89 & $>$24.56 & $>$23.49 & 20.57$\pm$0.05 & 19.62$\pm$0.08 & 19.22$\pm$0.08 & 19.33$\pm$0.13 \\
     CFHQSJ0210-0456 & $>$25.09 & $>$24.75 & $>$23.82 & 22.86$\pm$0.20 & $>$21.76 & $>$21.13 & $>$19.57 \\
     CFHQSJ0221-0802 & $>$22.72 & $>$23.34 & $>$22.93 & 22.44$\pm$0.24 & $>$21.69 & $>$20.50 & $>$20.20 \\
PSOJ036.5078+03.0498 & $>$24.68 & $>$24.34 & 22.69$\pm$0.16 & 20.54$\pm$0.05 & 19.34$\pm$0.07 & 19.61$\pm$0.10 & 19.36$\pm$0.16 \\
     CFHQSJ0227-0605 & $>$25.45 & $>$25.05 & $>$24.06 & 22.74$\pm$0.17 & 21.35$\pm$0.24 & $>$21.12 & $>$20.18 \\
PSOJ037.9706-28.8389 & $>$25.13 & $>$24.77 & 23.28$\pm$0.20 & 20.65$\pm$0.04 & 20.67$\pm$0.17 & $>$20.62 & $>$19.80 \\
      SDSSJ0239-0045 & $>$22.55 & $>$22.85 & $>$22.59 & $>$22.88 & $>$21.70 & $>$21.01 & $>$20.38 \\
PSOJ040.0159+17.5458 & $>$24.94 & $>$24.24 & 23.64$\pm$0.25 & 20.87$\pm$0.04 & 20.27$\pm$0.10 & 20.28$\pm$0.14 & $>$19.91 \\
PSOJ042.6690-02.9174 & $>$24.16 & $>$24.05 & 23.28$\pm$0.28 & 20.32$\pm$0.04 & 20.37$\pm$0.12 & 20.44$\pm$0.18 & $>$19.67 \\
PSOJ045.1840-22.5408 & $>$25.29 & 23.43$\pm$0.17 & 22.40$\pm$0.14 & 20.32$\pm$0.05 & 19.68$\pm$0.10 & 19.47$\pm$0.12 & 19.23$\pm$0.21 \\     
      SDSSJ0303-0019 & $>$25.36 & $>$25.29 & $>$23.98 & 21.17$\pm$0.05 & 20.92$\pm$0.15 & $>$20.92 & $>$19.98 \\
       VIKJ0305-3150 & $>$24.99 & $>$21.66 & $>$23.99 & 21.96$\pm$0.18 & 21.32$\pm$0.28 & $>$20.82 & $>$18.50 \\
     CFHQSJ0316-1340 & $>$25.43 & $>$25.21 & 23.98$\pm$0.27 & 21.43$\pm$0.05 & 20.79$\pm$0.22 & 20.65$\pm$0.26 & $>$20.39 \\
PSOJ049.2934-26.5543 & $>$24.85 & $>$24.47 & $>$23.80 & 20.72$\pm$0.05 & 21.39$\pm$0.27 & $>$21.26 & $>$18.84 \\
       VIKJ0328-3253 & $>$23.39 & $>$23.57 & 22.77$\pm$0.23 & 20.17$\pm$0.04 & 20.14$\pm$0.10 & 19.98$\pm$0.14 & $>$19.49 \\
PSOJ053.9605-15.7956 & $>$23.52 & $>$23.57 & $>$22.96 & 20.84$\pm$0.05 & 20.77$\pm$0.18 & 20.69$\pm$0.28 & 19.74$\pm$0.28 \\
PSOJ055.4244-00.8035 & $>$23.41 & $>$23.59 & 22.00$\pm$0.19 & 20.51$\pm$0.07 & 20.12$\pm$0.23 & 20.04$\pm$0.26 & $>$19.28 \\
PSOJ056.7168-16.4769 & $>$22.25 & $>$22.63 & $>$22.45 & 19.92$\pm$0.04 & 19.55$\pm$0.10 & 20.07$\pm$0.28 & 19.40$\pm$0.28 \\
      SDSSJ0353+0104 & $>$23.83 & $>$24.82 & $>$23.80 & 20.86$\pm$0.06 & 19.76$\pm$0.11 & 20.06$\pm$0.17 & 19.11$\pm$0.21 \\
PSOJ060.5529+24.8567 & $>$24.53 & $>$24.30 & $>$23.38 & 19.94$\pm$0.04 & 19.64$\pm$0.12 & 19.55$\pm$0.17 & 19.41$\pm$0.26 \\
PSOJ065.4085-26.9543 & $>$23.74 & $>$23.91 & 22.43$\pm$0.24 & 20.15$\pm$0.05 & 19.04$\pm$0.10 & 19.10$\pm$0.10 & $>$18.94 \\
PSOJ065.5041-19.4579 & $>$24.00 & $>$23.83 & $>$22.85 & 19.80$\pm$0.04 & 20.03$\pm$0.24 & 19.08$\pm$0.12 & 19.16$\pm$0.23 \\
PSOJ071.0322-04.5591 & $>$25.19 & $>$24.54 & 22.88$\pm$0.13 & 20.08$\pm$0.03 & 20.26$\pm$0.10 & 20.22$\pm$0.14 & 19.85$\pm$0.19 \\
PSOJ071.4507-02.3332 & $>$24.95 & 22.27$\pm$0.05 & 21.47$\pm$0.05 & 19.10$\pm$0.03 & 18.97$\pm$0.06 & 18.95$\pm$0.06 & 19.11$\pm$0.16 \\
       DESJ0454-4448 & $>$23.73 & $>$24.18 & 22.77$\pm$0.15 & 20.54$\pm$0.04 & 20.32$\pm$0.12 & 19.95$\pm$0.15 & 19.64$\pm$0.30 \\
PSOJ075.9356-07.5061 & $>$24.94 & $>$24.66 & 23.32$\pm$0.26 & 20.50$\pm$0.04 & 20.34$\pm$0.15 & 20.37$\pm$0.21 & $>$19.31 \\
PSOJ089.9394-15.5833 & $>$24.87 & $>$24.64 & 22.81$\pm$0.22 & 19.45$\pm$0.03 & 18.84$\pm$0.06 & 18.51$\pm$0.06 & 18.28$\pm$0.09 \\
PSOJ108.4429+08.9257 & $>$24.40 & $>$24.04 & 22.36$\pm$0.19 & 19.21$\pm$0.03 & 19.15$\pm$0.08 & 18.94$\pm$0.09 & 18.89$\pm$0.14 \\
SDSSJ0818+1722 & $>$24.23 & $>$24.12 & 21.71$\pm$0.09 & 19.54$\pm$0.04 & 19.04$\pm$0.09 & 18.59$\pm$0.09 & $>$17.46 \\
ULASJ0828+2633 & $>$22.64 & $>$23.11 & $>$22.58 & 20.58$\pm$0.08 & 20.07$\pm$0.17 & 20.10$\pm$0.25 & $>$18.57 \\
PSOJ127.2817+03.0657 & $>$24.09 & $>$23.75 & $>$23.29 & 20.69$\pm$0.06 & 20.73$\pm$0.23 & 20.13$\pm$0.22 & 19.76$\pm$0.24 \\
SDSSJ0836+0054 & $>$24.64 & 22.20$\pm$0.05 & 21.18$\pm$0.05 & 18.75$\pm$0.03 & 18.51$\pm$0.06 & 18.17$\pm$0.06 & 17.99$\pm$0.09 \\
VIKJ0839+0015 & $>$24.49 & $>$24.34 & $>$23.56 & 21.51$\pm$0.08 & 20.61$\pm$0.29 & $>$20.68 & $>$18.15 \\
SDSSJ0842+1218 & $>$22.39 & $>$22.55 & $>$21.71 & 19.79$\pm$0.07 & 19.48$\pm$0.11 & 19.35$\pm$0.13 & 18.94$\pm$0.29 \\
HSCJ0859+0022 & $>$24.60 & $>$24.52 & $>$23.70 & 23.06$\pm$0.27 & $>$20.11 & $>$19.81 & $>$17.41 \\
PSOJ135.3860+16.2518 & $>$24.20 & $>$23.91 & 22.79$\pm$0.27 & 20.73$\pm$0.07 & 20.51$\pm$0.18 & $>$20.49 & $>$19.72 \\
PSOJ135.8704-13.8336 & $>$24.86 & $>$24.36 & 23.20$\pm$0.23 & 20.25$\pm$0.04 & 20.88$\pm$0.22 & 20.42$\pm$0.25 & $>$19.93 \\
PSOJ157.9070-02.6599 & $>$24.23 & 22.28$\pm$0.09 & 22.53$\pm$0.15 & 20.16$\pm$0.04 & 20.03$\pm$0.12 & 20.38$\pm$0.22 & $>$19.23 \\
PSOJ159.2257-02.5438 & $>$24.85 & $>$24.63 & $>$23.93 & 20.41$\pm$0.05 & 19.74$\pm$0.12 & 19.81$\pm$0.18 & $>$18.65 \\
SDSSJ1044-0125 & $>$24.74 & 22.80$\pm$0.08 & 21.82$\pm$0.07 & 19.19$\pm$0.03 & 18.94$\pm$0.07 & 18.88$\pm$0.08 & $>$18.58 \\
CFHQSJ1059-0906 & $>$25.25 & $>$24.81 & $>$23.57 & 20.74$\pm$0.04 & 20.34$\pm$0.15 & 20.35$\pm$0.20 & $>$18.36 \\
PSOJ167.6415-13.4960 & $>$24.54 & $>$24.44 & $>$23.66 & 22.12$\pm$0.13 & 20.70$\pm$0.21 & 20.28$\pm$0.23 & $>$19.46 \\
PSOJ174.7920-12.2845 & $>$24.87 & 23.88$\pm$0.16 & 22.66$\pm$0.14 & 20.25$\pm$0.04 & 20.37$\pm$0.19 & $>$20.25 & $>$19.99 \\
PSOJ175.4091-20.2654 & $>$25.02 & 23.95$\pm$0.21 & 22.86$\pm$0.14 & 20.29$\pm$0.04 & 19.80$\pm$0.10 & 19.59$\pm$0.10 & $>$18.68 \\
VIKJ1152+0055 & $>$24.90 & $>$24.72 & $>$23.90 & 22.30$\pm$0.15 & 21.18$\pm$0.28 & $>$21.16 & $>$20.21 \\
PSOJ183.2991-12.7676 & $>$24.77 & 23.49$\pm$0.13 & 21.79$\pm$0.07 & 18.81$\pm$0.04 & 19.02$\pm$0.06 & 18.80$\pm$0.07 & 18.99$\pm$0.22 \\
PSOJ184.3389+01.5284 & $>$24.35 & $>$24.40 & $>$23.61 & 21.29$\pm$0.07 & $>$20.98 & $>$20.62 & $>$19.36 \\
PSOJ187.1047-02.5609 & $>$24.32 & $>$23.94 & $>$23.18 & 21.02$\pm$0.07 & 20.70$\pm$0.18 & 20.41$\pm$0.21 & $>$20.06 \\
PSOJ187.3050+04.3243 & $>$24.82 & $>$24.55 & $>$23.79 & 21.16$\pm$0.06 & 20.94$\pm$0.29 & $>$20.68 & $>$19.64 \\
SDSSJ1306+0356 & $>$24.84 & $>$24.48 & 23.04$\pm$0.21 & 19.71$\pm$0.04 & $>$19.02 & $>$19.15 & $>$17.61 \\
ULASJ1319+0950 & $>$25.31 & $>$24.50 & 22.74$\pm$0.12 & 20.16$\pm$0.04 & 19.60$\pm$0.08 & 19.53$\pm$0.11 & $>$19.50 \\
PSOJ209.2058-26.7083 & $>$24.39 & $>$23.57 & 21.98$\pm$0.10 & 19.41$\pm$0.03 & 19.20$\pm$0.08 & 19.20$\pm$0.10 & 19.32$\pm$0.22 \\
PSOJ210.8297+09.0474 & $>$24.26 & $>$23.60 & $>$22.74 & 20.39$\pm$0.07 & $>$20.17 & $>$20.02 & $>$19.99 \\
PSOJ210.8722-12.0094 & $>$25.01 & $>$24.64 & $>$23.30 & 21.04$\pm$0.06 & 20.96$\pm$0.23 & $>$20.27 & $>$19.78 \\
PSOJ212.2974-15.9865 & $>$24.61 & $>$24.24 & $>$23.74 & 21.20$\pm$0.07 & 20.53$\pm$0.16 & 20.34$\pm$0.19 & $>$20.03 \\
SDSSJ1411+1217 & $>$24.07 & $>$23.49 & $>$22.82 & 19.68$\pm$0.05 & 20.09$\pm$0.20 & 19.48$\pm$0.23 & $>$19.25 \\
PSOJ213.3629-22.5617 & $>$24.41 & $>$24.17 & $>$23.14 & 19.73$\pm$0.04 & 19.94$\pm$0.12 & 20.04$\pm$0.21 & $>$18.99 \\
PSOJ213.7329-13.4803 & $>$24.39 & $>$23.97 & 22.94$\pm$0.29 & 20.95$\pm$0.07 & 20.70$\pm$0.26 & 20.20$\pm$0.28 & $>$19.37 \\
PSOJ215.1514-16.0417 & $>$24.50 & 22.85$\pm$0.12 & 22.26$\pm$0.13 & 19.07$\pm$0.03 & 18.79$\pm$0.06 & 18.61$\pm$0.07 & 18.59$\pm$0.26 \\
PSOJ217.0891-16.0453 & $>$24.34 & $>$24.16 & 22.46$\pm$0.21 & 20.29$\pm$0.05 & 19.59$\pm$0.10 & 19.97$\pm$0.19 & $>$19.43 \\
PSOJ217.9185-07.4120 & $>$24.90 & $>$24.49 & 23.41$\pm$0.24 & 20.74$\pm$0.04 & 19.79$\pm$0.08 & 19.73$\pm$0.11 & 19.77$\pm$0.21 \\
CFHQSJ1509-1749 & $>$23.17 & $>$23.73 & $>$23.02 & 20.05$\pm$0.04 & 19.71$\pm$0.09 & 19.66$\pm$0.16 & 19.69$\pm$0.27 \\
PSOJ228.6871+21.2388 & $>$25.34 & $>$24.92 & $>$23.65 & 20.64$\pm$0.04 & 20.51$\pm$0.17 & 20.19$\pm$0.21 & $>$19.16 \\
PSOJ235.9450+17.0079 & $>$24.77 & 23.52$\pm$0.14 & 22.64$\pm$0.16 & 20.35$\pm$0.04 & 20.21$\pm$0.12 & 19.92$\pm$0.18 & $>$19.47 \\
PSOJ236.2912+16.6088 & $>$24.91 & $>$23.96 & $>$23.50 & 21.04$\pm$0.08 & $>$20.70 & $>$20.09 & $>$20.15 \\
PSOJ238.8510-06.8976 & $>$24.13 & $>$23.92 & $>$22.91 & 20.32$\pm$0.06 & 20.11$\pm$0.18 & 20.16$\pm$0.25 & $>$19.94 \\
PSOJ239.7124-07.4026 & $>$24.95 & $>$23.50 & 22.33$\pm$0.15 & 19.69$\pm$0.03 & 19.27$\pm$0.09 & 19.22$\pm$0.11 & 18.92$\pm$0.18 \\
PSOJ242.4397-12.9816 & $>$23.91 & $>$23.15 & $>$22.42 & 19.72$\pm$0.04 & 19.42$\pm$0.11 & 19.49$\pm$0.16 & 19.21$\pm$0.28 \\
PSOJ245.0636-00.1978 & $>$24.34 & $>$24.10 & $>$23.38 & 21.57$\pm$0.11 & 20.83$\pm$0.27 & $>$20.57 & $>$19.77 \\
PSOJ247.2970+24.1277 & $>$25.59 & $>$25.26 & $>$24.06 & 21.07$\pm$0.05 & 20.02$\pm$0.10 & 19.86$\pm$0.15 & $>$19.45 \\
PSOJ261.0364+19.0286 & $>$25.02 & $>$24.69 & 23.96$\pm$0.26 & 22.12$\pm$0.09 & 20.72$\pm$0.17 & 20.39$\pm$0.17 & $>$19.94 \\
PSOJ267.0021+22.7812 & $>$22.73 & $>$23.28 & $>$22.97 & 21.76$\pm$0.19 & $>$20.75 & $>$20.36 & $>$19.91 \\
PSOJ308.0416-21.2339 & $>$24.66 & $>$24.35 & $>$23.28 & 20.47$\pm$0.05 & 20.03$\pm$0.11 & 19.63$\pm$0.12 & 19.47$\pm$0.17 \\
PSOJ308.4829-27.6485 & $>$23.95 & 23.56$\pm$0.28 & 22.22$\pm$0.13 & 19.66$\pm$0.04 & 19.46$\pm$0.09 & 19.39$\pm$0.12 & 18.87$\pm$0.19 \\
SDSSJ2053+0047 & $>$23.50 & $>$23.71 & $>$23.31 & 21.21$\pm$0.07 & $>$20.93 & 20.73$\pm$0.25 & $>$19.92 \\
      SDSSJ2054-0005 & $>$23.67 & $>$24.02 & $>$23.01 & 20.67$\pm$0.04 & 20.33$\pm$0.11 & 20.04$\pm$0.15 & 19.31$\pm$0.22 \\
     CFHQSJ2100-1715 & $>$23.99 & $>$24.16 & $>$23.73 & 21.45$\pm$0.06 & 20.45$\pm$0.12 & 19.91$\pm$0.12 & 19.84$\pm$0.23 \\
PSOJ319.6040-10.9326 & $>$24.30 & $>$24.43 & 22.41$\pm$0.12 & 19.99$\pm$0.03 & 19.96$\pm$0.09 & 19.72$\pm$0.11 & 19.47$\pm$0.30 \\
SDSS211951.9-004020 & $>$24.75 & $>$24.66 & $>$23.45 & 22.30$\pm$0.17 & $>$20.65 & $>$20.08 & $>$18.72 \\
PSOJ320.8703-24.3604 & $>$23.89 & $>$24.05 & 22.84$\pm$0.27 & 20.27$\pm$0.05 & 20.19$\pm$0.12 & 19.42$\pm$0.10 & $>$19.64 \\
PSOJ323.1382+12.2986 & $>$23.79 & $>$24.58 & 23.40$\pm$0.26 & 20.29$\pm$0.04 & 19.32$\pm$0.07 & 19.20$\pm$0.10 & 19.07$\pm$0.22 \\
SDSSJ2147+0107 & $>$23.62 & $>$24.05 & $>$23.45 & 21.71$\pm$0.10 & $>$21.27 & $>$20.80 & $>$19.69 \\
PSOJ328.7339-09.5076 & $>$23.70 & $>$24.03 & $>$23.14 & 20.58$\pm$0.05 & 20.13$\pm$0.12 & 20.00$\pm$0.15 & $>$19.47 \\
IMSJ2204+0012 & $>$23.58 & $>$24.02 & $>$23.16 & 22.96$\pm$0.23 & $>$21.34 & $>$20.77 & $>$19.69 \\
HSCJ2216-0016 & $>$25.22 & $>$25.18 & $>$24.25 & 22.91$\pm$0.19 & $>$21.87 & $>$21.25 & $>$19.98 \\
SDSSJ2220-0101 & $>$24.83 & $>$23.73 & 22.18$\pm$0.10 & 20.82$\pm$0.05 & 20.34$\pm$0.14 & 20.07$\pm$0.16 & $>$20.28 \\
SDSSJ2228+0110 & $>$25.73 & $>$24.51 & $>$23.64 & 22.18$\pm$0.14 & 21.28$\pm$0.28 & $>$20.71 & $>$18.92 \\
CFHQSJ2229+1457 & $>$25.19 & $>$24.96 & $>$23.36 & 21.87$\pm$0.08 & $>$21.24 & $>$21.04 & $>$20.60 \\
PSOJ340.2041-18.6621 & $>$25.25 & $>$25.00 & 23.56$\pm$0.22 & 20.17$\pm$0.03 & 20.05$\pm$0.09 & 19.91$\pm$0.15 & 19.60$\pm$0.24 \\
SDSSJ2310+1855 & $>$24.71 & $>$24.25 & 21.76$\pm$0.13 & 19.24$\pm$0.04 & 18.63$\pm$0.06 & 18.73$\pm$0.10 & $>$19.99 \\
CFHQSJ2329-0403 & $>$24.00 & $>$24.30 & $>$23.36 & 21.90$\pm$0.17 & $>$20.21 & $>$19.59 & $>$17.72 \\
PSOJ357.8289+06.4019 & $>$24.57 & $>$23.78 & $>$23.66 & 21.43$\pm$0.09 & $>$21.12 & $>$20.85 & $>$20.02 \\
PSOJ359.1352-06.3831 & $>$23.87 & $>$23.57 & $>$22.36 & 19.90$\pm$0.05 & 19.89$\pm$0.15 & 19.76$\pm$0.20 & 18.91$\pm$0.21 \\
SDSSJ2356+0023 & $>$24.54 & $>$24.23 & $>$23.46 & $>$23.10 & $>$20.66 & $>$20.31 & $>$19.54 \\
\end{longtable}

\begin{table}[ht]
  \caption{Log of spectroscopic observations of QSO pair candidates in 2019.
    All are 900 sec exposures with grating R150 and a 0\farcs9 slit,
    except for the FIRE spectrum (1\asec\ slit).}
   \vspace{-0.2cm}
     \begin{tabular}{lcl}
     \hline
     \noalign{\smallskip}
     QSO candidate & Date / Time   & Instrument/ \\
                   & $\!\!$(MM-DD HH:MM)$\!\!$ & Telescope \\ 
      \noalign{\smallskip}
      \hline
      \noalign{\smallskip}
      IMSJ2204+0012        & 06-26 08:01 & FIRE/Baade$\!\!$ \\
      SDSSJ2054-0005\_4    & 09-03 01:34 & GMOS-S \\
      SDSSJ2054-0005\_3    & 09-03 02:04 & GMOS-S \\
      PSOJ261.0364+19.0286 & 09-03 08:59 & GMOS-N \\
      PSOJ267.0021+22.7812 & 09-03 09:30 & GMOS-N \\
      CFHQSJ1509-1749      & 09-04 00:03 & GMOS-S \\
      PSOJ011.3899+09.0325 & 09-06 14:22 & GMOS-N \\
      ULASJ0148+0600       & 09-06 14:52 & GMOS-N \\
      PSOJ009.7355-10.4316 & 09-19 05:34 & GMOS-S \\
      PSOJ045.1840-22.5408 & 09-22 04:38 & GMOS-S \\
      PSOJ071.0322-04.5591 & 10-29 05:29 & GMOS-S \\
      \noalign{\smallskip}
      \hline
      \noalign{\smallskip}
     \end{tabular}
   \label{speclog}
\end{table}

\begin{table*}[th]
  \tiny
  \caption{Pair candidates with existing ALLWISE channel W1/W2 photometry, not corrected for foreground-A$_{\rm V}$,
    all in the AB system. The ID column gives the best-fitting Le\,PHARE spectral template. The object marked
    in boldface had a QSO template as best-fit
    Le\,PHARE output, which is inconsistent with our GMOS spectrum - we therefore provide the next best-fit template, a low-redshift galaxy, as potential identification.}
    \vspace{-0.2cm}
      \setlength{\tabcolsep}{0.4em}
       \hspace*{-0.8cm}\begin{tabular}{@{}rlcccccccccl}
      \hline
      \noalign{\smallskip}
      Prime QSO & RA/Decl. (2000.0) & \gp & \rp & \ip & \zp & $J$ & $H$ & $K$ & W1 & W2 & ID \\  
      \noalign{\smallskip}
      \hline
      \noalign{\smallskip}
      CFHQSJ0227-0605 & ~~36.9473 $-$06.0495 & $>$25.56 & 24.26$\pm$0.17 & 21.64$\pm$0.05 & 20.11$\pm$0.04 & 18.91$\pm$0.06 & 18.65$\pm$0.06 & 18.84$\pm$0.13 & 19.67$\pm$0.10 & 20.38$\pm$0.33 & L0 \\
        HSCJ0859+0022 & 134.7528 +00.3927    & $>$24.90 & 23.30$\pm$0.10 & 20.88$\pm$0.04 & 19.29$\pm$0.03 & 18.05$\pm$0.06 & 17.82$\pm$0.06 & $>$17.72 & 18.67$\pm$0.05 & 19.26$\pm$0.16 & L0 \\ 
 PSOJ037.9706-28.8389 & ~~38.0071 $-$28.8519 & $>$25.02 & $>$24.59 & 23.27$\pm$0.21 & 21.75$\pm$0.08 & 20.27$\pm$0.12 & 19.99$\pm$0.14 & 19.70$\pm$0.23 & 20.29$\pm$0.15 & -- & M9 \\  
 PSOJ055.4244-00.8035 & ~~55.4110 $-$00.8079 & $>$23.72 & $>$23.52 & $>$23.25 & 21.71$\pm$0.19 & 20.08$\pm$0.24 & $>$20.26 & $>$19.21 & 20.04$\pm$0.15 & 19.95$\pm$0.30 & QSO\\
 PSOJ075.9356-07.5061 & ~~75.9660 $-$07.4760 & $>$25.59& $>$25.11 & 22.03$\pm$0.07 & 20.43$\pm$0.04 & 19.27$\pm$0.07 & 18.95$\pm$0.07 & 18.97$\pm$0.28 & 19.64$\pm$0.11 & $>$19.93 & L3 \\  
 PSOJ135.8704-13.8336 & 135.8657 $-$13.8220  & $>$25.08 & 23.44$\pm$0.11 & 21.13$\pm$0.04 & 19.43$\pm$0.03 & 17.91$\pm$0.05 & 17.62$\pm$0.05 & 17.86$\pm$0.08 & 18.29$\pm$0.04 & 18.78$\pm$0.10  & L3 \\  
 PSOJ340.2041-18.6621 & 340.2045 $-$18.6680  & $>$25.36 & $>$24.87 & 22.09$\pm$0.06 & 20.53$\pm$0.04 & 19.08$\pm$0.06 & 18.81$\pm$0.07 & 19.14$\pm$0.15 & 19.65$\pm$0.12 & 19.77$\pm$0.26 & L1 \\  
       SDSSJ0005-0006 & ~~~~1.4755 $-$00.0968& $>$25.06 & 23.76$\pm$0.18 & 21.40$\pm$0.05 & 19.67$\pm$0.03 & 18.14$\pm$0.05 & 17.94$\pm$0.06 & 17.88$\pm$0.08 & 18.67$\pm$0.06 & 18.79$\pm$0.11 & M9 \\  
  {\bf ULASJ0148+0600} & ~~27.1558 +06.0265  & $>$25.46 & $>$24.89 & $>$23.92 & 22.41$\pm$0.13 & 21.06$\pm$0.18 & 20.93$\pm$0.27 & $>$19.28 & 19.90$\pm$0.14 & $>$19.89 & QSO/Gal\\
     \noalign{\smallskip}
     \hline
   \end{tabular}
   \label{photWise}
\end{table*}

\begin{table*}[th]
  \small
  \caption{Pair candidates without ALLWISE photometry, not corrected for foreground-A$_{\rm V}$,
    all in the AB system. The ID column gives the best-fitting Le\,PHARE spectral template.
    The objects marked in boldface are those
    where the Le\,PHARE fit preferred a QSO high-z template, and we obtained
    optical spectroscopy.
    Except for PSOJ071.0322-04.5591 our spectra argue against a QSO
      nature, and we therefore provide the next best-fitting template,
      mostly a dwarf star (see text). For objects with a $*$ in the last column
     see Tab. \ref{photunWise}.}
    \vspace{-0.2cm}
     \setlength{\tabcolsep}{0.5em}
      \begin{tabular}{rlcccccccl}
      \hline
      \noalign{\smallskip}
      Prime QSO & RA/Decl. (2000.0) & \gp & \rp & \ip & \zp & $J$ & $H$ & $K$ & ID\\
      \noalign{\smallskip}
      \hline
      \noalign{\smallskip}
        {\bf CFHQSJ1509-1749} & 227.4413 $-$17.8303 & $>$23.47 & $>$23.94 & $>$23.32 & 22.31$\pm$0.16 & 21.28$\pm$0.27 & $>$21.01 & $>$20.32 & QSO/M8  \\
                DESJ0454-4448 & ~~73.5198 $-$44.8196& $>$23.65 & $>$24.21 & 23.49$\pm$0.26 & 21.60$\pm$0.08 & 20.05$\pm$0.10 & 20.41$\pm$0.20 & $\!\!$19.70$\pm$0.29 & L1\\
          {\bf IMSJ2204+0012} & 331.0878 +01.2344   & $>$23.79 & $>$24.20 & $>$23.66 & 22.09$\pm$0.11 & 20.95$\pm$0.21 & 20.37$\pm$0.21 & $>$19.71 & QSO/M8 \\
   {\bf PSOJ009.7355-10.4316} & ~~~~9.7185$-$10.4161& $>$24.52 & $>$24.36 & $>$23.17 & 21.79$\pm$0.12 & 20.85$\pm$0.28 & 20.54$\pm$0.30 & $>$19.99 & QSO/M9 \\
   {\bf PSOJ011.3899+09.0325} & ~~11.3462 +09.0813  & $>$23.84 & $>$23.84 & $>$23.51 & 21.54$\pm$0.09 & $>$21.67 & 20.96$\pm$0.33 & $>$20.13 & QSO/M6 \\
         PSOJ036.5078+03.0498 & ~~36.5245 +03.0958  & $>$24.79 & $>$24.69 & 23.30$\pm$0.28 & 21.60$\pm$0.09 & 20.06$\pm$0.10 & 20.02$\pm$0.14 & $>$20.08 & $*$\\
   {\bf PSOJ045.1840-22.5408} & ~~45.1936 $-$22.5280& $>$25.37 & $>$24.59 & $>$23.60 & 21.97$\pm$0.12 & 20.66$\pm$0.19 & 20.16$\pm$0.20 & $>$19.33 & QSO/M9 \\
   {\bf PSOJ071.0322-04.5591} & ~~71.0745 $-$04.5348& $>$24.87 & $>$24.38 & $>$24.08 & 22.29$\pm$0.13 & 21.22$\pm$0.20 & $>$20.94 & $\!\!$19.74$\pm$0.18 & $*$\\
         PSOJ071.4507-02.3332 & ~~71.4815 $-$02.3637& $>$25.08 & $>$24.75 & 23.25$\pm$0.18 & 21.65$\pm$0.07 & 20.45$\pm$0.11 & 20.52$\pm$0.18 & $>$20.11 & $*$\\
   {\bf PSOJ261.0364+19.0286} & 261.0269   +19.0228 & $>$25.23 & $>$24.85 & 23.90$\pm$0.23 & 22.23$\pm$0.09 & $>$21.48 & $>$20.84 & $>$19.89 & QSO/? \\
   {\bf PSOJ267.0021+22.7812} & 266.9979   +22.7922 & $>$23.07 & $>$23.51 & $>$23.14 & 21.35$\pm$0.09 & 20.85$\pm$0.23 & 20.28$\pm$0.25 & $>$19.86 & QSO/T6 \\
         PSOJ328.7339-09.5076 & 328.7633 $-$09.4779 & $>$23.80 & $>$24.10 & 23.25$\pm$0.32 & 21.63$\pm$0.09 & 20.44$\pm$0.14 & 20.03$\pm$0.13 & $>$19.38 & $*$\\
               SDSSJ0303-0019 & ~~45.8859 $-$00.3281& $>$25.72 & $>$25.16 & $>$24.49 & 22.89$\pm$0.14 & $>$22.02 & $>$21.53 & $>$20.06 & T8 \\  
            SDSSJ2053+0047\_1 & 313.3294   +00.7677 & $>$23.53 & $>$23.91 & $>$23.41 & 22.20$\pm$0.14 & 21.01$\pm$0.25 & 21.04$\pm$0.33 & $>$19.99 & M9 \\
            SDSSJ2053+0047\_3 & 313.3766   +00.7606 & $>$23.78 & $>$23.70 & $>$23.49 & 22.06$\pm$0.15 & 21.34$\pm$0.30& $>$20.91 & $>$20.34 & M5 \\
      {\bf SDSSJ2054-0005\_3} & 313.5334 $-$00.0914 & $>$24.00 & $>$24.25 & $>$23.71 & 21.81$\pm$0.10 & 20.82$\pm$0.18 & 20.44$\pm$0.25 & $>$19.73 & $*$ \\  
      {\bf SDSSJ2054-0005\_4} & 313.5450 $-$00.0525 & $>$24.00 & $>$24.25 & $>$23.71 & 22.16$\pm$0.12 & 20.98$\pm$0.20 & 20.79$\pm$0.28 & $>$19.98 & $*$ \\
      \noalign{\smallskip}
      \hline
   \end{tabular}
   \label{photnoWise}
\end{table*}

\begin{table*}[th]
  \caption{Pair candidates from Tab. \ref{photnoWise} with unWISE or CatWISE photometry (AB system). The ID columns give the best-fitting Le\,PHARE spectral template without (ID1) and with un/Cal-WISE photometry (ID2).}
    \vspace{-0.2cm}
      \begin{tabular}{rccccll}
      \hline
      \noalign{\smallskip}
    Prime QSO & \multicolumn{2}{c}{unWISE} & \multicolumn{2}{c}{CatWISE} & ID1 & ID2 \\
              & W1 & W2 & W1 & W2 & & \\
      \noalign{\smallskip}
      \hline
      \noalign{\smallskip}
   PSOJ036.5078+03.0498 & -- & -- & 20.96$\pm$0.16 & 21.03$\pm$0.32 & L2 & L0 \\
   {\bf PSOJ071.0322-04.5591} & 19.47$\pm$0.04 & 19.57$\pm$0.09 & 19.39$\pm$0.04 & 19.63$\pm$0.09 & QSO & low-z Gal\\
   PSOJ071.4507-02.3332 & 21.09$\pm$0.04 & -- & 21.37$\pm$0.21 & 21.45$\pm$0.45  & M9 & M9 \\
   PSOJ328.7339-09.5076 & 19.06$\pm$0.05 & 19.26$\pm$0.10 & 19.13$\pm$0.04 & 19.31$\pm$0.07 & M8 & low-z Gal \\
   {\bf SDSSJ2054-0005\_3} & 19.73$\pm$0.08 & -- & 19.80$\pm$0.06 & 20.46$\pm$0.17 & QSO & ? \\  
   {\bf SDSSJ2054-0005\_4} & -- & -- & 21.06$\pm$0.18 & 21.15$\pm$NULL & QSO & M8 \\ 
      \noalign{\smallskip}
      \hline
   \end{tabular}
   \label{photunWise}
\end{table*}

\section{Conclusions}

No candidate QSO pair has been found in a search of 116 spectroscopically
confirmed redshift 6 quasars within our 0.1--3.3 $h^{-1}$ cMpc search radius
with a companion brighter than M$_{1450}$ (AB)$< -26$ mag.
Although this result is consistent with
cosmological-scale galaxy evolution simulations (given the redshift,
search volume, and luminosity limits), the serendipitously discovered pairs
at z=4.26 \citep{Schneider+2000} and z=5.02 \citep{McGreer+2016} suggested
one pair every 40 QSOs, or 2.5\%, implying an  empirical
expectation of 3 pairs in our sample of 116 search targets.
At first glance, this result suggests little clustering at redshift 6.
However, with updated knowledge of the faint end of the luminosity
function, the $w_{\rm p}$ statistics returns more than an order of magnitude
higher value than inferred from the low-redshift clustered QSOs; thus,
given the still low statistics
in our search and the uncertainties (in particular in the unknown search
volumes) of the serendipitous detections,
we are not able to claim that the clustering at redshift 6 is indeed
smaller than up to redshift 5.

As a consequence of our search, we do find a higher than expected rate of
ultracool dwarfs. While the excess of L dwarfs is marginally consistent
with statistics and could be due to the accidental lack of a single
source about 1 mag fainter, this argument does not hold for the T dwarfs.
The observed excess of a factor of $\sim$20 suggests either (i) strong
surface density variations over the sky,
(ii) a lack of appropriate templates for other source types
(e.g. low-redshift red galaxies)
such that the T templates just happen to be the closest in shape to
the observed SEDs, or (iii) an under-estimate of the true rate of
T dwarfs. A larger survey at our depth would be needed to settle this
question.

\begin{acknowledgements}
We thank the referee for the detailed comments.
Part of the funding for GROND (both hardware as well as personnel)
was generously granted from the Leibniz-Prize to Prof. G. Hasinger
(DFG grant HA 1850/28-1). PS acknowledges support through the
Sofja Kovalevskaja Award from the Alexander von Humboldt Foundation
of Germany while at MPE. JG greatly acknowledges the GROND observers
Thomas Kr\"uhler, Tassilo Schweyer, Simon Steinmassl, Helmut Steinle
and Phil Wiseman. \\

We explicitly acknowledge the open attitude of providing the
detailed model predictions for dwarfs by \cite{RyanReid2016} for download,
which only enabled the creation of Fig. \ref{BD_Density}. \\

The reported spectroscopy is based on observations obtained at the international Gemini Observatory, a program of NSF's NOIRLab, which is managed by the Association of Universities for Research in Astronomy (AURA) under a cooperative agreement with the National Science Foundation on behalf of the Gemini Observatory partnership: the National Science Foundation (United States), National Research Council (Canada), Agencia Nacional de Investigación y Desarrollo (Chile), Ministerio de Ciencia, Tecnología e Innovación (Argentina), Ministério da Ciência, Tecnologia, Inovações e Comunicações (Brazil), and Korea Astronomy and Space Science Institute (Republic of Korea). \\

This paper includes data gathered with the 6.5 m Magellan Telescopes
located at Las Campanas Observatory, Chile. \\

This research has made use of the NASA/IPAC Infrared Science Archive,
which is operated by the Jet Propulsion Laboratory, California Institute
of Technology, under contract with the National Aeronautics and Space
Administration. \\

For photometric calibration, the Sloan Digital Sky Survey, PanSTARRS,
SkyMapper and 2MASS are acknowledged:
Funding for SDSS-III has been provided by the Alfred P. Sloan Foundation, the Participating Institutions, the National Science Foundation, and the U.S. Department of Energy Office of Science. The SDSS-III web site is http://www.sdss3.org/.
SDSS-III is managed by the Astrophysical Research Consortium for the Participating Institutions of the SDSS-III Collaboration including the University of Arizona, the Brazilian Participation Group, Brookhaven National Laboratory, Carnegie Mellon University, University of Florida, the French Participation Group, the German Participation Group, Harvard University, the Instituto de Astrofisica de Canarias, the Michigan State/Notre Dame/JINA Participation Group, Johns Hopkins University, Lawrence Berkeley National Laboratory, Max Planck Institute for Astrophysics, Max Planck Institute for Extraterrestrial Physics, New Mexico State University, New York University, Ohio State University, Pennsylvania State University, University of Portsmouth, Princeton University, the Spanish Participation Group, University of Tokyo, University of Utah, Vanderbilt University, University of Virginia, University of Washington, and Yale University. 
The Pan-STARRS1 Surveys (PS1) have been made possible through contributions of the Institute for Astronomy, the University of Hawaii, the Pan-STARRS Project Office, the Max-Planck Society and its participating institutes, the Max Planck Institute for Astronomy, Heidelberg and the Max Planck Institute for Extraterrestrial Physics, Garching, The Johns Hopkins University, Durham University, the University of Edinburgh, Queen's University Belfast, the Harvard-Smithsonian Center for Astrophysics, the Las Cumbres Observatory Global Telescope Network Incorporated, the National Central University of Taiwan, the Space Telescope Science Institute, the National Aeronautics and Space Administration under Grant No. NNX08AR22G issued through the Planetary Science Division of the NASA Science Mission Directorate, the National Science Foundation under Grant No. AST-1238877, the University of Maryland, and Eotvos Lorand University (ELTE).
The national facility capability for SkyMapper has been funded
through ARC LIEF grant LE130100104 from the Australian Research Council,
awarded to the University of Sydney, the Australian National University, Swinburne University of Technology, the University of Queensland, the University of Western Australia, the University of Melbourne, Curtin University of Technology, Monash University and the Australian Astronomical Observatory. SkyMapper is owned and operated by The Australian National University's Research School of Astronomy and Astrophysics. The survey data were processed and provided by the SkyMapper Team at ANU. The SkyMapper node of the All-Sky Virtual Observatory (ASVO) is hosted at the National Computational Infrastructure (NCI). Development and support the SkyMapper node of the ASVO has been funded in part by Astronomy Australia Limited (AAL) and the Australian Government through the Commonwealth's Education Investment Fund (EIF) and National Collaborative Research Infrastructure Strategy (NCRIS), particularly the National eResearch Collaboration Tools and Resources (NeCTAR) and the Australian National Data Service Projects (ANDS).
This publication makes use of data products from the Two Micron All Sky Survey, which is a joint project of the University of Massachusetts and the Infrared Processing and Analysis Center/California Institute of Technology, funded by the National Aeronautics and Space Administration and the National Science Foundation.
\end{acknowledgements}

{\it Facilities: Max Planck:2.2m (GROND), SDSS, PanSTARRS, SkyMapper, IRSA, ALLWISE, Magellan, Gemini}


\begin{thebibliography}{}

\bibitem[Arnouts et al.(1999)]{Arnouts+1999} Arnouts S., Cristiani S.,
  Moscardini L. et al. 1999, MNRAS 310, 540

\bibitem[Ayromlou et al.(2021a)]{Ayromlou+2021a} Ayromlou M., Nelson D., Yates R.M., et al. 2021, MNRAS 502, 1051
  
\bibitem[Ayromlou et al.(2021b)]{Ayromlou+2021b} Ayromlou M., Kauffmann G., Yates R.M., et al. 2021, MNRAS 505, 492
  
\bibitem[Ba\~nados et al.(2016)]{Banados+2016} Ba\~nados E., Venemans B.P., 
Decarli R. et al. 2016, ApJ Suppl. 227, 11 

\bibitem[Ba\~nados et al.(2018)]{Banados+2018} Ba\~nados E., Venemans B.P., 
Mazzucchelli C., et al. 2018, Nat. 553, 473

\bibitem[Begelman et al.(2006)]{2006MNRAS.370..289B} Begelman M.C., 
   Volonteri M., Rees M.J., 2006, MNRAS 370, 289
  
\bibitem[Bhowmick et al.(2019)]{Bhowmick+2019} Bhowmick A.K., DiMatteo T.,
  Eftekharzadeh S., Myers A.D., 2019, MNRAS 485, 2026
  
\bibitem[Bonoli et al.(2009)]{B09} Bonoli S., Marulli F., Springel V., et al., 2009, MNRAS, 396 423

\bibitem[Bromm \& Loeb(2003)]{BrommLoeb2003} Bromm V., Loeb A., 2003,
  ApJ 596, 34
  
\bibitem[Cappelluti et al.(2010)]{Cappelluti+2010} Cappelluti N.,
Ajello M., Burlon D. et al. 2010, ApJ 716, L209

\bibitem[Carnall et al.(2015)]{Carnall+2015} Carnall A.C., Shanks T.,
  Chehade B. et al. 2015, MNRAS 451, L16

\bibitem[Chambers et al.(2016)]{Chambers+2016} Chambers K.C., Magnier E.A.,
  N. Metcalfe N. et al. 2016, arXiv:1612.05560

\bibitem[Chehade et al.(2016)]{Chehade+2016} Chehade B., Shanks T.,
 Findlay J. et al. 2016, MNRAS 459, 1179

\bibitem[Connor et al.(2020)]{Connor+2020} Connor T., Ba\~nados E.,
  Mazzucchelli C. et al. 2020, ApJ 900, 189
 
\bibitem[Costa et al.(2014)]{Costa+2014} Costa  T., Sijacki D., Trenti M.,
  Haehnelt M.G., 2014, MNRAS 439, 2146

\bibitem[Croton et al.(2006)]{C06} Croton D.J., Springel V., White S.D.M.,
  et al., 2006, MNRAS 365, 11

\bibitem[Das et al.(2021)]{Das+2021} Das A., Schleicher D.R.G.,
  Leigh N.W.C., Boekholt T.C.N., 2021, MNRAS 503, 1051
  
\bibitem[Davis \& Peebles(1983)]{DavisPeebles1983} Davis M., Peebles P.J.E.,
  1983, ApJ 267, 465

\bibitem[Decarli et al.(2017)]{Decarli+2017} Decarli R., Walter F.,
  Venemans B.P. et al. 2017, Nat 545, 457 

\bibitem[DeGraf \& Sijacki(2017)]{DeG17} DeGraf C., Sijacki D., 2017,
  MNRAS, 466 3331
  
\bibitem[Devecchi \& Volonteri(2009)]{2009ApJ...694..302D} Devecchi B.,
  Volonteri M., 2009, ApJ 694, 302

\bibitem[Devecchi et al.(2012)]{Devecchi2012} Devecchi B., Volonteri M.,
    Rossi E.M., et al. 2012, MNRAS 421, 1465

\bibitem[Dijkstra et al.(2008)]{2008MNRAS.391.1961D} Dijkstra M.,
    Haiman Z., Mesinger A., Wyithe J.S.B., 2008, MNRAS 391, 1961
    
\bibitem[Di Matteo et al.(2017)]{DiMatteo+2017} Di Matteo T., Croft R.A.C.,
    Feng Y., et al., 2017, MNRAS 467, 4243
    
\bibitem[Djorgovski et al.(2003)]{Djorgovski+2003} Djorgovski S.G.,
    Stern D., Mahabal A.A., Brunner R., 2003, ApJ 596, 67

\bibitem[Duras et al.(2020)]{D20} Duras F., Bongiorno A., Ricci F.,
    et al., 2020, A\&A, 636 73
 
\bibitem[Eftekharzadeh et al.(2017)]{Eftekharzadeh+2017}
Eftekharzadeh S., Myers A.D., Hennawi J.F. et al. 2017, MNRAS 468, 77

\bibitem[Eisenstein et al.(2011)]{Eisenstein+2011} Eisenstein D.J.,
  Weinberg D.H., Agol E. et al. 2011, AJ 142, 72

\bibitem[Fan et al.(1999)]{Fan+1999} Fan X., Strauss M.A., Schneider D.P.,
  et al. 1999, AJ 118, 1
  
\bibitem[Fang(1989)]{Fang1989} Fang L.Z., 1989, Int. J. Mod. Phys. A, vol. 4,
  p. 3477 (DOI: 10.1142/S0217751X89001394)

\bibitem[Greiner et al.(2008)]{gbc08}
Greiner J., Bornemann W., Clemens C., et al. 2008, PASP 120, 405

\bibitem[Habouzit et al.(2016a)]{2016MNRAS.456.1901} Habouzit M.,
  Volonteri M., Latif M., et al. 2016a, MNRAS 456, 1901

\bibitem[Habouzit et al.(2016b)]{Habouzit+2016} Habouzit M.,
  Volonteri M., Latif M., Dubois Y., Peirani S., 2016b, MNRAS 463, 529 

\bibitem[Habouzit et al.(2019)]{Habouzit+2019} Habouzit M.,
  Volonteri M., Somerville R.S., et al. 2019, MNRAS 489, 1206
  
\bibitem[Habouzit et al.(2021)]{H21} Habouzit M., Li Y., Somerville R.S., et al., 2021, MNRAS 503, 1940

\bibitem[Haiman \& Hui(2001)]{Haiman+2001} Haiman Z., Hui L., 2001, ApJ 547, 27

\bibitem[Haehnelt \& Nusser(1999)]{HaehneltNusser1999} Haehnelt M.,
  Nusser A., 1999, in ``Evolution of large scale structure: from recombination
  to Garching, Proc. MPA-ESO cosmology conf., August 1998, eds.
  A.J. Banday et al., p. 375

\bibitem[Hennawi et al.(2006)]{Hennawi+2006} Hennawi J.F., Strauss M.A.,
Oguri M. et al. 2006, AJ 131, 1 

\bibitem[Hennawi et al.(2010)]{Hennawi+2010} Hennawi J.F., Myers A.D., 
Shen Y. et al. 2010, ApJ 719, 1672

\bibitem[Henriques et al.(2020)]{He20} Henriques B.M.B., Yates R.M.,
  Fu J. et al., 2020, MNRAS 491, 5795

\bibitem[Hopkins et al.(2007)]{Hopkins+2007} Hopkins P.F., Lidz A.,
  Hernquist L. et al. 2007, ApJ 662, 110

\bibitem[Husband et al.(2013)]{Husband+2013} Husband K., Bremer M.N.,
  Stanway E.R et al. 2013, MNRAS 432, 2869
  
\bibitem[Ilbert et al.(2006)]{Ilbert+2006} Ilbert O., Arnouts S.,
  McCracken H.J. et al. 2006, A\&A 457, 841

\bibitem[Inayoshi et al.(2020)]{Inayoshi+2020} Inayoshi K., Visbal E.,
  Haiman Z., 2020, ARAA 58, 27
  
\bibitem[Johnson et al.(2013)]{Johnson+2013} Johnson J.L., Whalen D.J.,
  Li H., Holz D.E., 2013, ApJ 771, 116
  
\bibitem[Jiang et al.(2009)]{Jiang+2009} Jiang L., Fan X., Bian F. 
et al. 2009, AJ 138, 305
  
\bibitem[Jiang et al.(2016)]{Jiang+2016} Jiang L., McGreer I.D., Fan X. 
et al. 2016, ApJ 833, 222

\bibitem[Kakazu et al.(2010)]{Kakazu+2010} Kakazu Y., Hu E.M., Liu M.C.
  et al. 2010, ApJ 723, 184

\bibitem[Kashikawa et al.(2007)]{Kashikawa+2007} Kashikawa N., Kitayama T.,
  Doi M. et al., 2007, ApJ 663, 765

\bibitem[Katz et al.(2015)]{Katz+2015} Katz H., Sijacki D., Haehnelt M.G.,
  2015, MNRAS 451, 2352

\bibitem[Kellogg(2017)]{Kellogg2017} Kellogg K., 2017, Electronic Thesis
  and Dissertation Repository, 4857 (https://ir.lib.uwo.ca/etd/4857)

\bibitem[Kim et al.(2009)]{Kim+2009} Kim S., Stiavelli M., Trenti M.
  et al. 2009, ApJ 695, 809
  
\bibitem[Kim et al.(2015)]{Kim+2015} Kim Y., Im M., Jeon Y. et al. 2015,
  ApJ 813, L35

\bibitem[Kr\"uhler et al.(2008)]{kkg08} Kr\"uhler T., K\"{u}pc\"{u} Yolda\c{s}
 A., Greiner J.,   et al. 2008, ApJ 685, 376

\bibitem[K\"{u}pc\"{u} Yolda\c{s} et al.(2008)]{kkg08b} K\"{u}pc\"{u} 
Yolda\c{s} A., Kr\"uhler T., Greiner J., 
 et al. 2008, AIP Conf. Proc., 1000, 227

\bibitem[Krumpe et al.(2012)]{Krumpe+2012} Krumpe M., Miyaji T., 
Coil A.L., Aceves H., 2012, ApJ 746, 1

\bibitem[Latif \& Volonteri(2015)]{2015MNRAS.452.1026L} Latif M.A.,
  Volonteri M., 2015, MNRAS 452, 1026

\bibitem[Lesniewska \& Michalowski(2019)]{LesnMichal2019}
  Le\'sniewska A., Michalowski M.J., 2019, A\&A 624, L13
  
\bibitem[Lodato \& Narayan(2006)]{2006MNRAS.371.1813L} Lodato G.,
  Narayan P., 2006, MNRAS 371, 1813

\bibitem[Loeb \& Rasio(1994)]{1994ApJ...432...52L} Loeb A., Rasio F.A.,
  1994, ApJ 432, 52

\bibitem[Lupi et al.(2019)]{Lupi+19} Lupi A., Volonteri M., Decarli R., 
  et al. 2019, MNRAS, 488, 4004

\bibitem[Madau \& Rees(2001)]{MadauRees2001} Madau P., Rees M.J., 2001,
  ApJ 551, L27

\bibitem[Madau et al.(1999)]{Madau+1999} Madau P., Haardt F., Rees M.J.,
  1999, ApJ 514, 648

\bibitem[Mainzer et al.(2011)]{Mainzer+2011} Mainzer A., Bauer J., Grav T.,
  et al. 2011, ApJ 731, 53
  
\bibitem[Marshall et al.(2020)]{M20} Marshall M. A., Ni Y., Di Matteo T. et al., 2020, MNRAS 499, 3891

\bibitem[Marinacci et al.(2018)]{Marinacci+18} Marinacci F., Vogelsberger M., Pakmor R. et al., 2018, MNRAS 480, 5113

\bibitem[Marocco et al.(2021)]{Marocco+2021} Marocco F., Eisenhardt P.R.M.,
  Fowler J.W. et al. 2021, ApJS 253, 8
  
\bibitem[Matsuoka et al.(2016)]{Matsuoka+2016} Matsuoka Y., Onoue M.,
  Kashikawa N. et al. 2016, ApJ 828, 26

\bibitem[Matsuoka et al.(2018)]{Matsuoka+2018} Matsuoka Y., Strauss M.A.,
  Kashikawa N. et al. 2018, ApJ 869, 150

\bibitem[Mazzucchelli et al.(2017a)]{Mazzucchelli+2016} Mazzucchelli C., 
    Ba\~nados E., Decarli R. et al. 2017a, ApJ 834, 83
    
\bibitem[Mazzucchelli et al.(2017b)]{Mazzucchelli+2017} Mazzucchelli C., 
Ba\~nados E., Venemans B.P. et al. 2017b, ApJ 849, 91

\bibitem[McGreer et al.(2016)]{McGreer+2016} McGreer I.D., Eftekharzadeh S., 
Myers A.D., Fan X., 2016, AJ 151, 61

\bibitem[Naiman et al.(2018)]{Naiman+18} Naiman J. P., Pillepich A., Springel V. et al. 2018, MNRAS 477, 1206

\bibitem[Neeleman et al.(2019)]{Neeleman+2019} Neeleman M., Ba\~nados E.,
  Walter F. et al. 2019, ApJ 882, 10
	
\bibitem[Nelson et al.(2018)]{Nelson+18} Nelson D., Pillepich A., Springel V. et al. 2018, MNRAS 475, 624

\bibitem[Omukai et al.(2008)]{2008ApJ...686..801O} Omukai K., Schneider R.,
  Haiman Z., 2008, ApJ 686, 801
  
\bibitem[Pillepich et al.(2018)]{P18} Pillepich A., Springel V., Nelson D.
  et al., 2018, MNRAS 473, 4077

\bibitem[Planck Collaboration(2016)]{Planck2015} Planck Collaboration XIII,
  2016, A\&A 594, A13
  
\bibitem[Porciani et al.(2007)]{Porciani+2004} Porciani C., 
Magliocchetti M., Norberg P., 2007, MNRAS 355, 1010

\bibitem[Reed et a.(2015)]{Reed+2015} Reed S.L., McMahon R.G., Banerji M.
  et al. 2015, MNRAS 454, 3952

\bibitem[Regan \& Haehnelt(2009)]{2009MNRAS.393..858R} Regan J.A.,
  Haehnelt M.G., 2009, MNRAS 393, 858

\bibitem[Ross et al.(2013)]{Ross+2013} Ross N.P., McGreer I.D., White M.
  et al., 2013, ApJ 773, 14

\bibitem[Ryan \& Reid(2016)]{RyanReid2016} Ryan R.E., Reid I.N., 2016,
  AJ 151, 92

\bibitem[Schlafly \& Finkbeiner(2011)]{Schlafly+11} 
   Schlafly E.F., Finkbeiner D.P., 2011, ApJ 737, 103

\bibitem[Schlafly et al.(2019)]{Schlafly+2019} Schlafly E.F.,
   Meisner A.M., Green G.M., 2019, ApJS 240, 30
   
\bibitem[Schneider et al.(2000)]{Schneider+2000} Schneider D.P.,
Fan X., Strauss M.A. et al. 2000, AJ 120, 2183

\bibitem[Scolnic et al.(2015)]{Scolnic+2015} Scolnic D., Casertano S.,
  Riess A., et al. 2015, ApJ 815, 117

\bibitem[Shen et al.(2007)]{Shen+2007} Shen Y., Strauss M.A., Oguri M. et al.
2007, AJ 133, 2222

\bibitem[Shen et al.(2010)]{Shen+2010} Shen Y., Hennawi J.F., Shankar F.
   et al. 2010, ApJ 719, 1693

\bibitem[Shen et al.(2020)]{Shen+2020} Shen X., Hopkins P.F.,
  Faucher-Gigue\`re C.-A., et al. 2020, MNRAS 495, 3252 
  
\bibitem[Sheth et al.(2001)]{Sheth+2001} Sheth R.K., Mo H.J., Tormen G., 2001,
MNRAS 323, 1

\bibitem[Silverman et al.(2020)]{S20} Silverman J.D., Tang S., Lee K.-G., 
et al. 2020, ApJ 899, 154

\bibitem[Skrutskie et al.(2006)]{Skrutskie+2006} Skrutskie M.F., Cutri R.M., Stiening R. et al. 2006, AJ 131, 1163.

\bibitem[Songaila(2004)]{Songaila+2004} Songaila A., 2004, AJ 127, 2598

\bibitem[Spaans \& Silk(2006)]{2006ApJ...652..902} Spaans M., Silk J., 2006,
  ApJ 652, 902

\bibitem[Springel et al.(2005)]{S05} Springel V., White S.D.M., Jenkins
 A. et al., 2005, Nature 435, 629

\bibitem[Springel et al.(2018)]{Springel+18} Springel V., Pakmor R., Pillepich A. et al., 2018, MNRAS 475, 676
  
\bibitem[Tenneti et al.(2019)]{T19} Tenneti A., Wilkins S.M., Di Matteo T. et al., 2019, MNRAS 483, 1388
  
\bibitem[Tody(1993)]{Tody1993} Tody, D. 1993, ASP Conf. Ser. 52, 
 Astronomical Data Analysis Software and Systems II, ed. R.J. Hanisch,
 R.J.V. Brissenden, \& J. Barnes, p. 173

\bibitem[Utsumi et al.(2010)]{Utsumi+2010} Utsumi Y., Goto T., Kashikawa N.,
  et al. 2010, ApJ 721, 1680
 
\bibitem[Valiante et al.(2017)]{Valiante+2017} Valiante R., Agrarval B.,
  Habouzit M., Pezzulli E., 2017, PASA 34, e031

\bibitem[Valiante et al.(2018)]{Valiante+2018} Valiante R., Schneider R.,
  Graziani L., Zappacosta L., 2018, MNRAS 474, 3825
  
\bibitem[Venemans et al.(2013)]{Venemans+2013} Venemans B.P., Findlay J.R.,
  Sutherland W.J. et al. 2013, ApJ 779, 24

\bibitem[Venemans et al.(2015)]{Venemans+2015} Venemans B.P.,
  Verdoes Kleijn G.A., Mwebaze J. et al. 2015, MNRAS 453, 2259

\bibitem[Visbal et al.(2014)]{Visbal+2014} Visbal E., Haiman Z.,
  Bryan G.L., 2014, MNRAS  445, 1056

\bibitem[Vito et al.(2021)]{Vito+2021} Vito F., Brandt W.N., Ricci F.
  et al. 2021, A\&A 649, A133
  
\bibitem[Volonteri(2010)]{Volonteri2010} Volonteri M., 2010,
  A\&ARv 18, 279
  
\bibitem[Volonteri et al.(2003a)]{Volonteri+2003} Volonteri M., Haardt F.,
  Madau P., 2003a, ApJ 582, 559

\bibitem[Volonteri et al.(2003b)]{Volonteri+2003b} Volonteri M., Madau P.,
    Haardt F.. 2003b, ApJ 594, 661
  
\bibitem[Volonteri et al.(2016)]{Volonteri+2016} Volonteri M., Dubois Y.,
  Pichon C., Devriendt J., 2016, MNRAS 460, 2979

\bibitem[Wang et al.(2021)]{Wang+2021} Wang F., Yang J., Fan X. et al.
  2021, ApJ 907, L1
  
\bibitem[Willott et al.(2007)]{Willott+07} Willott C.J., Delorme P., Omont A.
  et al. 2007, AJ 134, 2435

\bibitem[Willott et al.(2009)]{Willott+09} Willott C.J., Delorme P., Reyle C.
  et al. 2007, AJ 137, 3541
  
\bibitem[Willott et al.(2010)]{Willot2010} Willott C.J., Albert L.,
  Arzoumanian D. et al. 2010, AJ 140, 546

\bibitem[Wolf et al.(2018)]{Wolf+2018} Wolf C., Onken C.A., Luvaul L.C.
  et al. 2018, PASA 35, 10

\bibitem[Wright et al.(2010)]{Wright+2010} Wright E.L., Eisenhardt P.R.M.,
  Mainzer A.K. et al., 2010, AJ 140, 1868

\bibitem[Wyithe et al.(2005)]{Wyithe+2005} Wyithe J.S.B., Loeb A., Carilli C.,
  2005, ApJ 628, 575
  
\bibitem[Yang et al.(2020)]{Yang+2020} Yang J., Wang F., Fan X. et al.
  2020, ApJ 897, L14
  
\bibitem[Yue et al.(2014)]{Yue2014} Yue B., Ferrara A., Salvaterra R., et al.
  2014, MNRAS 440, 1263
  
\bibitem[Zeimann et al.(2011)]{Zeimann+2011} Zeimann G.R., White R.L.,
  Becker R.H. et al. 2011, ApJ 736, 57
  
\end{thebibliography}
\end{document}